\documentclass[11pt,a4paper]{article}
\pdfoutput=1
\usepackage[utf8]{inputenc}
\usepackage[a4paper, total={16.5cm, 23cm}]{geometry}
\usepackage{graphicx}
\usepackage{cancel}
\usepackage{textcomp}
\usepackage{cite}
\usepackage{bm}
\usepackage{times}
\usepackage{color}
\usepackage{hyperref}
\usepackage{setspace}
\usepackage{paralist}
\usepackage{multirow}
\usepackage{soul}
\usepackage{bbm}
\usepackage{enumitem}

\usepackage[dvipsnames]{xcolor}
\usepackage{slashed}
\usepackage{comment}
\usepackage{epstopdf}
\usepackage{array}
\usepackage{tikz}

\usepackage{rotating}
\usepackage{framed}

\usepackage{placeins}
\usepackage{amsmath} 
\usepackage{amsthm} 
\usepackage{amssymb}	
\usepackage{pifont}
\usepackage{url}

\definecolor{Darkgreen}{RGB}{30,120,30}

\newcommand{\myComment}[1]{}

\newcommand{\abs}[1]{\left| #1 \right|} 
 
\newcommand{\ket}[1]{\left| #1 \right>} 
\newcommand{\bra}[1]{\left< #1 \right|} 

\definecolor{Darkgreen}{RGB}{30,120,30}

\let\baraccent=\= 
\renewcommand{\=}[1]{\stackrel{#1}{=}} 
\newcommand{\xmark}{\ding{55}}%
\newcommand{\cmark}{\ding{51}}%

\theoremstyle{definition}

\theoremstyle{remark}

\def\beqn{\begin{eqnarray}}
\def\eeqn{\end{eqnarray}}
\def\no{\nonumber}
\def\LR{{L \atop R}}
\def\dsp{\displaystyle}
\def\bc{\begin{center}}
\def\ec{\end{center}}
\def\be{\begin{equation}}
\def\ee{\end{equation}}
\def\bea{\begin{eqnarray}}
\def\eea{\end{eqnarray}} 

\def\gev{\ensuremath{\mathrm{Ge\kern -0.1em V}}}
\def \Re{\mathcal{R}\mathrm{e}}                   
\def \Im{\mathcal{I}\mathrm{m}}                   

\def \azeL{{{\cal A}_0^L}}
\def \azeR{{{\cal A}_0^R}}
\def \apaL{{{\cal A}_\parallel^L}}
\def \apaR{{{\cal A}_\parallel^R}}
\def \apeL{{{\cal A}_\perp^L}}
\def \apeR{{{\cal A}_\perp^R}}
\def \azeTL{{{\cal A}_{T\,0}^L}}
\def \azeTR{{{\cal A}_{T\,0}^R}}
\def \apaTL{{{\cal A}_{T\,\|}^L}}
\def \apaTR{{{\cal A}_{T\,\|}^R}}
\def \apeTL{{{\cal A}_{T\,\perp}^L}}
\def \apeTR{{{\cal A}_{T\,\perp}^R}}

\def\thD{{\ensuremath{\theta_D}}}
\def\thl{{\ensuremath{\theta_\tau}}}

\def\rd {\mathcal{R}_{D}}
\def\rdst {\mathcal{R}_{D^*}}
\def\rdrdst {\mathcal{R}_{D^{(*)}}}
\def\rJpsi{\mathcal{R}_{J/\psi}}
\def\ptau{\mathcal{P}_{\tau}^{D^*}}
\def\ptauint{\bar{\mathcal{P}}_{\tau}^{D^*}}
\def\FLint{\bar{F}_{L}^{D^*}}

\hypersetup{
    pdfnewwindow=true,      
    colorlinks=true,       
    linkcolor=black,          
    citecolor=blue,        
    filecolor=blue,      
    urlcolor=blue           
}

\makeatletter
\newcommand\xleftrightarrow[2][]{%
  \ext@arrow 9999{\longleftrightarrowfill@}{#1}{#2}}
\newcommand\longleftrightarrowfill@{%
  \arrowfill@\leftarrow\relbar\rightarrow}
\makeatother

\begin{document}

\thispagestyle{empty}
\hfill\mbox{IFIC/20-14, \, FTUV/20-1404, \, SI-HEP-2020-10 } \\
\vspace{3cm}

\begin{center}

{\huge\sc 
The role of right-handed neutrinos\\[10pt] in $b \to c \tau \bar{\nu}$ anomalies
}

\vspace{1cm}

{\sc Rusa Mandal${}^a$, Clara Murgui${}^b$, Ana Peñuelas${}^b$ and Antonio Pich${}^b$
}

\vspace*{.7cm}

{\sl
${}^a$ Theoretische Physik 1, Naturwissenschaftlich-Technische Fakultät,
\\ Universität Siegen,  57068 Siegen, Germany

${}^b$
Departament de F\'\i sica Te\`orica, IFIC, Universitat de Val\`encia -- CSIC,\\
Parque Científico,  Catedrático José Beltrán 2, E-46980 Paterna, Spain
}

\end{center}

\date{\vspace{-5ex}}

\begin{abstract}

Motivated by the persistent anomalies reported in the $b\to c\tau\bar{\nu}$ data, we perform a general model-independent analysis of these transitions, in the presence of light right-handed neutrinos. We adopt an effective field theory approach and write a  low-energy effective Hamiltonian, including all possible dimension-six operators. The corresponding Wilson coefficients are determined through a numerical fit to all available experimental data.
In order to work with a manageable set of free parameters, we define eleven well-motivated scenarios, characterized by the different types of new physics that could mediate these transitions, and analyse which options seem to be preferred by the current measurements. The data exhibit a clear preference for new-physics contributions, and good fits to the data are obtained in several cases. However, the current measurement of the longitudinal $D^*$ polarization in $B\to D^*\tau \bar\nu$ cannot be easily accommodated within its experimental $1\sigma$ range. A general analysis of the three-body $B\to D \tau \bar\nu$ and four-body $B\to D^*(\to D\pi)\tau \bar\nu$ angular distributions is also presented. The accessible angular observables are studied in order to assess their sensitivity to the different new physics scenarios. Experimental information on these distributions would help to disentangle the dynamical origin of the current anomalies.

\end{abstract}

\newpage

\newpage

\section{Introduction}
\label{sec:Intro}

Intriguing hints of discrepancies between the 
measured data
and the Standard Model (SM) predictions have been observed in $B$ decays by several experimental collaborations \cite{Pich:2019pzg,Bifani:2018zmi}. Such observations can be regarded as indirect evidence of physics beyond the SM
and thus have drawn immense attention by the scientific community in the last few years. Among these decays, the $b \to c \tau \bar{\nu} $ modes are of special interest. 
In spite of being a semileptonic charged-current channel, which proceeds at tree-level in the SM, three different experiments have reported sizeable tensions
in the ratios of branching fractions ($\mathcal{B}$) \cite{Lees:2012xj,Lees:2013uzd,Aaij:2015yra,Huschle:2015rga,Hirose:2016wfn,Hirose:2017dxl,Aaij:2017uff,Aaij:2017deq,Abdesselam:2019dgh}
\bea
\label{eq:RD}
\rdrdst \equiv \frac{ {\mathcal B}(B\to D^{(*)}\tau \bar{\nu})}{ {\mathcal B}(B\to D^{(*)}\ell \, \bar{\nu})}\,,
\eea
with $\ell=e$ or $\mu$, and 
\bea
\label{eq:defrjpsi}
\rJpsi \equiv \frac{ {\mathcal B}(B_c\to J/\psi\, \tau \bar{\nu})}{ {\mathcal B}(B_c\to J/\psi\, \mu \, \bar{\nu})}\,.
\eea
measured in Ref.~\cite{Aaij:2017tyk}.
These ratios are particularly clean probes of New Physics (NP) due to the cancellation of the leading uncertainties inherent in individual $\mathcal{B}$ predictions.

The latest world averages of $\rdrdst$ measurements,
performed by the Heavy Flavour Averaging Group (HFLAV)~\cite{Amhis:2019ckw}, 
\bea
\rd^{\rm ave}=	0.340 \pm 0.027 \pm 0.013
\qquad \text{and}\qquad
\rdst^{\rm ave}	=0.295 \pm 0.011 \pm 0.008\, ,
\eea
deviate at the $3.1\sigma$ level (considering their correlation of $-0.38$) from the arithmetic average of SM predictions~\cite{Bigi:2016mdz,Bernlochner:2017jka,Bigi:2017jbd,Jaiswal:2017rve} quoted
by HFLAV: $\rd^{\rm SM}= 0.299 \pm 0.003$ ($1.4\sigma$) and \ $\rdst^{\rm SM}	= 0.258 \pm 0.005$ ($2.5\sigma$). 
Using more updated form factors (FFs) \cite{Jung:2018lfu}, we get
\begin{equation}
\rd^{\rm SM}=  0.302 \pm 0.004  \qquad \text{and} \qquad \rdst^{\rm SM}= 0.258\, {}^{+\, 0.006}_{-\,0.005}\, ,  
\end{equation}
which slightly increases the tension to $3.2\sigma$.
The measured ratio $\rJpsi = 0.71 \pm 0.17 \pm 0.18$~\cite{Aaij:2017tyk} is also $1.7\sigma$ larger than its SM prediction,  $\rJpsi^{\rm SM} \approx  0.25-0.28$~\cite{Anisimov:1998uk,Kiselev:2002vz,Ivanov:2006ni,Hernandez:2006gt,Huang:2007kb,Issadykov:2018myx,Wang:2008xt,Hu:2019qcn,Leljak:2019eyw,Azizi:2019aaf,Tran:2018kuv}.
Moreover, the recent measurement of the longitudinal polarization of the $D^{*-}$ meson in $B^0\to D^{*-}\tau^+ \bar{\nu}$,  $\bar{F}_L^{D^*} = 0.60\pm 0.08\pm 0.04$ , differs also from its SM value by $1.6\sigma$ \cite{Abdesselam:2019wbt}.

These experimental facts suggest a surprisingly large violation of lepton-flavour universality, and have triggered a large number of detailed phenomenological studies trying to determine the most plausible NP explanation. A quite complete list of relevant references can be found in Ref.~\cite{Murgui:2019czp}, where an exhaustive analysis of all available data has been accomplished with a model-independent effective field theory (EFT) approach, assuming only the SM particle content and symmetries in order to define the basis of allowed low-energy operators. A global fit to all data, with a good statistical quality, has been obtained in terms of the four possible Wilson coefficients; however, the fit does not allow to clearly identify a potential mediator of the underlying NP interaction~\cite{Murgui:2019czp}. Moreover, the experimental value of $\bar{F}_L^{D^*}$ cannot be accommodated within $1\sigma$~\cite{Murgui:2019czp}.

Light right-handed neutrinos (RHNs) 
have been suggested \cite{Ligeti:2016npd,Asadi:2018wea,Greljo:2018ogz,Robinson:2018gza,Azatov:2018kzb,Heeck:2018ntp,Asadi:2018sym,Babu:2018vrl,Bardhan:2019ljo,Shi:2019gxi,Gomez:2019xfw, Dutta:2017xmj,Dutta:2017wpq, Dutta:2013qaa, Dutta:2016eml, Becirevic:2016yqi, Cline:2015lqp} 
as a possibility to evade the current phenomenological constraints on the EFT operators containing 
left-handed neutrino (LHN) fields. Sterile neutrinos are singlets under the SM gauge group and, therefore, their properties are not linked to any charged electroweak partners. Moreover, the existing limits from the neutrino sector do not constrain significantly the scale of $\nu_R$ operators beyond what is probed in $b\to c \tau \bar{\nu}$ transitions. In order not to disrupt the measured $B\to D^{(*)} \tau \bar{\nu}$ invariant-mass distributions \cite{Lees:2013uzd,Huschle:2015rga}, one just needs to assume the $\nu_R$ fields to be light, $m_{\nu_R}\lesssim O(100)$~MeV, which also helps to avoid other cosmological and astrophysical limits. Neglecting neutrino masses, there is no interference between the two neutrino chiralities, and the decay probability becomes an incoherent sum of $\nu_L$ and $\nu_R$ contributions:
$\mathcal{B}(b\to c  \tau \bar{\nu}) = \mathcal{B}(b\to c  \tau \bar{\nu}_L) + \mathcal{B}(b\to c  \tau \bar{\nu}_R)$. Therefore, it is not difficult to increase the predicted rates towards the experimentally favoured range. However, a large $\nu_R$ contribution requires the corresponding Wilson coefficients to be large, of the order of the SM $\nu_L$ interaction, because the rates are quadratic in the $\nu_R$ transition amplitude.

 Previous works considering RHNs in $B \to D^{(*)} \tau \bar{\nu}$ decays~\cite{Ligeti:2016npd,Asadi:2018wea,Greljo:2018ogz,Robinson:2018gza,Azatov:2018kzb,Heeck:2018ntp,Asadi:2018sym,Babu:2018vrl,Bardhan:2019ljo,Shi:2019gxi,Gomez:2019xfw, Dutta:2017xmj,Dutta:2017wpq, Dutta:2013qaa, Dutta:2016eml, Becirevic:2016yqi, Cline:2015lqp} have focused on reproducing the integrated rates, most of them within particular scenarios of NP. All phenomenological analyses need to rely on the underlying assumption that the differential decay distributions, and hence the experimental acceptances, are not significantly modified by the NP contributions. While this assumption is unavoidable, in the absence of direct access to the data, none of the previous studies have included the measured $q^2$ distributions in their fits. This shape information has been shown to play an important role, discarding many proposed solutions with $\nu_L$ fields~\cite{Murgui:2019czp,Sakaki:2014sea,Freytsis:2015qca,Bhattacharya:2016zcw,Celis:2016azn}, and could be expected to be even more relevant for those solutions based on RHNs, since they induce distortions in the rates that are quadratic in NP contributions.

We aim to improve the situation in this paper, by extending the EFT analysis of Ref.~\cite{Murgui:2019czp} to a basis of dimension-six operators that includes light RHNs. In our fit procedure, we consider all observables measured for $B \to D^{(*)} \tau \bar{\nu}$ decays until date; including the data for binned differential distributions with respect to the lepton-neutrino invariant-mass squared, the $D^*$ longitudinal polarization fraction $\bar{F}_L^{D^*}$, the lepton polarization asymmetry $\ptauint{}$ and the experimental results for $\rdrdst$. The last ratios have been recently altered, reducing the tension with the SM and making a fresh re-analysis necessary. We 
also study the differential three-body $B\to D \tau \bar\nu$ decay distribution and
derive the four-body angular distribution of the $B \to D^{*}(\to D \pi) \tau \bar{\nu}$ decay for the most general dimension-six Hamiltonian. By identifying the possible high-scale NP mediators which can generate the operators involving RHNs, we predict several angular observables that can be tested at the experiment.

 The rest of the paper is organized as follows. In Section~\ref{sec:theory} the most general effective Hamiltonian for our analysis is described, and expressions of the relevant observables are written in terms of the Wilson coefficients.  In Section~\ref{sec:interp} the experimental status of the $b \to c$ transitions is interpreted from an EFT approach, by looking at the effect that individual Wilson coefficients may produce in the relevant observables. In addition, all possible NP mediators that can effectively generate a $b \to c  \tau \bar{\nu}_R$  transition, and the corresponding Wilson coefficients that will arise at low energies after their integration, are listed. 
 In Section~\ref{sec:fits} the  results of our fits are presented and discussed. We consider
 different scenarios, originated by the integration of the relevant NP mediators, and compare their fitted results with the SM case.  Section~\ref{sec:predctions} contains the predicted angular coefficients of the $B\to D \tau \bar\nu$ and $B \to D^{*}(\to D \pi) \tau \bar{\nu}$ distributions for the best fit scenarios, including the forward-backward asymmetries ${\cal A}_{FB}^{D^{(*)}}$, the $\tau$ polarization asymmetries ${\cal P}_{\tau}^{D^{(*)}}$,
and the integrated longitudinal polarization fraction $F_L^{D^*}$.
Finally, conclusions are exposed in Section~\ref{sec:conclu}.
Many technical details, such as hadronic matrix elements, FFs, and the full set of relevant helicity amplitudes, are compiled in several appendices.

\section{Theoretical framework and observables}
\label{sec:theory}

\subsection{Effective field theory}

Including RHN fields, the
most general dimension-six effective
Hamiltonian relevant for $ b \to c \tau \bar{\nu} $ transitions can be written, 
at the bottom quark-mass scale, as
\begin{equation}
\label{eq:Lag}
{\cal H}_\text{eff}\, =\, \frac{ 4 G_F V_{cb}}{\sqrt{2}}\left( {\cal O}_{LL}^V + \sum_{\substack{X=S,V,T \\ 
A,B=L,R}} C_{AB}^X\; {\cal O}_{AB}^X \right),
\end{equation}
with the ten four-fermion operators:
\begin{eqnarray}\label{eq:def-operators}
{\cal O}^V_{AB} &\equiv & \left(\bar{c}\, \gamma^\mu P_A b\right)\left(\bar{\tau} \gamma_\mu P_B \nu\right)\, ,
\no\\
{\cal O}^{S}_{AB} &\equiv & \left(\bar{c}\, P_A b\right)\left(\bar{\tau} P_B \nu\right)\, ,
\no\\
{\cal O}^T_{AB} & \equiv & \delta_{AB}\;\left(\bar{c}\,\sigma^{\mu \nu} P_A b\right)\left(\bar{\tau} \sigma_{\mu \nu} P_A \nu\right)\, ,
\end{eqnarray}
which are invariant under $SU(3)_C\otimes U(1)_{\mathrm{em}}$. Tensor operators with different lepton and quark chiralities vanish identically.\footnote{This is a direct consequence of the Dirac-algebra identity 
$\gamma_5\sigma^{\mu\nu} = -\frac{i}{2}\,\varepsilon^{\mu\nu\alpha\beta}\sigma_{\alpha\beta}$, which implies
$\sigma_{\mu\nu}\otimes \sigma^{\mu\nu}\gamma_5 = \sigma_{\mu\nu}\gamma_5\otimes \sigma^{\mu\nu}$ and 
$\sigma_{\mu\nu}\gamma_5\otimes \sigma^{\mu\nu}\gamma_5 = \sigma_{\mu\nu}\otimes \sigma^{\mu\nu}$.
We use the convention $\varepsilon^{0123}=-\varepsilon_{0123}=-1$. 
}
The SM charged-current contribution to $\mathcal{O}^V_{LL}$, from a $W_\mu$ exchange, has been explicitly added to Eq.~\eqref{eq:Lag}, so that $C_{AB}^X=0$ in the SM. Any non-zero contribution to these Wilson coefficients is then a manifestation of NP beyond the SM.
We are assuming that NP contributions are only present in operators involving charged leptons of the third generation. This is well justified, since potential NP effects have been shown to be negligible in $ b \to c  \ell  \bar{\nu}$ transitions~\cite{Jung:2018lfu}.

In the subsequent sections, we present the analytic expressions of observables constructed from different decay modes containing the $b\to c  \tau \bar{\nu}$ quark-level transition. 
The presence of RHN only modifies the leptonic currents; therefore, the decay amplitudes can be given in terms of the same hadronic FFs
used for $\nu_L$ operators, that were already listed in Ref.~\cite{Murgui:2019czp}. 
For completeness, we compile them again in the Appendix~\ref{app:FF}.
We also list in Appendix~\ref{app:HelAmp} the whole set of relevant helicity amplitudes, which we have obtained following the standard helicity formalism for semileptonic $B$ decays \cite{Hagiwara:1989cu,Hagiwara:1989zt,Korner:1989qb}. We have checked that our expressions reproduce the available results for LHNs, and that they satisfy the correct parity relations between the $\nu_L$ and $\nu_R$ transition amplitudes.

\subsection{Observables}

The relevant observables for our analysis can be classified into those involving $B \to D$  and $B \to D^*$ semileptonic transitions. We also consider the leptonic decay $B_c  \to \tau \bar{\nu}$ since it can constrain certain Wilson coefficients.

\subsubsection[\texorpdfstring{$B \to D  \, \tau \bar{\nu}$}{dum}]{\texorpdfstring{\boldmath $B \to D \tau \bar{\nu}$}{dum}}

The differential distribution of the decay $B \to D \tau \bar{\nu}$ can be written as
\be
\label{eq:Ddistang}
\frac{d\Gamma (B \to D \tau \bar{\nu})}{dq^2 \, d\cos{\theta_\tau}}
\; =\;  \frac{G_F^2 V_{cb}^2}{256 \, m_B^3  \pi^3 }\;  q^2 \, \lambda_D^{1/2} (q^2)
\left(1-\frac{m_\tau^2}{q^2}\right)^2\left\{
J_0(q^2) + J_1(q^2)\,\cos{\theta_\tau} +J_2(q^2)\,\cos^2{\theta_\tau}
\right\} ,
\ee 
where $q^2~\equiv~(p_\tau+p_{\bar{\nu}})^2$, $\theta_\tau$ is the polar angle of the $\tau$ momentum
in the rest frame of the $\tau \bar{\nu}$ pair, with respect to the 
$z$-axis defined by the momentum of the $D$  meson in the $B$ rest frame, and we have introduced the shorthand notation for the K\"{a}llen function 
\be 
\lambda_{D^{(*)}}(q^2) \equiv \lambda(m_B^2,m_{D^{(*)}}^2, q^2)=m_B^4+ m_{D^{(*)}}^4+ q^4 - 2 m_B^2m_{D^{(*)}}^2 - 2 m_{D^{(*)}}^2 q^2- 2 m_B^2 \, q^2\, .
\ee

The coefficient functions of the different angular dependences 
are given by
\beqn\label{eq:Jfun}
J_0(q^2) & = & \Big| \mathcal{\tilde A}_0^L - \frac{2 m_\tau }{\sqrt{q^2}}\, 
\mathcal{\tilde A}_T^{L}\Big|^2
+\frac{m_\tau^2}{q^2}\, \Big| \mathcal{\tilde A}_t^L + \frac{\sqrt{q^2}}{m_\tau}\, 
\mathcal{\tilde A}_S^{L}\Big|^2\, +\, (L \leftrightarrow R)\, ,
\no\\
J_1(q^2) & = & \frac{2 m_\tau^2}{q^2}\;\Re\bigg[\Big( \mathcal{\tilde A}_0^L - \frac{2 \sqrt{q^2}}{m_\tau}\, \mathcal{\tilde A}_T^{L}\Big) \Big( \mathcal{\tilde A}_t^{L*} + \frac{\sqrt{q^2}}{m_\tau}\, \mathcal{\tilde A}_S^{L*}\Big)\bigg]\, +\, (L \leftrightarrow R)\, ,
\no\\
J_2(q^2) & = & -\left( 1- \frac{m_\tau^2}{q^2}\right)\,\left( |\mathcal{\tilde A}_0^L|^2 - 4\, |\mathcal{\tilde A}_T^L|^2\right) \, +\, (L \leftrightarrow R)\, ,
\eeqn
where
\beqn \label{eq:tildeAfun}
\mathcal{\tilde A}_0^L &\!\! = &\!\! \left( 1 + C_{LL}^V + C_{RL}^V\right)\, H_{V,0}^s\, ,
\hskip 2.0cm
\mathcal{\tilde A}_0^R \, =\, \left( C_{LR}^V + C_{RR}^V\right)\, H_{V,0}^s\, ,
\no\\
\mathcal{\tilde A}_t^L &\!\! = &\!\! \left( 1 + C_{LL}^V + C_{RL}^V\right)\, H_{V,t}^s\, ,
\hskip 2.1cm
\mathcal{\tilde A}_t^R \, = \,\left( C_{LR}^V + C_{RR}^V\right)\, H_{V,t}^s\, ,
\no\\
\mathcal{\tilde A}_S^L &\!\! = &\!\! \left(  C_{RL}^S + C_{LL}^S\right)\, H_{S}^s\, ,
\hskip 2.9cm
\mathcal{\tilde A}_S^R \, =\, \left( C_{RR}^S + C_{LR}^S\right)\, H_{S}^s\, ,
\no\\
\mathcal{\tilde A}_T^L &\!\! = &\!\!  2\, C_{LL}^T \, H_{T}^s\, ,
\hskip 4.3cm
\mathcal{\tilde A}_T^R \, =\,   2\, C_{RR}^T\, H_{T}^s\, .
\eeqn
The hadronic helicity amplitudes $H^{s}_{V,0}$, $H^s_{V,t}$, $H^s_S$ and $H^s_T$ are functions of $q^2$, and their explicit expressions are given in the Appendix~\ref{app:FF}. 
The LHN contributions to Eq.~\eqref{eq:Jfun} are in full agreement with Ref.~\cite{Becirevic:2019tpx}.
Notice that the vector and scalar Wilson coefficients only appear in the combinations
$C_{LX}^V+C_{RX}^V$ and $C_{LX}^S + C_{RX}^S$, regardless of the neutrino chirality $X$.

Integrating over $\cos{\theta_\tau}$, one obtains~\cite{Asadi:2018sym}
\beqn
\label{eq:Ddist}
\lefteqn{\frac{d\Gamma}{dq^2}(B \to D \tau \bar{\nu})\; =\;  \frac{G_F^2 V_{cb}^2}{192 m_B^3 \pi^3 }\;  q^2 \, \lambda_D^{1/2} (q^2)
\left(1-\frac{m_\tau^2}{q^2}\right)^2} && 
\nonumber \\
 &\quad\times\!\! & \left\lbrace \Bigl(	 |1+C^V_{LL}+C^V_{RL}|^2 + |C^V_{LR}+C^V_{RR}|^2 	\Bigr) \left[	(H^{s}_{V,0})^2  \left(\frac{m_\tau^2}{2q^2} + 1 \right) + \frac{3 m_\tau^2}{2q^2} (H^s_{V,t})^2	 	\right] \right.  
\nonumber  \\
&&\mbox{} + \frac{3}{2}\, (H^s_S)^2 \Bigl( |C^S_{RL}+C^S_{LL}|^2 +|C^S_{RR}+C^S_{LR}|^2 \Bigr)  +8\, \Bigl( |C^T_{LL}|^2 + |C^T_{RR}|^2\Bigr) (H^s_T)^2 \left(	1+ \frac{2m_\tau^2}{q^2}	\right) \nonumber\\
&&\mbox{} + 3\, \Re \Bigl[ (1+C^V_{LL}+C^V_{RL})\, (C^{S}_{RL}+C^{S}_{LL})^{*} + (C^V_{LR}+C^V_{RR})\, (C^S_{RR}+C^S_{LR})^* \Bigr]\; \frac{m_\tau}{\sqrt{q^2}} H_S^s H^s_{V,t} \nonumber \\
&&\left.\mbox{} -  12\, \Re \Bigl[	(1+C^V_{LL}+C^V_{RL})\, C^{T*}_{LL}	+ (C^V_{RR}+C^V_{LR})\, C^{T*}_{RR}\Bigr]\; \frac{m_\tau}{\sqrt{q^2}} H^s_T H^s_{V,0} \right\rbrace .
\eeqn

The linear term in the $\thl$ distribution given in Eq.~\eqref{eq:Ddistang} can be accessed via the forward-backward asymmetry, traditionally defined as
\begin{equation}
\label{eq:DAFB}
{\cal A}_{FB}^{D}\, =\,\frac{1}{d\Gamma/dq^2}\; \Big[ \int_0^1-\dsp \int_{-1}^0\Big]\, d\cos{\thl}\; 
\frac{d^2 \Gamma}{d q^2d\cos{\thl}}\,=\frac{1}{2} \frac{J_1(q^2)}{J_0(q^2)+ \frac{1}{3}J_3(q^2)}\, ,
\end{equation}
and the $\tau$ polarization asymmetry can be constructed as
\begin{equation}
\label{eq:PtauD}
    \mathcal{P}_\tau^D = \frac{ d\Gamma_{\lambda_{\tau} = 1/2} / dq^2  - d\Gamma_{\lambda_{\tau} = -1/2}/dq^2}{d\Gamma/dq^2}\,,
\end{equation}
where the decomposition of the amplitude in $\tau$ helicity states is given in the Appendix~\ref{app:totamp}.  
\subsubsection[\texorpdfstring{$B \to D^* \tau \bar{\nu}$}{dum2}]{\texorpdfstring{\boldmath $B \to D^* \tau \bar{\nu}$}{dum2}}
\begin{figure}[ht]
    \centering
    \includegraphics[width=0.5\linewidth]{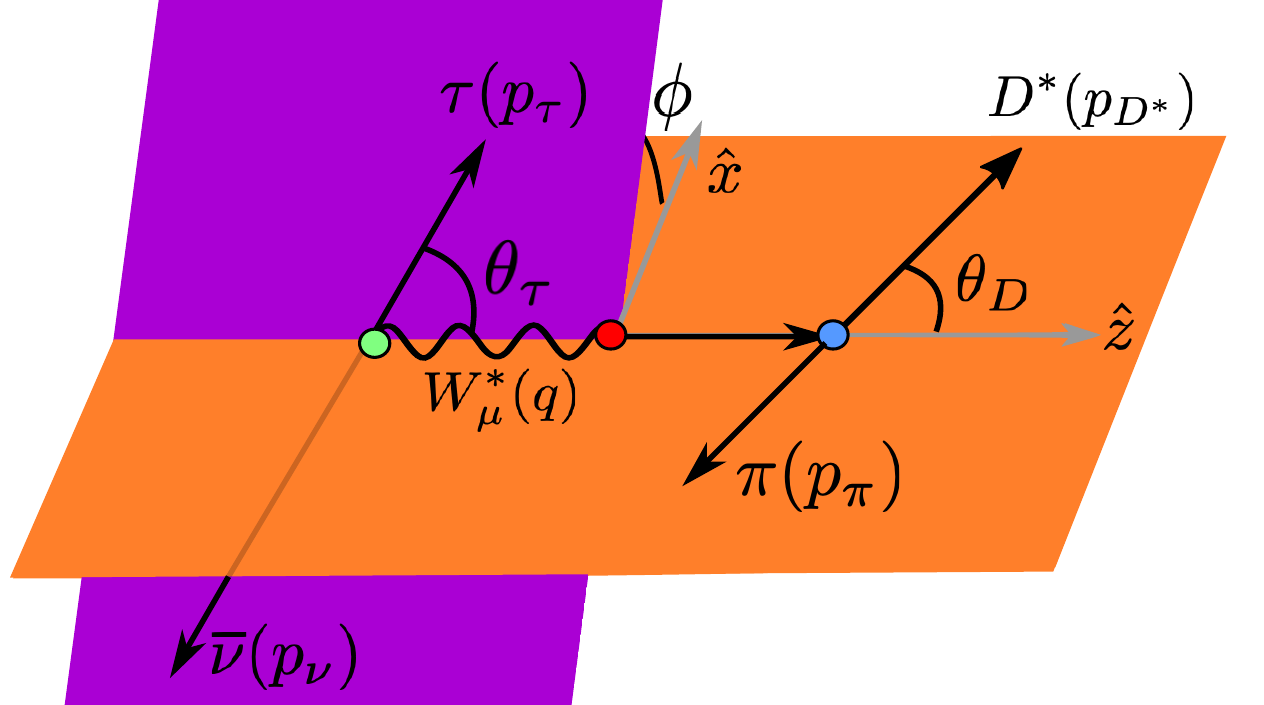}
    \caption{Schematic representation of the kinematical variables for the $B \to D^{*} (\to D \, \pi) \tau \bar{\nu}$ process.}
    \label{fig:kinematics}
\end{figure}

%
The vector meson $D^*$ in the final state provides additional observables compared to the previous case. The angular analysis of a four-body final state,
namely $B \to D^*(\to D \, \pi) \tau \bar{\nu}$, further allows us to construct a multitude of observables that can be extracted from data \cite{Duraisamy:2013pia,Duraisamy:2014sna,Becirevic:2016hea,Alonso:2016gym,Ligeti:2016npd,Becirevic:2019tpx,Hill:2019zja,Aebischer:2019zoe}. 
The differential decay distribution of the transition process $B(p_B)\to D^*(p_{D^*})\,\tau (p_\tau)\,\bar{\nu} (p_{\bar{\nu}})$, with $D^*(p_{D^*})\to D(p_D)\,\pi(p_\pi)$ on the mass shell, 
can be expressed in the form \cite{Becirevic:2019tpx}:
\beqn
\label{eq:dGamma}
\lefteqn{  \frac{d^4\Gamma (B\to D^*\tau \bar{\nu})}{dq^2\, d\cos\thl\,	d\cos\thD\, d\phi}
 \,\equiv\, I(q^2,\thl,\thD,\phi)} &&
 \no\\[3pt]
 &\qquad = \;\dfrac{9}{32\pi}\!\!\!\! & \Big\{ I_1^s\, \sin^2{\thD} + I_1^c\, \cos^2{\thD} 
+ \left( I_2^s\, \sin^2{\thD} + I_2^c\, \cos^2{\thD}\right) \cos{2\thl}
\no \\ &&\mbox{} 
+ \left( I_3\, \cos{2\phi} + I_9\, \sin{2\phi} \right) \sin^2{\thD} \sin^2{\thl}  
+ \left( I_4\, \cos{\phi} + I_8\, \sin{\phi} \right) \sin{2\thD} \sin{2\thl}
\no\\ &&\mbox{} 
+ \left( I_5\,\cos{\phi} + I_7\,\sin{\phi}  \right) \sin{2\thD} \sin{\thl}  
+ \left( I_6^s\, \sin^2{\thD} +  I_6^c\, \cos^2{\thD}\right) \cos{\thl} \Big\}\, .
\eeqn
In addition to the lepton-pair invariant-mass squared $q^2=(p_\tau+p_{\bar{\nu}})^2$, we use as kinematic variables the three angles $\phi$, $\thl$ and $\thD$, which are defined as follows. Taking as positive $z$- axis the direction of the $D^*$ momentum in the $B$ rest frame, $\theta_\tau$ and $\theta_D$ are the polar angles of the $\tau$ and the final $D$  meson in the $\tau\nu$ and $D\pi$ rest frames, respectively. The azimuth $\phi$ is the angle between the decay planes formed by $\tau\nu$ and $D\pi$. See Fig.~\ref{fig:kinematics} for a visual representation of these kinematical variables.\\

Measuring this four-dimensional distribution is obviously a major experimental challenge, since the subsequent $\tau$ decay involves one ($\tau\to\nu_\tau\! +\!\text{hadrons}$) or two ($\tau\to\nu_\tau \, \ell \, \bar\nu_\ell$) additional neutrinos, making difficult to reconstruct the $\tau$ direction. Some information can be recovered by measuring the distribution of the secondary $\tau$ decay \cite{Ligeti:2016npd,Alonso:2016gym,Hill:2019zja}, but we refrain to enter here into this type of technical (but important) details.

The angular coefficients $I_i$'s are functions of $q^2$ that encode both short- and long-distance physics contributions. They can be written in terms of the hadronic FFs given in Appendix~\ref{app:FF}.
Using the global normalization
\be
\label{NF}
N_F \, =\,  \frac{G^2_F\, |V_{cb}|^2}{2^{7}\, 3 \pi^3 m^3_B}\; q^2\, \lambda^{1/2}_{D^*}(q^2)   \left( 1-\frac{m_\tau^2}{q^2}\right)^2 \, \mathcal{B}(D^*\rightarrow D\pi)\, ,
\ee
where $\mathcal{B}(D^*\rightarrow D \, \pi)$ is the branching fraction of the $D^*$ decay into
$D  \, \pi$ states described in Appendix~\ref{app:DtoDpi}. The expressions for the angular coefficients are:
\bea
\label{eq:Idef}
I^c_1 &\!\!\! = &\!\!\!  N_F\,\Big[ 2\, \Big( 1 + \frac{m^2_\tau}{q^2} \Big) \Big(|\azeL|^2 +4\, |\azeTL|^2\Big) -  \frac{16 m_\tau}{\sqrt{q^2}}\,\Re [\azeL \azeTL^{\!\!\!\!*}\,]  +  \frac{4 m^2_\tau}{q^2}\, | 
A_{tP}^L|^2 + \left( L \to R \right) \Big]\, , 
\no\\ 
I^s_1 &\!\!\! = &\!\!\!  N_F\,\Big[ \frac{1}{2} \Big(3 +  \frac{m^2_\tau}{q^2} \Big) \Big(|\apeL|^2 + |\apaL|^2  \Big)+ 2\, \Big(1 +  \frac{3m^2_\tau}{q^2} \Big) \Big(|\apeTL|^2 + |\apaTL|^2  \Big)  
\no\\
&&\hskip .4cm\mbox{} -  8\,\frac{m_\tau}{\sqrt{q^2}}\:\Re [\apeL \apeTL^{\!\!\!\!*} + \apaL \apaTL^{\!\!\!\!*}\,]+ \left( L \to R \right) \Big] \, ,  
\no \\
I^c_2 &\!\!\! = &\!\!\! -2\, N_F\, \Big(1 -  \frac{m^2_\tau}{q^2} \Big)\,  \Big(|\azeL|^2 - 4\,|\azeTL|^2  + \left( L \to R \right) \Big), 
\no\\ 
I^s_2 &\!\!\! = &\!\!\! \frac{1}{2}\,  N_F\, \Big(1 -  \frac{m^2_\tau}{q^2} \Big)\,\Big(|\apeL|^2 + |\apaL|^2 - 4\,\big(|\apeTL|^2 + |\apaTL|^2 \big)  + \left( L \to R \right) \Big)\, , 
\no\\ 
I_3 &\!\!\! = &\!\!\! N_F\, \Big(1 -  \frac{m^2_\tau}{q^2} \Big)\, \Big(|\apeL|^2 - |\apaL|^2 -4\, \big(|\apeTL|^2 - |\apaTL|^2 \big)+ \left( L \to R \right) \Big)\, , \no \\
I_4 &\!\!\! = &\!\!\! \sqrt{2}\, N_F\, \Big(1 -  \frac{m^2_\tau}{q^2} \Big)\, 
\Re [\azeL^{}\apaL^* -4\, \azeTL^{}\apaTL^{\hspace{-2mm}*} + \left( L \to R \right)]
\, , \no
\eea

\bea 
I_5 &\!\!\! = &\!\!\! 2 \sqrt{2}\, N_F\, \Big[\Re[\big( \azeL^{} -2\, \frac{m_\tau}{\sqrt{q^2}}\, \azeTL \big)\, \big( \apeL^* -2 \,\frac{m_\tau}{\sqrt{q^2}}\, \apeTL^{\hspace{-3mm}*}~ \big) -\left( L \to R \right) ]  
\no \\
&&\hskip 1.45cm\mbox{} - \frac{m^2_\tau}{q^2}\,  \Re[ {A^{L}}^*_{\!\!\!\!tP}\, \big( \apaL - 2\,\frac{\sqrt{q^2}}{m_\tau} \apaTL\big)+\left( L \to R \right)  ]\Big]\, ,
\no\\ 
I^c_6 &\!\!\! = &\!\!\!   N_F\, \frac{8 m^2_\tau}{q^2}\, \Re [{A^{L}}^*_{\!\!\!\!tP}\, \big( \azeL - 2\,\frac{\sqrt{q^2}}{m_\tau}\, \azeTL\big) + \left( L \to R \right) ] \, ,  
\no\\ 
I^s_6 &\!\!\! = &\!\!\! 4\, N_F\, \Re [\big( \apaL^{} -2\, \frac{m_\tau}{\sqrt{q^2}}\, \apaTL \big) \big( \apeL^* -2 \frac{m_\tau}{\sqrt{q^2}} \apeTL^{\hspace{-3mm}*}~ \big) -\left( L \to R \right) ] \, , \nonumber\\ 
I_7 &\!\!\! = &\!\!\! - 2 \sqrt{2}\, N_F\, \Big[\Im [\big( \azeL^{} -2\, \frac{m_\tau}{\sqrt{q^2}}\, \azeTL\big)\, \big( \apaL^* -2\, \frac{m_\tau}{\sqrt{q^2}}\, \apaTL^{\hspace{-2mm}*}~ \big) -\left( L \to R \right)] 
\no \\
&&\hskip 1.2cm\mbox{} + \frac{m^2_\tau}{q^2}\; \Im [  {A^{L}}^*_{\!\!\!\!tP}\, \big( \apeL - 2\,\frac{\sqrt{q^2}}{m_\tau}\, \apeTL\big) + \left( L \to R \right) ]\Big]\, , 
\no\\ 
I_8 &\!\!\! = &\!\!\! \sqrt{2}\, N_F\, \Big(1 -  \frac{m^2_\tau}{q^2} \Big)\; \Im [\azeL^{*}\apeL-4\, \azeTL^{{\hspace{-2mm}*}} \,\apeTL+ \left( L \to R \right) ]\, , 
\no\\ 
I_9 &\!\!\! = &\!\!\!  2\, N_F\, \Big(1 -  \frac{m^2_\tau}{q^2} \Big)\; \Im [\apaL^{}\apeL^* -4\, \apaTL^{}\apeTL^{\hspace{-3mm}*}+ \left( L \to R \right)] \, .
\eea
In the above expressions, the $\mathcal{A}^{L,R}_\lambda$ denote the transversity amplitudes, which are the projections of the total decay amplitude into the explicit polarization basis. The contribution of the RHN transitions to the angular coefficients is equivalent to the LHN ones, i.e. $(L \to R)$, up to a sign that depends on 
the relation between right-handed and left-handed leptonic transversity amplitudes. 
In the SM, the decay $B\to D^*\tau \bar{\nu}$ can be described by a total of four transversity amplitudes that correspond to one longitudinal ($\mathcal{A}_0$) and two transverse ($\mathcal{A}_{\perp,\|}$) directions, and a time-like component ($\mathcal{A}_t$) for the virtual vector boson decaying into the $\tau \bar{\nu}$ pair. However, with the inclusion of 
RHNs, we must distinguish the left and right chiralities of the leptonic current; thus, we get in total eight amplitudes: $\mathcal{A}^{L,R}_{0,\perp,\|,t}$. Now, in presence of the NP operators given in Eq.~\eqref{eq:def-operators}, the (axial)vector contributions can be incorporated in the above mentioned eight transversity amplitudes, modified by the presence of the new Wilson coefficients. Nevertheless, the (pseudo)scalar and tensor operators induce eight further amplitudes (four for each neutrino chirality):
two (pseudo)scalar amplitudes $\mathcal{A}_P^{L,R}$ and six tensor transversities $\mathcal{A}^{L,R}_{T\,0,T\,\perp,T\,\|}$. 
Thus, with the most general dimension-six Hamiltonian in Eq.~\eqref{eq:Lag}, the decay $B \to D^*(\to D \, \pi) \tau \bar{\nu}$ can be described by a total of sixteen tranversity amplitudes. Their explicit dependence on the hadronic helicity amplitudes, compiled in Appendix~\ref{app:FF}, and the Wilson coefficients is listed below,
\begin{eqnarray}
\label{eq:BtoDsAs}
&&{\cal A}_0^L = H_{V,0} \ (1+C^V_{LL}-C^V_{RL}) ,  \quad \hspace{2.9cm}
{\cal A}_0^R = H_{V,0} \ (C^V_{LR}-C^V_{RR}), \no \\
&&{\cal A}_\parallel^L = \frac{1}{\sqrt{2}} \left( H_{V,+} + H_{V,-} \right) \ (1+C^V_{LL}-C^V_{RL}), \quad\hspace{.5cm}
{\cal A}_\parallel^R = \frac{1}{\sqrt{2}} \left( H_{V,+} + H_{V,-} \right) \ (C^V_{LR}-C^V_{RR}), \nonumber \\
&&{\cal A}_\bot^L = \frac{1}{\sqrt{2}} \left( H_{V,+} - H_{V,-} \right) \ (1+C^V_{LL}+C^V_{RL}), \quad \hspace{.5cm}
{\cal A}_\bot^R = \frac{1}{\sqrt{2}} \left( H_{V,+} - H_{V,-} \right) \ (C^V_{LR}+C^V_{RR}), \nonumber \\
&&{\cal A}_t^L = H_{V,t} \ (1+C^V_{LL}-C^V_{RL}), \quad \hspace{3cm}
{\cal A}_t^R = H_{V,t} \ (C^V_{LR}-C^V_{RR}), \nonumber \\
&&{\cal A}_P^L = H_S  \ (C^S_{RL}-C^S_{LL}), \quad \hspace{3.8cm}
{\cal A}_P^R = H_S  \ (C^S_{RR}-C^S_{LR}), \nonumber \\
&&{\cal A}_{T0}^L = 2 H_{T,0} \ C^T_{LL}, \quad \hspace{4.7cm}
{\cal A}_{T0}^R = -2 H_{T,0} \ C^T_{RR}, \nonumber \\
&&{\cal A}_{T\parallel}^L  = \sqrt{2} \left(H_{T, +} -H_{T, -}  \right) \ C^T_{LL}, \quad \hspace{2.5cm}
{\cal A}_{T\parallel}^R  = - \sqrt{2} \left(H_{T, +} -H_{T, -}  \right) \ C^T_{RR}, \nonumber\\
&&{\cal A}_{T \bot}^{L,R}  = \sqrt{2} \left(H_{T, +} + H_{T, -}  \right) \ C^{T,R}_{LL} \, ,
\end{eqnarray}
where the $t$ and the $P$ amplitudes arise in the $B \to D^*$ observables combined as
\be
\label{tp_comb}
{\cal{A}}_{tP}^{L,R} \, =\, \Big( {\cal{A}}_t^{L,R}  + \frac{\sqrt{q^2}}{m_\tau}\, {\cal{A}}_P^{L,R} \Big)\, .
\ee
With these definitions, the left-handed contributions to the angular coefficients in Eq.~(\ref{eq:Idef}) are in agreement with Ref.~\cite{Becirevic:2019tpx,Becirevic:2016hea}.
Performing the angular integrations in Eq.~\eqref{eq:dGamma}, one easily obtains the differential distribution with respect to $q^2$, given by
\bea
\label{eq:gammaF}
\frac{d\Gamma}{dq^2} \,\equiv\, \Gamma_f\, =\, {\frac14} \left(3I_{1}^c + 6I_{1}^s - I_{2}^c - 2I_{2}^s  \right) \, ,
\eea
which written explicitly in terms of the different Wilson coefficients takes the following form:
\beqn 
  \label{eq:GammaDstar}
 \no
\lefteqn{\frac{d \Gamma (\bar{B} \to D^* \tau \bar{\nu})}{d q^2} \; =\;
\frac{G_F^2 \abs{V_{cb}}^2}{192 \, \pi^3 m_B^3}\; q^2\, \lambda^{1/2}_{D^*} (q^2)\left( 1- \frac{m_{\tau}^2}{q^2} \right)^2 {\cal B}(D^* \to D \pi) } && 
\\ \no & \hskip .5cm &
\times\, \left\lbrace \left( \abs{1+C^V_{LL}}^2 + \abs{C^V_{RL}}^2 + \abs{C^V_{LR}}^2 + \abs{C^V_{RR}}^2 \right)  \left( 1 + \frac{m_{\tau}^2}{2q^2} \right) \left(H_{V,+}^2+H_{V,-}^2 \right) \right.  
\\ \no && +\, \left( \abs{1+C^V_{LL}-C^V_{RL}}^2 + \abs{C^V_{LR}-C^V_{RR}}^2 \right)  \left[   \left( 1 + \frac{m_{\tau}^2}{2q^2} \right) \, H_{V,0}^2  + \frac{3}{2}\frac{m_{\tau}^2}{q^2} \, H_{V,t}^2  \right] 
\\ \no && 
-\, 4\, \Re \left [ \left( 1 + C^{V}_{LL} \right) C^{V*}_{RL} + C^V_{LR} \, C^{V*}_{RR} \right]   \left( 1 + \frac{m_{\tau}^2}{2q^2} \right)  H_{V, +} H_{V, -}
\\ \no&& 
+\, \frac{3}{2}\,\left( \abs{C^{S}_{RL} - C^{S}_{LL}}^2 + \abs{C^{S}_{RR} - C^{S}_{LR}}^2 \right) H_S^2 
\\ \no && 
+\, 8\, \left( \abs{C^T_{LL}}^2 + \abs{C^T_{RR}}^2 \right) \left( 1 + \frac{2 m_{\tau}^2}{q^2}  \right) \left( H_{T,+}^2  + H_{T, -}^2 + H_{T, 0}^2 \right) 
\\ \no && 
+\, 3\, \Re \left[ \left( 1+ C^{V}_{LL} - C^{V}_{LR} \right) \left(C^{S}_{RL} - C^{S}_{LL} \right)^* + \left( C^V_{LR} -  C^V_{RR} \right) \left( C^S_{RR} - C^S_{LR}\right)^* \right] \frac{m_{\tau}}{\sqrt{q^2}}\, H_S H_{V,t} 
\\ \no && 
-\, 12\, \Re \left[\left( 1+ C^{V}_{LL} \right) C^{T*}_{LL}
+ C^{V}_{RR}\, C^{T*}_{RR} \right] \frac{m_{\tau}}{\sqrt{q^2}}\,\left( H_{T, 0} H_{V, 0} + H_{T,+} H_{V, +}  - H_{T, -} H_{V, -}
\right)
\\ && \left.
+\, 12\,\Re \left[C^{V}_{RL}\, C^{T*}_{LL} + C^{V}_{LR} \, C^{T*}_{RR} \right] \frac{m_{\tau}}{\sqrt{q^2}} \left( H_{T, 0} H_{V, 0} + H_{T,+} H_{V, -}  - H_{T, -} H_{V, +}
\right) \right\rbrace \, .
\eeqn 

Differential distributions with respect to a single angle,
which can be obtained by integrating two angles at a time, are also of special interest. These are 
\begin{align}
	\frac{d^2\Gamma }{ dq^2 \, d\cos{\thl}}\, &=\, \frac{3}{8}\, \big[ \left( I_{1}^c + 2I_{1}^s - I_{2}^c - 2I_{2}^s \right) + \left( I_{6}^c + 2I_{6}^s \right) \cos{\thl} + \left( 2I_{2}^c + 4I_{2}^s \right) \cos^2{\thl}\big] \, , 
\\	\label{eq:dqdthD}
	\frac{d^2\Gamma }{ dq^2 \, d\cos{\thD}}\, &=\, \frac{3}{8}\, \big[  \left( 3I_{1}^s - I_{2}^s \right) + \left( 3I_{1}^c - I_{2}^c - 3 I_{1}^s + I_{2}^s \right) \cos^2\thD \big] 
\no \\
	&=\, \frac{3}{4}\, \Gamma_f\, \big[  F_T^{D^*} \sin^2{\thD} + 2\, F_L^{D^*} \cos^2{\thD} \big]\, , 
\\ 	\label{eq:dqdPhi}
	\frac{d^2\Gamma }{ dq^2 \, d\phi}\, &=\, \frac{1}{8\pi}\, \big[ \left( 3I_{1}^c + 6I_{1}^s - I_{2}^c - 2I_{2}^s \right) +4 I_3 \cos{2\phi}+ 4 I_9 \sin{2\phi} \big] 
\no \\
	&=\, \frac{1}{2\pi}\, \Gamma_f\, \big[ 1 + A_3 \cos{2\phi}+ A_9 \sin{2\phi} \big]\, .
\end{align}
In the following we define several observables constructed from the coefficients of various
angular dependences.
The distribution with respect to $\cos{\thD}$ in Eq.~\eqref{eq:dqdthD} provides the longitudinal and transverse polarization fractions for the $D^*$ meson, defined as~\cite{Becirevic:2019tpx, Becirevic:2016hea}
\begin{align}
F_L^{D^*}\, =\, \frac{3 I_{1}^c -I_{2}^c}{ 3I_{1}^c + 6I_{1}^s - I_{2}^c - 2I_{2}^s }\, , 
\qquad \text{and} \qquad F_T^{D^*}\, =\, \frac{2(3 I_{1}^s -I_{2}^s)}{ 3I_{1}^c + 6I_{1}^s - I_{2}^c - 2I_{2}^s }\, ,
\label{eq:FLD}
\end{align}
which satisfy that $F_L^{D^*} + F_T^{D^*} = 1$.
Notice that these quantities are functions of $q^2$. The Belle measurement mentioned before in Section~\ref{sec:Intro}, named as $\bar{F}_L^{D^*}$, refers to the $q^2$-integrated polarization. We define the $q^2$-integrated observables as follows,
\begin{equation}
    \bar{{\cal O}} \equiv \frac{1}{\Gamma} \int_{q^2_\text{min}}^{q^2_\text{max}}{dq^2 \, {\cal O}[q^2] \, \Gamma_f[q^2]},
\end{equation}
where $\Gamma$ is the total decay width and the $q^2$ dependence of the observables has been written explicitly.
The angular coefficients $I_3$ and $I_9$ can simply be extracted by measuring the terms proportional to $\cos2\phi$ and $\sin2\phi$ in Eq.~\eqref{eq:dqdPhi},
\bea
A_3\, =\, \dsp\frac{I_3}{\Gamma_f}\, , \qquad \text{and}\qquad  A_9\, =\,  \dsp\frac{I_9}{\Gamma_f}\, ,
\eea
respectively. 
Furthermore, we define several asymmetries starting with the well-known forward-backward asymmetry, defined as
\begin{equation}
\label{eq:AFB}
{\cal A}_{FB}^{D^*}\, =\,\frac{1}{\Gamma_f}\; \Big[ \int_0^1-\dsp \int_{-1}^0\Big]\, d\cos{\thl}\; 
\frac{d^2 \Gamma}{d q^2d\cos{\thl}}\, .
\end{equation}
The coefficients $I_4$ and $I_5$  in Eq.~\eqref{eq:dGamma} can be extracted with the two angular asymmetries:
\begin{align}
  \label{eq:A4}
A_{4}\, =\, & \frac{1}{\Gamma_f}\;
\Big[ \int_{-\pi/2}^{\pi/2}- \int_{\pi/2}^{3\pi/2} 	\Big] d\phi\;
	\Big[ \int_0^1 - \int_{-1}^0 \Big] d\cos{\thD}\; 
	\Big[\int_0^1 -\int_{-1}^0 \Big] d\cos{\thl}\;
\frac{d^4\Gamma}{dq^2d\cos{\thl} d\cos{\thD} d\phi} \, ,
\no\\  
A_{5}\, =\, & \frac{1}{\Gamma_f}\;
\Big[ \int_{-\pi/2}^{\pi/2} - \int_{\pi/2}^{3\pi/2}	\Big] d\phi\;
	\Big[\int_0^1 -\int_{-1}^0 \Big] d\cos{\thD}\;
	\int_{-1}^1 d\cos{\thl}\;
	\frac{d^4\Gamma}{dq^2d\cos{\thl} d\cos{\thD} d\phi}\, .
\end{align}

One can further define the following two observables, 
\begin{align}
\label{eq:A7} 
A_{7}\, =\, & \frac{1}{\Gamma_f}\;
\Big[ \int_{0}^{\pi}-\int_{\pi}^{2\pi} \Big] d\phi \;
	\Big[\int_0^1 -\int_{-1}^0 \Big] d\cos{\thD}\; 
	\int_{-1}^1 d\cos{\thl}\; \frac{d^4\Gamma}{dq^2d\cos{\thl} d\cos{\thD} d\phi}\, ,
\no\\
A_{8}\, =\, & \frac{1}{\Gamma_f}\;
\Big[ \int_{0}^{\pi} - \int_{\pi}^{2\pi}\Big] d\phi \;
	\Big[ \int_0^1 - \int_{-1}^0 \Big] d\cos{\thD}  \;
	\Big[\int_0^1  -\int_{-1}^0 \Big] d\cos{\thl}\;
	\frac{d^4\Gamma}{dq^2 d\cos{\thl} d\cos{\thD} d\phi}\, ,
\end{align}
which are non-vanishing only if NP induces a complex contribution to the amplitude. This holds true for the coefficient $A_9$ as well.
These asymmetries are simply related to the angular coefficients in~\eqref{eq:dGamma}:
\begin{eqnarray}
\label{eq:obsfac}
A_4= \frac{2}{\pi}\, \frac{I_4}{\Gamma_f}\, ,
\qquad
A_5= \frac{3}{4}\, \frac{I_5}{\Gamma_f}\, ,
\qquad
{\cal A}_{FB}^{D^*}= \frac{3}{8}\, \frac{I_6^c+ 2I_6^s}{\Gamma_f}\, , 
\qquad
A_7= \frac{3}{4}\, \frac{I_7}{\Gamma_f}\, ,
\qquad
A_8= \frac{2}{\pi}\, \frac{I_8}{\Gamma_f}\, .
\end{eqnarray}

Finally, the total branching ratio can be decomposed in terms of the $\tau$ polarization, giving rise to another observable: the lepton polarization asymmetry, defined as
\begin{equation}
\label{eq:PtauDstar}
    \ptau{} = \frac{ d\Gamma_{\lambda_{\tau} = 1/2} / dq^2  - d\Gamma_{\lambda_{\tau} = -1/2}/dq^2}{\Gamma_f} . 
\end{equation}
%

\subsubsection[\texorpdfstring{$B_c \to \tau \bar{\nu}$}{dum3}]{\texorpdfstring{\boldmath $B_c \to \tau \bar{\nu}$}{dum3}}

Another interesting but yet not observed decay is $B_c \to \tau \bar{\nu}$ for which the branching ratio can be written as
\beqn
\mathcal{B}(B_c \to \tau \bar{\nu})/\mathcal{B}(B_c \to \tau \bar{\nu})_{\text{SM}} &=& \, \abs{1+ C_{LL}^V - C_{RL}^V + \frac{m_{B_c}^2}{m_{\tau}(m_b+m_c)}\, (C_{RL}^S - C_{LL}^S)}^2 
\no\\
&&+ \abs{C_{RR}^V - C_{LR}^V + \frac{m_{B_c}^2}{m_{\tau}(m_b+m_c)}\, (C_{LR}^S - C_{RR}^S)}^2\, .
\label{eq:Bctaunu}
\eeqn
The first line contains the usual contribution from LHNs \cite{Murgui:2019czp}, while the $\nu_R$ contribution in the second line is readily obtained with a parity transformation.

\section{Interpreting the anomalies}
\label{sec:interp}
%
This section is devoted to study the origin of the observed experimental deviations from the SM predictions. We show from a theoretical perspective the implications of new physics in the observables involving $b \to c$ transitions and discuss the possible ultraviolet (UV) scenarios that could give rise to such anomalies in the context of $b \to c$ processes involving both left- and right-handed neutrinos. 
\subsection{Fit-independent results}
%
\label{sec:IndepFit}

The Wilson coefficients introduced in Eq.~\eqref{eq:Lag} encode all NP contributions that can enter in $b \to c$ transitions at dimension-six operator level, also in the presence of sterile light RHNs. Therefore, the landscape of possibilities generating the anomalies can be classified by the impact of these ten parameters on the measurable observables. To get a general idea about the sensitivity to the different Wilson coefficients,
we quote the numerical expressions of several observables that have already been measured.
These expressions have been obtained setting the FFs
at their central values and, therefore, ignoring the uncertainties and correlations among the different numerical factors. The complete analytical expressions, with a proper account of hadronic uncertainties, will be used instead in the data fits that we will present in Section~\ref{sec:fits}.
The observables $\mathcal{R}_D$ and $\rdst{}$ are normalized to their SM predictions: 
\begin{eqnarray}
   \no
{\cal R}_{D} / {\cal R}_{D}^\text{SM} &\!\! \approx &\!\! \left(|1+C^V_{LL} + C^V_{RL}|^2+|C^V_{LR}+ C^V_{RR}|^2\right) + 1.037 \left(|C^S_{LL}+C^S_{RL}|^2+|C_{LR}^S+C^S_{RR}|^2\right)
\\[3pt] \no  & + &\!\! 
 0.939 \left(|C^{T}_{LL}|^2+|C^T_{RR}|^2\right) + 1.171\, \Re \left[ (1+C^V_{LL} + C_{RL}^{V})\, C^{T*}_{LL} + (C^V_{LR} + C_{RR}^{V})\, C^{T*}_{RR} \right] 
 \\[3pt] \no & + &\!\!
  1.504 \,  \Re \left[(1+C^V_{LL} + C_{RL}^{V}) (C^{S*}_{LL} +  C^{S*}_{RL}) + (C^V_{LR} + C_{RR}^{V}) (C^{S*}_{LR} +  C^{S*}_{RR} ) \right], 
\label{eq:RDRDSM}
\end{eqnarray}
and
\begin{eqnarray}
\no
    {\cal R}_{D^*} / {\cal R}_{D^*}^\text{SM} &\!\! \approx &\!\!\! \left(|1+C^V_{LL}|^2+|C^V_{RL}|^2+|C^V_{LR}|^2+|C^V_{RR}|^2\right) + 0.037 \left(|C^S_{RL}-C^S_{LL}|^2+|C_{RR}^S-C^S_{LR}|^2\right) 
\\[3pt] \no  & + &\!\! \!
   17.378 \left(|C^{T}_{LL}|^2+|C^T_{RR}|^2\right) -1.781 \,  \Re \left[(1+C^V_{LL})\, C_{RL}^{V*}+C^V_{LR}\, C^{V*}_{RR}\right]
 \\[3pt] \no & + &\!\!\!
 5.748 \, \Re \left[ C_{RL}^V C^{T*}_{LL} +C^V_{LR} C^{T*}_{RR} \right] - 5.130 \, \Re \left[(1+C^V_{LL})\, C^{T*}_{LL} +C^V_{RR}\, C^{T*}_{RR}\right]
\\[3pt]  & + &\!\!\!
0.114 \, \Re \left[ (1+C^V_{LL}-C_{RL}^V)\, (C^{S*}_{RL}-C^{S*}_{LL})+(C^V_{RR}-C^V_{LR})\, (C_{LR}^{S*}-C^{S*}_{RR})\right]  .
   \label{eq:RDstRDSM}
\end{eqnarray}
For the $q^2$-integrated polarization observables $\ptauint{}$ and $\FLint{}$, we show their numerical values multiplied by $\rdst$:
\begin{eqnarray}
\no
\ptauint{}\! \times {\cal R}_{D^*} &\!\! 
\approx &\!\! 
-0.128 \left( |1+C^V_{LL}|^2+|C_{RL}^V|^2-|C^V_{RR}|^2-|C^V_{LR}|^2\right)
+0.282 \left( |C^{T}_{LL}|^2-|C^{T}_{RR}|^2\right)
\\[3pt] \no &\!\! +&\!\!
0.010 \left( |C^S_{RL}-C^S_{LL}|^2-|C^S_{RR}-C^S_{LR}|^2\right) 
+ 0.221 \, \Re \left[ (1+C^V_{LL})\, C_{RL}^{V*}-C^{V*}_{RR}\, C^V_{LR}\right]
\\[3pt] \no &\!\! +&\!\!
0.442 \, \Re \left[ (1+C^V_{LL})\, C^{T*}_{LL}-C^{V*}_{RR}\, C^{T}_{RR}\right]
- 0.592 \, \Re \left[ C_{RL}^V\, C^{T*}_{LL}-C^{V*}_{LR}\, C^{T}_{RR} \right]
\\[3pt]  &\!\! +&\!\!
0.030\, \Re \left[ (1+C^V_{LL}-C_{RL}^V)\, (C^{S*}_{RL}-C^{S*}_{LL})+(C^{V*}_{RR}-C^{V*}_{LR})\, (C_{RR}^S-C^S_{LR})\right] , 
\label{eq:PDRDst}
\end{eqnarray}
and
\begin{eqnarray}
\no
\FLint{}\! \times {\cal R}_{D^*} &\!\!\approx&\!\! 0.120 \left( 
|1+C_{LL}^V-C_{RL}^V|^2     
+|C_{RR}^V-C_{LR}^V|^2\right) 
\\[3pt] \no &\!\! +&\!\! 
0.010 \left( |C^S_{RL} - C^S_{LL}|^2 + |C^S_{RR} - C_{LR}^S|^2\right)
+ 0.869 \left( |C^{T}_{LL}|^2 + |C^T_{RR} |^2 \right)
\\[3pt] \no &\!\! +&\!\!
0.030\, \Re \left[ (1+C_{LL}^V-C_{RL}^V)     
\, (C^{S*}_{RL} - C^{S*}_{LL}) - (C^V_{RR} - C^V_{LR})\, (C_{RR}^{S*} - C^{S*}_{LR})\right]  
\\[3pt] &\!\! -&\!\!
0.525 \, \Re \left[ (1+C_{LL}^V-C_{RL}^V)     
\, C^{T*}_{LL}+ (C^V_{RR} -C^V_{LR} )\, C^{T*}_{RR} \right] .
\label{eq:FLDRDst}
\end{eqnarray}

\FloatBarrier 

\begin{figure}[ht]
\centering
\includegraphics[width=0.24\linewidth]{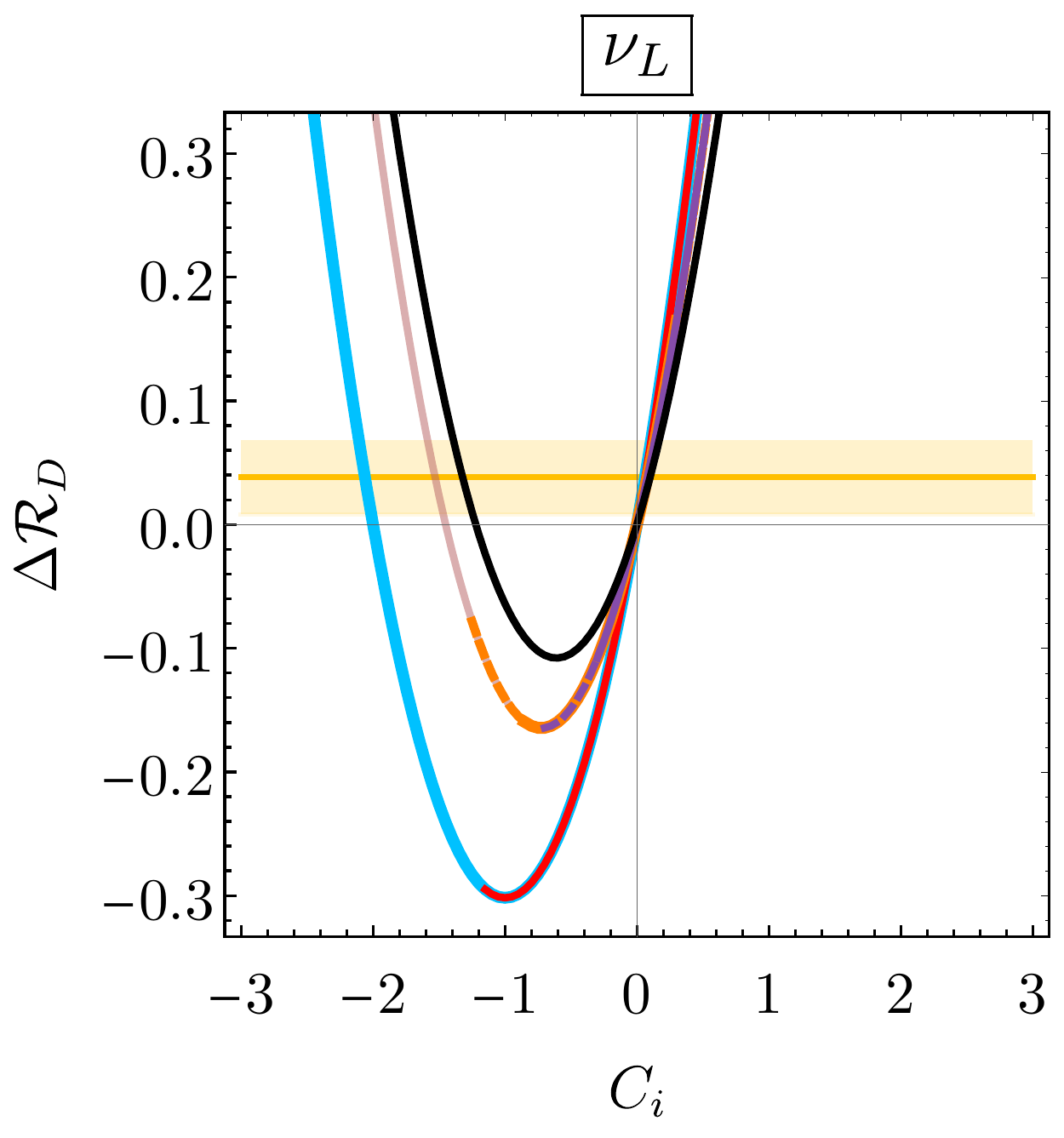}
\includegraphics[width=0.24\linewidth]{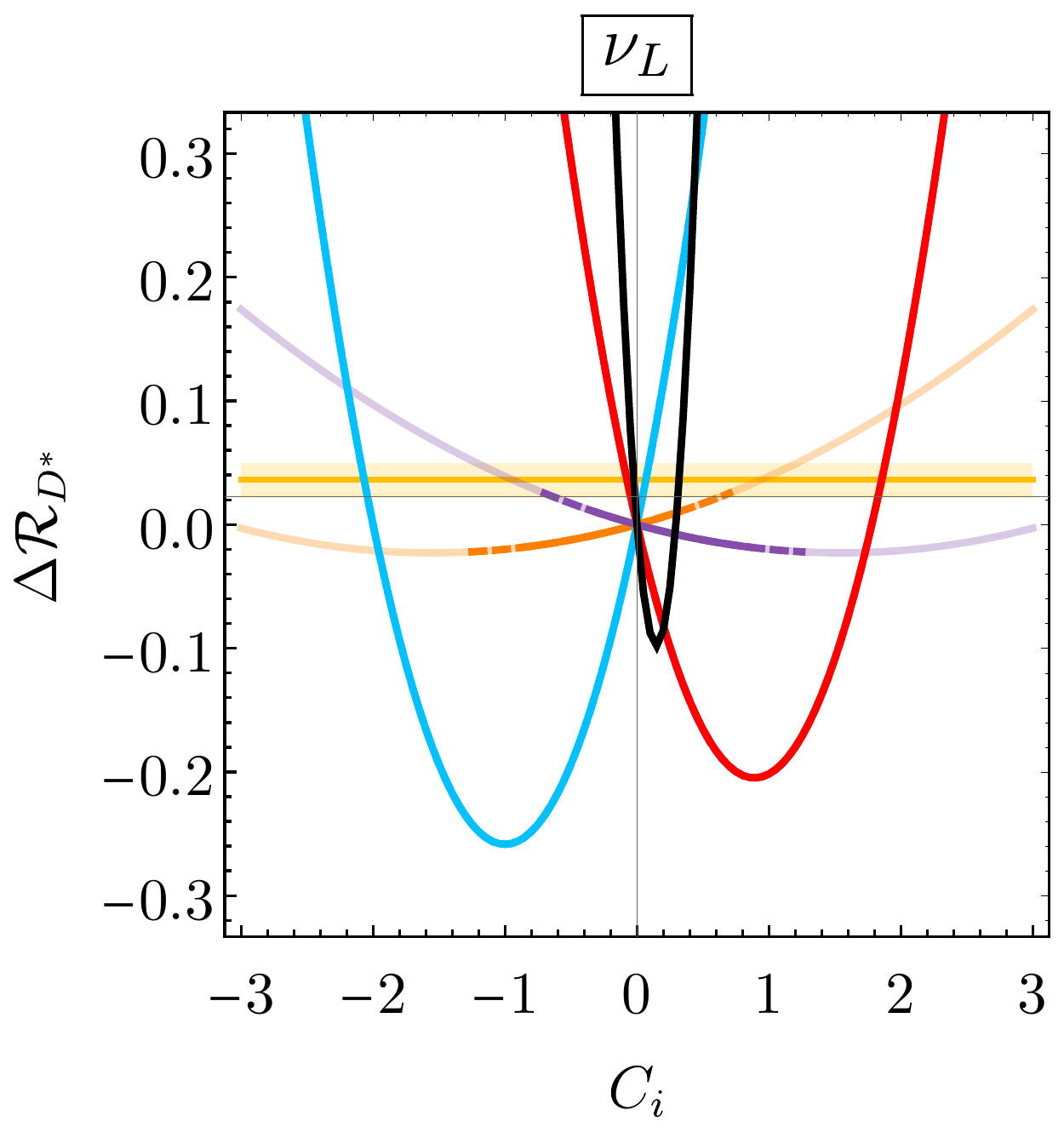}
\includegraphics[width=0.24\linewidth]{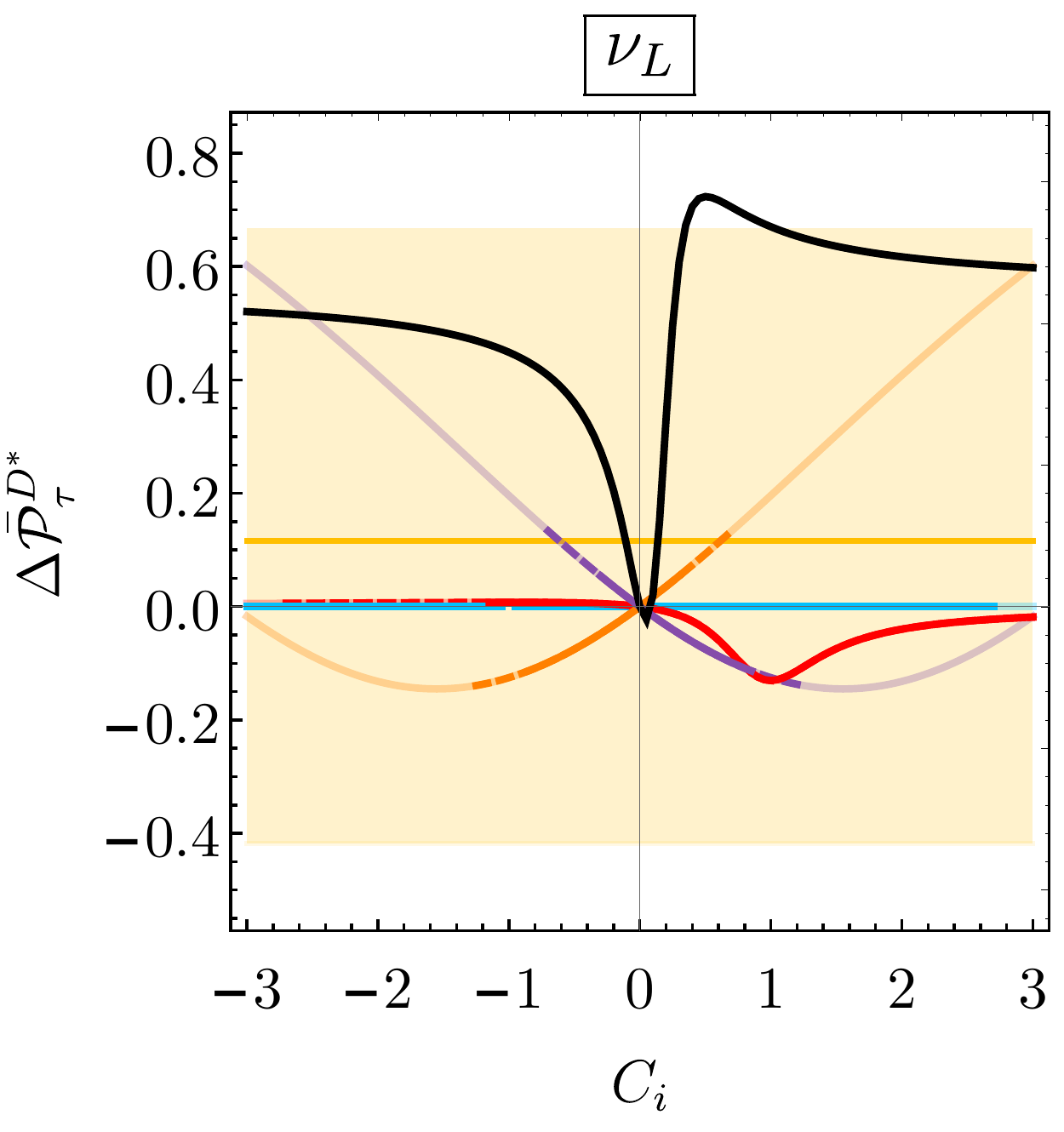}
\includegraphics[width=0.24\linewidth]{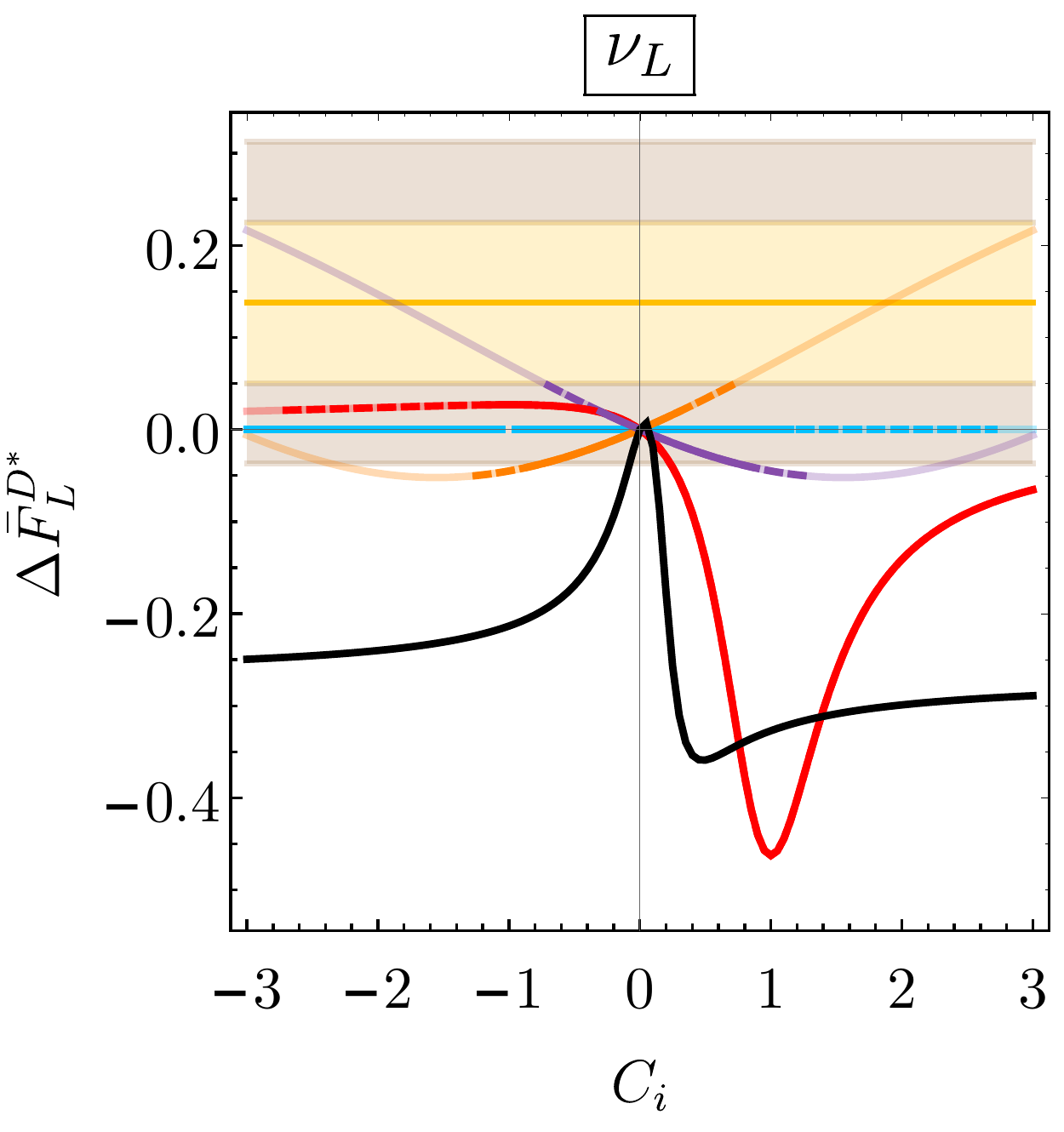}\\
\includegraphics[width=0.24\linewidth]{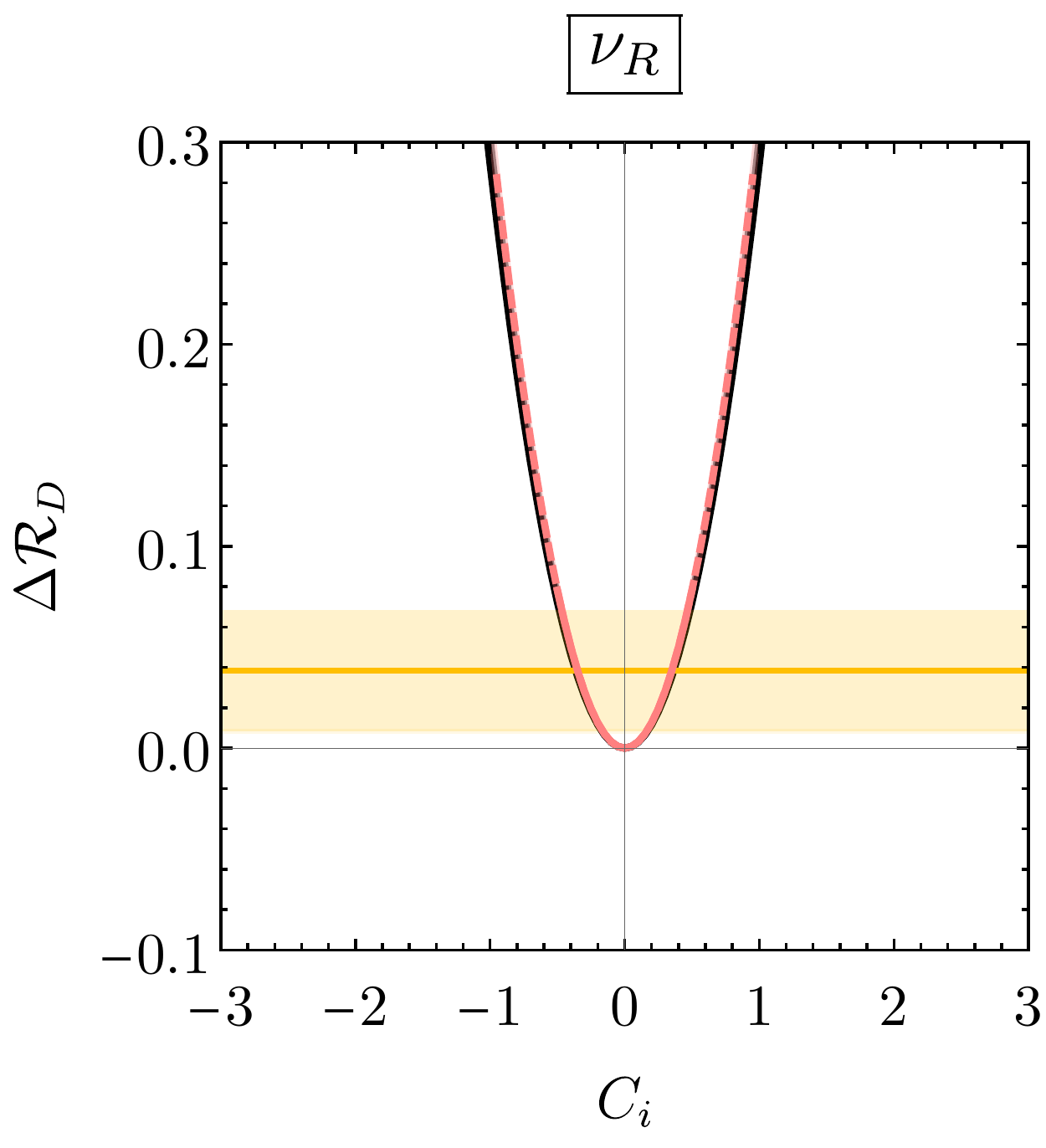}
\includegraphics[width=0.24\linewidth]{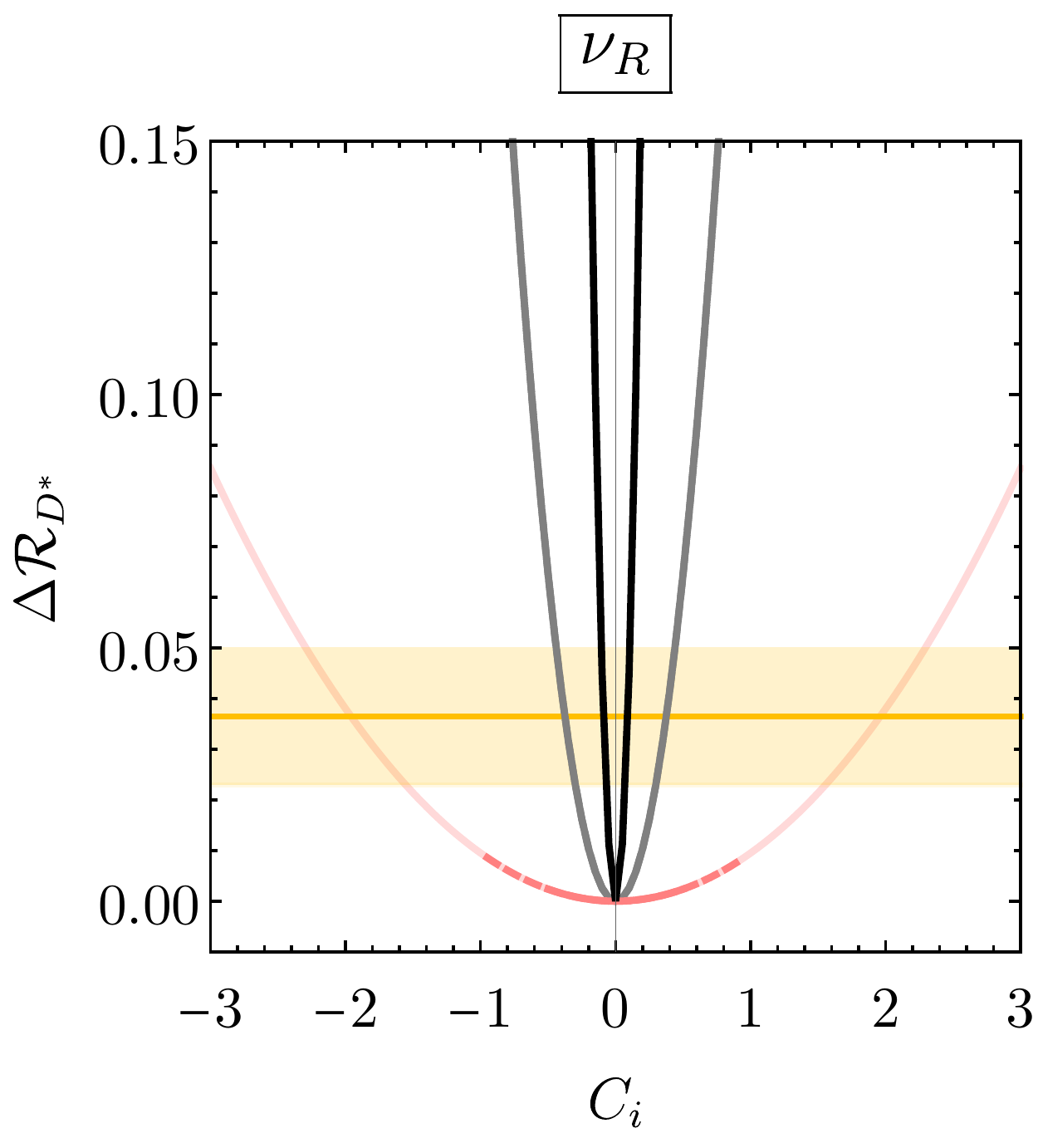}
\includegraphics[width=0.24\linewidth]{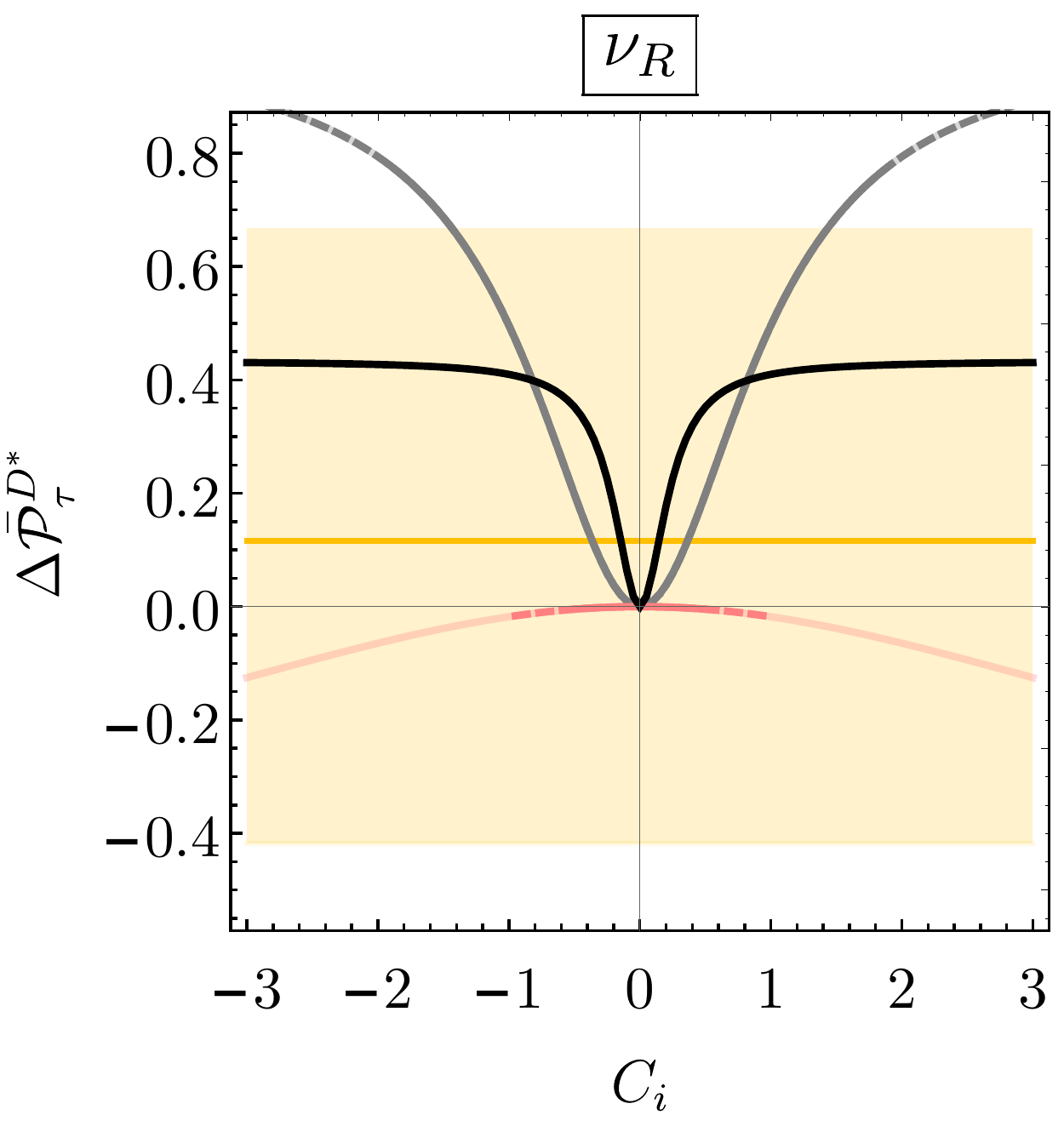}
\includegraphics[width=0.24\linewidth]{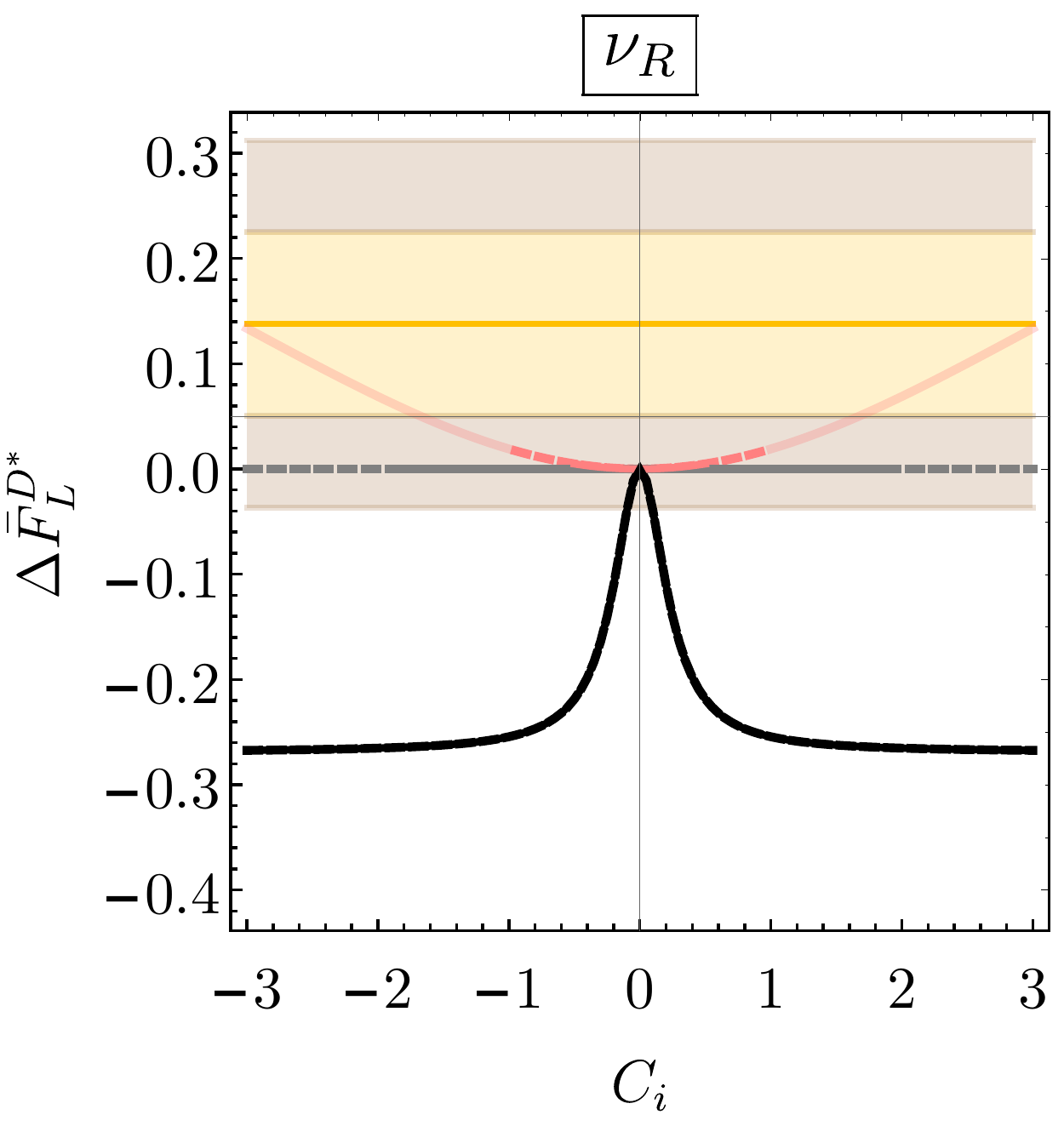}\\
 \includegraphics[width=0.5\linewidth]{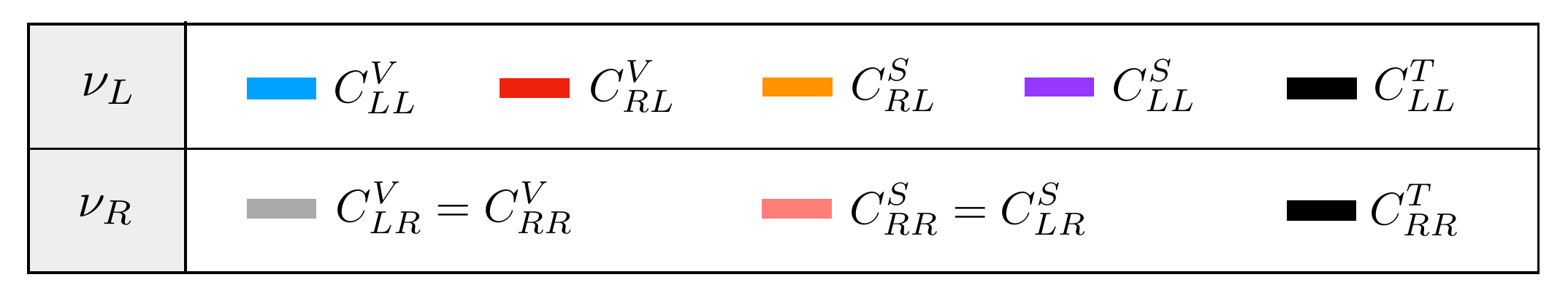}
\caption{Individual contributions of the Wilson coefficients involving LHNs (upper panels) and RHNs (lower panels). The solid (dashed) lines show the parameter space allowed by the constraint $\mathcal{B}(B_c \to \tau  \, \bar{\nu}) < 10 \% \, (30\%) $, whereas the fainted lines show the predictions without taking into account this constraint.}
\label{fig:IndFitPlot}
\end{figure}

With the above expressions of the four observables, namely $\rd$, $\rdst$, $\ptauint$ and $\FLint$, we analyse the modifications induced by each individual Wilson coefficient on the SM predictions. The corresponding shifts are shown in Fig.~\ref{fig:IndFitPlot}, both for the $\nu_L$ (upper panels) and $\nu_R$ (lower panels) EFT operators. The experimental central values of the observables are displayed as yellow lines whereas bands of the same colour are used for their $1\sigma$ uncertainties. For $ \FLint{}$ we also indicate the $2\sigma$ uncertainty with brown bands. The solid (dashed) lines show the parameter space allowed by the constraint $\mathcal{B}(B_c \to \tau \bar{\nu}) < 10 \% \,  (30\%) $. The fainted lines show the ranges for each Wilson coefficient without imposing the constraint from the leptonic branching ratio ${\cal B}(B_c \to \tau \bar{\nu})$.

Different Wilson coefficients could help to reproduce the measured values of $\rd{}$ and $\rdst{}$. However, the scalar coefficients would need to take values that are already excluded by $\mathcal{B}(B_c \to \tau \bar{\nu})$, leaving vector and axial-vector contributions as the preferred options to fit the experimental results. The large uncertainties in the $\bar{\mathcal{P}}_{\tau}^{D^*}$ measurement make almost any shift in the Wilson coefficients to be in agreement with the experimental value, 
being the only exceptions large shifts in the vector Wilson coefficients $C^V_{LR,RR}$ and a positive increment of $C^T_{LL}$.

Looking at the dependence of these observables on the RHN contributions, one observes that all of them are symmetric under the exchanges $C^V_{LR} \leftrightarrow C^V_{RR}$ and $C^S_{LR} \leftrightarrow C^S_{RR}$. In particular, $\FLint{}$ is insensitive to any (single) $\nu_R$ vector contribution because the dependence on the corresponding Wilson coefficient exactly cancels, since it is defined as a ratio as Eq.~(\ref{eq:FLD}) shows. This does not hold true for $C^V_{RL}$, since there is an interference between this NP operator and the SM contribution. 

It is particularly challenging to reproduce the experimental value of $\FLint{}$,
regardless of the type of NP contribution; the $\pm 1\sigma$ band cannot be reached varying any of the Wilson coefficients individually \cite{Blanke:2018yud, Asadi:2019xrc}. Negative non-zero values of $C^V_{RL}$ can only slightly increase the predicted longitudinal $D^*$ polarization, while the changes induced by the tensor Wilson coefficients go in the opposite direction of the experimental value, decreasing the SM predictions.
The only contributions that would help are the scalar ones, but for values of their Wilson coefficients that are already excluded by the constraint $\mathcal{B}(B_c \to \tau \bar{\nu})< 30\%$.

\FloatBarrier

\subsection{UV Physics}
%
Once the impact of individual Wilson coefficients in $B \to D^{(*)} \tau \bar{\nu}$ observables is understood, the following step is to extend the analysis to the combined effect of several coefficients that are present in these transitions simultaneously. The most general EFT Hamiltonian in Eq.~\eqref{eq:Lag} includes 10 Wilson coefficients, which in general can be complex. Even assuming them to be real, a 10-parameter fit 
would become unstable. Moreover, its interpretation in terms of NP mediators and UV completions might be unrealistic. Instead, we consider particular cases, described in Section~\ref{sec:scenarios}.  Most of them are motivated from the ``simplified model" scenarios. In this context, ``simplified" refers to a single new mediator particle that can be integrated out to contribute to  one or more of the effective operators entering into the $b\to c  \tau \bar{\nu}$ transitions. As the main purpose of this work is to explore the effect of light RHNs, we single out those mediators that can contribute to the $b \to c$ transitions and involve a gauge-singlet RHN.

These NP fields can be classified into scalars, vector bosons and leptoquarks, as listed in Table~\ref{table:NPQN}. Since in most cases both right-
and left-handed neutrino operators are generated simultaneously after a given mediator is integrated out, we will explore both the effect of considering only the right-handed contributions as well as the scenarios in which the full set of operators is generated. 
Unlike in previous references discussing the role of RHNs in $b \to c$ anomalies~\cite{Robinson:2018gza, Asadi:2018sym}, we also include a fit to the Wilson coefficients that will appear if NP is mediated through the leptoquark $\tilde{V}_2\sim (\bar{3},2,-1/6)$.

\begin{table}[ht]
    \centering
    \renewcommand\arraystretch{1.2}
    \begin{tabular}{|c| c | c | c | c |}
    \hline
        Spin & Q.N. & Nature &  $\nu_L$-WET  & $\nu_R$-WET \\
        \hline
0 & $S_1 \sim (\bar{3},1,1/3)$ & LQ &  $C_{LL}^V$, $C_{LL}^S$, $C_{LL}^T$ &  $C_{RR}^V$, $C_{RR}^S$, $C_{RR}^T$ \\
0 & $\Phi \sim (1,2,1/2)$ & SB & $C_{LL}^S$, $C_{RL}^S$ & $C_{LR}^S$, $C_{RR}^S$ \\
0 & $\tilde{R}_2 \sim (3,2,1/6)$ & LQ &  %
--  & $C_{RR}^S$, $C_{RR}^T$ \\
1 & $U_1^\mu \sim (3,1,2/3)$ & LQ  & $C_{LL}^V$, $C_{RL}^S$ & $C_{RR}^V$, $C_{LR}^S$\\
1 & $\tilde{V}_2^\mu \sim (\bar{3},2,-1/6)$ & LQ  & 
-- & $C_{LR}^S$\\
1 & $V^\mu \sim (1,1,-1)$ & VB &  
-- & $C_{RR}^V$ \\
\hline
    \end{tabular}
    \caption{Spin, $SU(3)_C \otimes SU(2)_L \otimes U(1)_Y$ quantum numbers and nature (LQ = leptoquark, SB = scalar boson, VB = vector boson) of the possible candidates to mediate $b \to c$ transitions involving $\nu_R \sim (1,1,0)$. The fourth and fifth columns list the operators with left-handed and right-handed neutrinos, respectively, generated by the integration of the correspondent mediator.}
    \label{table:NPQN}
\end{table}

\section{Fit Results}
\label{sec:fits}
%
Under the assumption that NP enters only in the third generation of leptons and that Wilson coefficients are real, we have performed fits in different scenarios of the most general dimension-six Hamiltonian, taking into account all experimental data available nowadays. We start by listing the inputs used in the fit, and then we describe the motivated scenarios, based on the previous section, that we are considering. Finally, the results obtained by performing global fits in each of the scenarios are interpreted.
%
\subsection{Numerical input of the fit}

For our fits we will use the most recent world-average values of $\rd{}$ and $\rdst{}$ from Ref.~\cite{Amhis:2019ckw}, 
including a correlation of $-0.38$ between them. The value of the $q^2$-integrated $\tau$ polarization, $\ptauint{}$, measured recently by Belle \cite{Hirose:2016wfn} and the longitudinal $D^*$ polarization, $ \FLint{}$, measured by BaBar \cite{Abdesselam:2019wbt} are also taken into account. Finally we consider the $q^2$ distributions of the $D$ and $D^*$ meson \cite{Lees:2013uzd, Huschle:2015rga}, summarized in Table~9 of Ref.~\cite{Murgui:2019czp}. The different experimental inputs used in the fits are collected in Table~\ref{table:t}.

\begin{table}[ht]
    \centering
    \begin{tabular}{ |c | c | c | c |}
    \hline
        Observable & Experimental Value & Reference & Comments  \\
        \hline
$\rd{}$ & $0.340 \pm 0.027 \pm 0.013$ & \cite{Amhis:2019ckw}
& \multirow{2}{*}{$\rd{}$  and  $\rdst{}$ correlation of $-0.38$}\\
 $\rdst{}$ & $0.295 \pm 0.011 \pm 0.008$ & \cite{Amhis:2019ckw}
 & \\
$\ptauint{}$ & $-0.38 \pm 0.51^{+ 0.21}_{-0.16}$ & \cite{Hirose:2016wfn} &   \\
$ \FLint{}$ & $ 0.60 \pm 0.08 \pm 0.035 $ & \cite{Abdesselam:2019wbt} &  \\
$D^{\phantom{*}}$ differential $q^2$ dist. &  & \cite{Lees:2013uzd, Huschle:2015rga} & \\
$D^*$ differential $q^2$ dist. & & \cite{Lees:2013uzd, Huschle:2015rga} & \\
$\mathcal{B}(B_c \to \tau \bar{\nu})$ & $ \leq 10\%, \, 30\%$ &  \cite{Alonso:2016oyd, Celis:2016azn, Beneke:1996xe, Akeroyd:2017mhr} & \\
\hline
    \end{tabular}
    \caption{Experimental inputs used in our fits.} 
    \label{table:t}
\end{table}
The upper bound for the leptonic decay rate  $\mathcal{B}(B_c \to \tau \bar{\nu})$ is taken to be either 30\% or 10\%. The first limit is derived from the $B_c$ lifetime \cite{Alonso:2016oyd, Celis:2016azn, Beneke:1996xe}, while a stronger bound of 10\% is obtained from the LEP data at the $Z$ peak \cite{Akeroyd:2017mhr}.\footnote{ 
Note, however, that the 10\% bound assumes the probability of a $b$ quark hadronizing into a $B_c$ meson to be the same in LEP, Tevatron and LHCb, which exhibit very different transverse momenta. This has been proved to be an inaccurate approximation for $b$-baryons \cite{Amhis:2019ckw}. 
Since the dominant contribution to the $B_c$ decay width comes from the decay of the $c$ quark, the 30\% limit could also be relaxed to about 60\% \cite{Blanke:2018yud} by lowering the charm mass used in the lifetime analysis \cite{Beneke:1996xe}.}
In our analyses the stronger $10\%$ limit is first assumed in the fit and, in those cases where the $10\%$ bound is saturated the fit is repeated by relaxing it to 30\%.
As Eq.~\eqref{eq:Bctaunu} shows, the $B_c\to\tau \bar{\nu}$ limit constrains the splittings between the $C^V_{LL(RR)}$ and $C^V_{RL(LR)}$ and, specially, between the $C^S_{RL(LR)}$ and $C^S_{LL(RR)}$ Wilson coefficients.\\ 

For the FFs, we follow the same approach as in Ref.~\cite{Murgui:2019czp}. Using heavy quark effective field theory~\cite{Neubert:1993mb,Manohar:2000dt}, we adopt the Boyd, Grinstein and Lebed (BGL) parametrization \cite{Boyd:1994tt,Boyd:1995sq,Boyd:1997kz}, including corrections of order $\alpha_s$, $\Lambda_{\text{QCD}}/m_{b,c}$~\cite{Bernlochner:2017jka} and partly $\Lambda_{\text{QCD}}^2/m_c^2$~\cite{Jung:2018lfu}. We also include the cubic term in the expansion of the leading Isgur-Wise function in powers of the conformal-mapping variable $z$~\cite{Isgur:1989ed,Bordone:2019vic}. The inputs of the FFs, listed in Table~1 of Ref.~\cite{Murgui:2019czp}, haven been obtained from lattice quantum chromodynamics (QCD) \cite{Na:2015kha,Lattice:2015rga,Bailey:2014tva,Harrison:2017fmw}, light-cone sum rules \cite{Faller:2008tr} and QCD sum rules \cite{Neubert:1992wq,Neubert:1992pn,Ligeti:1993hw}, without making use of experimental data. See Ref.~\cite{Jung:2018lfu} for more details on the FF parameters.
As in Ref.~\cite{Murgui:2019czp}, we allow the FF parameters to fluctuate around these input values, which are considered as pseudo-observables with their corresponding $\chi^2_{\mathrm{FF}}$ taken into account in the fits. Since this theoretical $\chi^2$ gives a very small contribution to the total $\chi^2$ of the fits, we will not discuss it again and refer to Ref.~\cite{Murgui:2019czp} for additional technical details.

\subsection{Scenarios and fit results}
\label{sec:scenarios}
%
As previously mentioned, by adding RHN, the set of operators increases from 5 to 10. The large number of free parameters makes difficult to perform a global fit to the full basis of operators.  Instead, we will work in different motivated scenarios that arise by integrating out a single NP mediator and, therefore, contribute to small subsets of operators at the $m_b$ scale. Possible candidates, their quantum numbers and the operators generated once the given mediator is integrated out are listed in Table~\ref{table:NPQN}. The last two columns show the operators involving left-handed and right-handed neutrinos. Following previous works, we consider scenarios that only take into account the contributions from RHN operators, labelling them with the letter ``{\it a}"  \cite{Robinson:2018gza}, while ``{\it b}" scenarios also contain the LHN operators that are generated in the presence of the corresponding mediators. In addition, we define \textit{Scenarios 1} and \textit{2}, which correspond to consider only right-handed operators, with and without the SM-like contributions, respectively. The set of scenarios that we are going to analyse and the operators involved in each case are:
\begin{itemize}
    \item[1)]  RHN + SM-like contribution: \ $\mathcal{O}^{V}_{LL}\, , \mathcal{O}^{V}_{LR}\, , \mathcal{O}^{V}_{RR}\, , \mathcal{O}^{S}_{LR}\, , \mathcal{O}^{S}_{RR}\, , \mathcal{O}^{T}_{RR}\,$,
    \item[2)] RHN: \ $\mathcal{O}^V_{LR}\, ,$
    $\mathcal{O}^{V}_{RR}\, , \mathcal{O}^{S}_{LR}\, , \mathcal{O}^{S}_{RR}\, , \mathcal{O}^{T}_{RR}\,$,
    \item[3)] $V^{\mu}$: \ $\mathcal{O}^{V}_{RR}\,$,
    \item[4a)] $\Phi$: \ $\mathcal{O}^{S}_{LR}\, , \mathcal{O}^{S}_{RR}\,$,
    \item[4b)] $\Phi$: \ $\mathcal{O}^{S}_{LL}\, , \mathcal{O}^{S}_{RL}$ \ and \ $ \mathcal{O}^{S}_{LR}\, , \mathcal{O}^{S}_{RR}\,$,
    \item[5a)] $U_{1}^\mu$: \ $\mathcal{O}^{V}_{RR}\, , \mathcal{O}^{S}_{LR}\,$,
    \item[5b)] $U_{1}^\mu$: \ $\mathcal{O}^{V}_{LL}\, , \mathcal{O}^{S}_{RL}$ \ and \ $ \mathcal{O}^{V}_{RR}\, , \mathcal{O}^{S}_{LR}\,$,
    \item[6)] $\tilde{R}_2$: \ $\mathcal{O}^{S}_{RR}\, ,  \mathcal{O}^{T}_{RR}$ \ with \ $C^{S}_{RR} = 4 r \, C^{T}_{RR}\,$,
    \item[7a)] $S_1$: \ $\mathcal{O}^V_{RR}\, , \mathcal{O}^S_{RR}\, , \mathcal{O}^{T}_{RR}$ \ with \ $C^S_{RR} = - 4 r \, C^{T}_{RR}\,$,
    \item[7b)] $S_1$: \ $\mathcal{O}^V_{LL}\, , \mathcal{O}^S_{LL}\, , \mathcal{O}^{T}_{LL}$ \ and \ $\mathcal{O}^V_{RR}\, , \mathcal{O}^S_{RR}\, , \mathcal{O}^{T}_{RR}$ \ with \ $C^S_{LL} = - 4 r \, C^{T}_{LL}$ \ and \ $C^S_{RR} = - 4 r \, C^{T}_{RR}\,$,
    \item[8)] $\tilde{V}_{2}^\mu$: \ $\mathcal{O}^{S}_{LR}\,$. 
\end{itemize}
\textit{Scenarios 3, 6} and \textit{8} do not generate any left-handed operator, making the ``{\it a}'' and ``{\it b}'' labelling unnecessary. In \textit{Scenarios 6, 7a} and \textit{7b}, where scalar and tensor couplings arise at the NP scale, the renormalization-group running between $\Lambda_\text{NP} \sim 1$ TeV and the scale $m_b$ generates the factor $r \approx 2$. \textit{Scenarios 3} to \textit{7} have been also studied at Ref.~\cite{Robinson:2018gza}.

Within each scenario we will perform a standard $\chi^2$ fit to the data. There are 60 experimental degrees of freedom (d.o.f.), 4 corresponding to $\rdrdst{},\, \FLint{}$ and $\ptauint{}$, and 56 to the binned $q^2$ distributions. Therefore, the number of d.o.f. of our fits is $60 - N_{\text{WC}} - 1 = 59- N_{\text{WC}}$, where $N_\text{WC}$ is the number of Wilson coefficients entering in the fit.

All solutions resulting from our fits will present up to three \textit{flipped} minima with degenerate $\chi^2$ values. The first flipped minimum is obtained by reversing the sign of the LHN Wilson coefficients while keeping the right-handed Wilson coefficients untouched: 
\begin{equation}
    C^{V'}_{LL} = - 2 - C^{V}_{LL}\, ,  \, \qquad C^{X'}_{iL} = - C^{X}_{iL} \, ,  \, \qquad C^{X'}_{iR} =  C^{X}_{iR}\,, 
\label{signflip}
\end{equation}
for $X = S,V, T$ and $i = L, R$, except for  $C^{V}_{LL}$. The second flipped minimum is obtained reversing  only the right-handed coefficients, 
\begin{equation}
   C^{X'}_{iL} =  C^{X}_{iL} \, ,  \, \qquad C^{X'}_{iR} =  - C^{X}_{iR}\, , 
    \label{signflip2}
\end{equation}
for $X = S,V,T$ and $i = L, R$, and the last one flipping both left and right Wilson coefficients,
\begin{equation}
    C^{V'}_{LL} = - 2 - C^{V}_{LL}\, ,  \, \qquad C^{X'}_{iL} = - C^{X}_{iL} \, ,  \, \qquad C^{X'}_{iR} =  - C^{X}_{iR}\, ,  
        \label{signflip3}
\end{equation}
for $X = S,V, T$ and $i = L, R$,  except for $C^{V}_{LL}$. 
From now on, we will only discuss
the minimum which is closest to the SM scenario.

In the following subsections, we will present the fitted solutions for each considered scenario. 
Whenever some uncertainties are marked
with the symbol ${}^{\dagger}$ (i.e., $C^V_{RR} = -0.69^{+0.64^{\dagger}}_{-0.44}$), this indicates that the $\chi^2$ distribution has fallen to another minimum. In these cases, the uncertainty is defined as the range between the central value and the point in which the $\chi^2$ falls to  the other minimum.
To complete the discussion, it is interesting to see the predicted values of the different observables within each fitted scenario. This information is given in Fig.~\ref{fig:preds} and in Table~\ref{tab:predictions}, where the numerical predictions are marked either with a green tick (\textcolor{Darkgreen}{\cmark})  if they agree with the experimental value at $1 \sigma$ or with a red cross (\textcolor{red}{\xmark}) if they do not agree.  All minima are in agreement with all experimental observables at the $2\sigma$ level.

\subsubsection{SM fit}
%
The SM fit, 
where all the Wilson coefficients are set to zero, i.e. $C^X_{AB}=0$, gives us the following $\chi^2$:
\begin{equation}
\chi^{2}_{\text{SM}}/{\text{d.o.f.}} = 52.87/59,
\end{equation}
corresponding to a 69.95\% probability ($p$-value, defined below). The ``apparent" good quality of the fit, i.e. $\chi^2_\text{SM}/ {\rm d.o.f.} <1$, might be surprising since it contrasts with the approximately $3\sigma$ discrepancy claimed in the
${\cal R}_D$ and ${\cal R}_{D^*}$ measurements.
This can be understood by looking at the split up contributions of the fit inputs. Considering only the contribution of the $q^2$ distributions we find that $\chi^{2}_{\text{SM}}(q^2 \, \text{distributions})/{\text{d.o.f.}} = 36.77/56$, while
$\chi^{2}_{\text{SM}}(\rdst{},  \FLint{}, \mathcal{\bar{P}}_{\tau}^{D^*})/{\text{d.o.f.}} = 16.1/4$, %
corresponding to a $2.98\sigma$ tension for the later. Taking into account only the $\chi^2$ value of $\rdrdst{}$ we obtain
$13.36$ for 2 d.o.f., recovering the well-known $3.2\sigma$ tension.

The last results suggest an overestimation of the absolute $\chi^2$ value, which is introduced while considering in the fit multiple inputs with large uncertainties as, in our case, the $q^2$ distributions for the $B$-meson semileptonic decays. 
The goodness of a fit is usually characterized through the $p$-value, defined as
\be p(\chi^2_{\text{min}},n)\, \equiv \, \int_{\chi^2_{\text{min}}}^\infty dz\;\chi^2(z,n)\, ,
\ee
where  $\chi^2(z,n)$ is the $\chi^2$ probability distribution function with $n$ d.o.f..
Larger $p$-values correspond to better explanations of the experimental data than lower ones.
In order to quantify the quality of our fit, it is convenient to introduce  another parameter 
called {\it Pull} that compares any fitted solution with the SM results. 
This statistical measure is defined as the probability in units of $\sigma$ corresponding to the difference $\Delta \chi_i^2 \equiv \chi^2_\text{SM} - \chi^2_i$, assuming that $\Delta \chi_i^2$ follows a $\chi^2$ distributed function with $\Delta n_i \equiv n_\text{SM} - n_i $ d.o.f., where the label $i$ refers to the $i$th scenario. The translation from probability to sigmas is done by associating such probability to the one corresponding to a {\it Pull} number of standard deviations in a normal distribution with $\Delta n_i$ d.o.f.,\footnote{A probability of $(68.3\%,\, 95.5\%,\, 99.7\%)$ equals to $(1\sigma,\, 2\sigma,\, 3\sigma)$, respectively.} i.e. \cite{Capdevila:2017iqn,Capdevila:2018jhy}
\begin{equation}
    \text{Pull}_\text{SM}\, \equiv\, \text{prob}(\Delta \chi_i^2, \Delta n_i)[\sigma]\, =\, \sqrt{2}\, \text{Erf}^{-1}\big[ \text{CDF}(\Delta \chi_i^2,\Delta n_i)\big],
\end{equation}
where $\text{CDF}(\Delta \chi_i^2, \Delta n_i) \equiv 1 - p(\Delta \chi_i^2, \Delta n_i)$ is the $\chi^2$-cumulative distribution function evaluated at $\Delta \chi_i^2$ for $\Delta n_i$ d.o.f..

In Table~\ref{tab:pulls} we display the $\text{Pull}_{\text{
SM}}$ values of the different fitted minima, together with their corresponding $p$-values, for all the scenarios analysed. In order to better quantify how favourable are the fitted scenarios with respect to the SM regarding the different observables entering in the fit, we also include their pull for the particular pieces of the $\chi^2$, splitting it into three contributions: the polarization observables $\ptauint$ and $\bar{F}_L^{D^*}$, the ratios ${\cal R}_D$ and ${\cal R}_{D^*}$ and the $q^2$-distributions of the $B \to D^{(*)} \tau \bar{\nu}$ decay. In the former we ignore the FF contribution to the $\chi^2$. As we can see in Table~\ref{tab:pulls}, all scenarios exhibit a sizeable improvement with respect to the SM $p$-value.

\subsubsection[\texorpdfstring{Scenario 1: $\nu_R$ +   SM-like}{dum2}]{\texorpdfstring{\boldmath Scenario 1: $\nu_R$ +   SM-like}{dum2}}
%
Considering only RHN operators and the SM-like contribution, i.e. $C^V_{LL}$, and imposing an upper bound for ${\cal B}(B_c \to \tau \bar{\nu})$ of 10\%, we find two different solutions: a global minimum and a local one with a slightly higher $\chi^2$, i.e.
\beqn
\no
&\chi^2/\text{d.o.f.}& = 37.26/53 \, , \\ \no
&C^{V}_{LL} &=-0.36^{+0.34}_{-0.64^{\dagger}} ,\phantom{0}\qquad  C^V_{LR} =\phantom{-}  1.10^{+0.46}_{-0.50}\, , \qquad   C^V_{RR} = \phantom{-} 0.031^{+0.14}_{-0.17}\, , 
\\ 
&C^S_{RR} &= -0.03^{+0.18}_{-0.60}\, , \qquad\;\; C^S_{LR} = -0.29^{+0.31}_{-0.53}\, , \qquad C^T_{RR} = -0.105^{+0.066}_{-0.084}\, ,
\eeqn
and 
\beqn
\no
&\chi^2/\text{d.o.f.}& = 38.86/53 \, , \\ \no
&C^{V}_{LL} &= -0.13^{+0.10}_{-0.82}\,, \qquad  C^V_{LR} = -0.09^{+0.29}_{-0.27}\, ,\phantom{0,} \qquad   C^V_{RR} = -0.69^{+0.64^{\dagger}}_{-0.44}, \\
&C^S_{RR} &=\phantom{-} 0.34^{+0.37}_{-0.56^{\dagger}}, \qquad C^S_{LR} = -0.030^{+0.74}_{-0.18^{\dagger}} , \qquad C^T_{RR} = -0.006^{+0.239^{\dagger}}_{-0.082} .
\eeqn
Shifting the Wilson coefficients up to $1.2\sigma$, the global minimum 
becomes compatible with a solution in which the only non-vanishing Wilson coefficients are $C^V_{LR}$ and $C^T_{RR}$. As it can be seen in Fig.~\ref{fig:IndFitPlot}, both $C^V_{LR}$ and $C^T_{RR}$ help to reproduce the experimental value of $\rd{}, \rdst{}$ and $\ptauint{}$. For $ \FLint{}$ it is a combination of several operators that helps.  
In the  local minimum, the dominant contribution comes from $C^V_{RR}$.

As it can be seen in Table~\ref{tab:predictions}, both minima saturate the $\mathcal{B}(B_c \to \tau \bar{\nu}) \leq 10\%$ constraint. Thus, relaxing it to be up to a 30\%, we find
\beqn
\no
&\chi^2/\text{d.o.f.}& = 36.42/53 \, , \\ \no
&C^{V}_{LL} &=-0.50^{+0.41}_{-0.49^{\dagger}}, \qquad  C^V_{LR} = \phantom{-} 1.34^{+0.25}_{-0.60}\, , \qquad   C^V_{RR} = \phantom{-} 0.204^{+0.298}_{-0.020}\, , \\
&C^S_{RR} &= -0.22^{+0.27^{\dagger}}_{-0.27}, \qquad C^S_{LR} = -0.92^{+0.22^{\dagger}}_{-0.15} , \qquad C^T_{RR} = -0.123^{+0.069}_{-0.077}\, ,
\eeqn
and
\beqn
\no
&\chi^2/\text{d.o.f.}& = 38.54/53 \, , \\ \no
&C^{V}_{LL} &=-0.15^{+0.21}_{-0.86}\, , \qquad  C^V_{LR} = -0.15^{+0.31^{\dagger}}_{-0.17^{\dagger}}, \qquad   C^V_{RR} = -0.69^{+0.70}_{-0.42}\, , \\
&C^S_{RR} &=  \phantom{-} 0.59^{+0.38^{\dagger}}_{-0.41}, \qquad C^S_{LR} = -0.24^{+0.61}_{-0.13^{\dagger}} , \qquad C^T_{RR} = \phantom{-}0.007^{+0.114}_{-0.087}\, .
\eeqn
The value of the $\chi^2/{\text{d.o.f.}}$ slightly improves in this case, whereas the scalar Wilson coefficients are further away from the SM limit. 

In both cases one can see that most of the Wilson coefficients have large uncertainties.
This can be understood from the fact that a large set of variables to fit allow for larger correlations among them, which in turn allows wider ranges for the Wilson coefficients considered.
The global and local minima have in fact quite close values of $\chi^2$/d.o.f., and the $\chi^2$ distribution in the region between them is rather flat. Thus, when evaluating their $1\sigma$ variations, one minimum falls often into the other one, as indicated by the ${}^\dagger$ symbols.

This scenario is the most general, in the sense that the preferred $C_{LL}^V$ solution without considering RHNs~\cite{Murgui:2019czp} is included in the fit, together with all
possible contributions generated as a consequence of having RHNs. No specific NP scenario has been assumed in here.

\subsubsection[\texorpdfstring{Scenario 2: $\nu_R$ }{dum2}]{\texorpdfstring{\boldmath Scenario 2: $\nu_R$}{dum2}}
%
In this scenario we consider solely the contribution to $b \to c$ processes coming from the presence of RHNs in the theory. Again, this assumption is very general and model independent, in the sense that no specific types of NP mediators are assumed. 

As in the previous scenario, with the constraint $\mathcal{B}(B_c \to \tau \bar{\nu}) \leq 10\%$, a global and a local minimum are obtained: 
\beqn
\no
&\chi^2/\text{d.o.f.}& = 38.54/54 \, , \\ \no
&C^V_{LR}& = 0.52^{+0.13}_{-0.16}\, , \qquad   C^V_{RR} = \phantom{-} 0.06^{+0.15}_{-0.22}\, , \\
&C^S_{RR} &= 0.04^{+0.35}_{-0.66}\, , \qquad C^S_{LR} = -0.35^{+0.72}_{-0.16}\, , \qquad C^T_{RR} = -0.057^{+0.080}_{-0.058}\, ,
\eeqn
and
\beqn
\no
&\chi^2/\text{d.o.f.}& = 39.05/54  \, , \\ \no
&C^V_{LR}& = \phantom{+}0.07^{+0.30}_{-0.30^{\dagger}}\, , \qquad   C^V_{RR} = 0.42^{+0.11}_{-0.21}\, , \\
&C^S_{RR} &= - 0.32^{+0.74}_{-0.21}\, , \, \qquad C^S_{LR} = 0.10^{+0.20^{\dagger}}_{-0.68} , \qquad C^T_{RR} = 0.004^{+0.080}_{-0.088}\, .
\eeqn
By shifting all the Wilson coefficients within their $1\sigma$ uncertainties, the global minimum is compatible with a solution in which the only non-zero coefficient is $C^V_{LR}$. 
This coincides with the fit dealing only with the LHN operators where the global minimum was compatible with a global shift of the SM-like operator (i.e. $C_{LL}^V \neq 0$ ) \cite{Murgui:2019czp}. In other words, $C^V_{LR}$ plays a similar role as
the $\nu_L$ Wilson coefficient modifying the SM contribution.  
In the local minimum, the main contributions to the observables are coming from $C^V_{RR}$.

Since the previous fit saturates the leptonic $B_c$ decay bound, we list below the minima obtained after relaxing such constraint to $\mathcal{B}(B_c \to \tau \bar{\nu}) \leq 30\%$: 
\beqn
\no
&\chi^2/\text{d.o.f.}& = 38.33/54 \, ,  \\ \no
&C^V_{LR}& = 0.47^{+0.16}_{-0.20}\, , \qquad   C^V_{RR} = \phantom{-} 0.10^{+0.21}_{-0.23}\, , \\
&C^S_{RR} &= 0.28^{+0.24}_{-0.97}\, , \qquad C^S_{LR} = -0.59^{+0.80}_{-0.17}\, , \qquad C^T_{RR} = -0.054^{+0.081}_{-0.058}\, ,
\eeqn
and
\beqn
\no
&\chi^2/\text{d.o.f.}& = 38.80/54 \, ,  \\ \no
&C^V_{LR}& = \phantom{-} 0.12 \pm 0.30\, , \qquad   C^V_{RR} = 0.38^{+0.13}_{-0.20}\, , \\
&C^S_{RR} &= -0.57^{+0.57^{\dagger}}_{-0.28}, \qquad \hspace{0.3cm} C^S_{LR} = 0.33^{+0.20^{\dagger}}_{-0.48^{\dagger}} , \qquad C^T_{RR} = -0.006^{+0.081}_{-0.091}\, .
\eeqn
Similarly to the previous scenario, when relaxing the leptonic decay bound, the $\chi^2$ experiences an improvement and the scalar Wilson coefficients further depart from the SM limit.

\FloatBarrier

\subsubsection[\texorpdfstring{Scenario 3: $V_{\mu}$ }{dum2}]{\texorpdfstring{\boldmath Scenario 3:  $V_{\mu}$}{dum2}}
%
The mediator $V^\mu \sim (1,1,-1)$ only involves interactions with RHN regarding $ b \to c $ transitions. Note that we call it $V^\mu$ instead of the usual nomenclature $W'^\mu$ in order to distinguish it from the $SU(2)$ triplet which does couple to the LHNs. Therefore, this scenario induces exclusively $b \to c \tau \bar{\nu}_R$ interactions, and particularly the $V^\mu$ only contributes to the vector Wilson coefficient $C_{RR}^V$.\\

\begin{figure}[ht]
    \centering
    \includegraphics[scale=0.55]{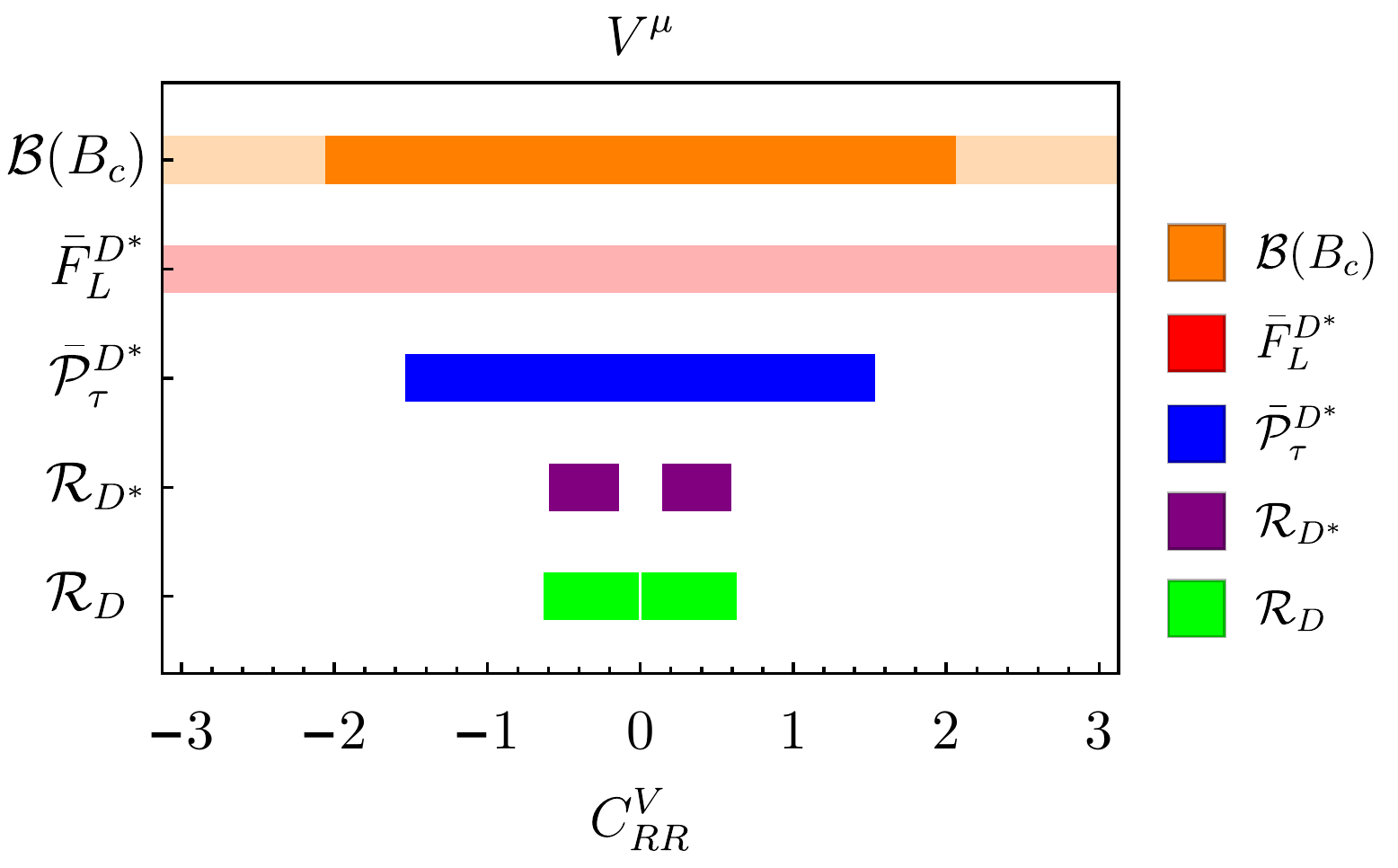}
    \caption{Fit-independent plot of \textit{Scenario 3}.  Dark colours indicate the allowed regions satisfying the experimental constraints at $1\sigma$ and the 10\% upper limit on $\mathcal{B}(B_c \to \tau \bar{\nu})$, for a given value of $C^V_{RR}$. 
    There is no allowed region for $ \FLint{}$ at $1\sigma$.
    The lighter orange and red shaded areas correspond to the more relaxed 30\% bound on the leptonic $B_c$ decay and the $2\sigma$ region for $ \FLint{}$, respectively.%
    }
    \label{fig:case3}
\end{figure}

The global fit gives us the minimum value for this Wilson coefficient together with its $\chi^2$:
\beqn
\no
&\chi^2/\text{d.o.f.}& = 39.50/58\, , \\
&C^{V}_{RR}& = 0.370^{+0.051}_{-0.059}\, .
\eeqn
Given that in this case our model depends on a single Wilson coefficient, we can study the regions of the parameter space that reproduce the different experimental observables included in the global fit from a fit-independent perspective, as shown in Fig.~\ref{fig:case3}. This figure shows that no region of common overlap can be found at $1\sigma$. This agrees with  Fig.~\ref{fig:IndFitPlot}, which showed that the shift of a single Wilson coefficient with respect to the SM scenario does not modify the $ \FLint{}$ prediction. We also indicate in Fig.~\ref{fig:case3} the parameter space allowed when relaxing the experimental constraint on $\FLint{}$  to $2\sigma$ and taking ${\cal B}(B_c\to \tau \bar{\nu})\le 30\%$. As expected, in that context we find full agreement with the experiment.

\FloatBarrier

\subsubsection[\texorpdfstring{Scenario 4a: $\Phi$ }{dum2}]{\texorpdfstring{\boldmath Scenario 4a:  $ \Phi$}{dum2}}

Considering that the mediator $\Phi \sim (1,2,1/2)$, with the same quantum numbers as the SM Higgs, is responsible for the NP interactions, and assuming that only right-handed Wilson coefficients appear at the low-energy scale, two different minima with the same $\chi^2$ value, 
\beqn
\no
&\chi^2/\text{d.o.f.}& = 49.93/57\, , \\
&C^S_{RR}& = 0.46^{+0.05}_{-0.18}\,  , \qquad C^S_{LR}= -0.06^{+0.19}_{-0.07}\,  , 
\eeqn
and
\beqn
\no
&\chi^2/\text{d.o.f.}& = 49.93/57\, , \\
&C^S_{RR}& = 0.06^{+0.07}_{-0.19}\, , \qquad
C^S_{LR} = -0.46^{+0.18}_{-0.05}\,  , 
\eeqn
are found.
As one can see, they correspond to degenerate solutions, flipping the values of $C^S_{LR}$ and $C^S_{RR}$. This can be easily understood by looking at the expressions of $B \to D$ and $B \to D^*$ listed in Eqs.~\eqref{eq:Ddist} and \eqref{eq:GammaDstar}, respectively.  
These observables depend on the absolute values of the right-handed scalar and pseudoscalar combinations of Wilson coefficients when the vector coefficients are switched off, and therefore 
remain invariant under the exchange $C^S_{LR} \leftrightarrow C^S_{RR}$. The same is true for the $D^*$ polarization observables that, as shown in Eqs.~\eqref{eq:PDRDst} and \eqref{eq:FLDRDst}, are blind to a sign flip of the combination $C^S_{RR}-C^S_{LR}$.
As Table~\ref{tab:predictions} shows, these minima saturate the $\mathcal{B}(B_c \to \tau \bar{\nu})\le 10\%$ bound. Relaxing this constraint to $\mathcal{B}(B_c \to \tau \bar{\nu}) \leq 30\%$, the minima read
\beqn
\no
&\chi^2/\text{d.o.f.}& = 44.49/57\, , \\
&C^S_{RR}& = 0.297^{+0.074}_{-0.096}\, , \qquad C^S_{LR}= -0.673^{+0.091}_{-0.053}\, , 
\eeqn
and
\beqn
\no
&\chi^2/\text{d.o.f.}& = 44.49/57\, , \\
&C^S_{RR}& = 0.673^{+0.053}_{-0.091}\,  , \qquad
C^S_{LR} = -0.297^{+0.096}_{-0.074}\, ,
\eeqn
where, as expected, the pseudoscalar combination of Wilson coefficients increases its value and the $\chi^2$ slightly improves.

\begin{figure}[ht]
    \centering
    \includegraphics[scale=0.45]{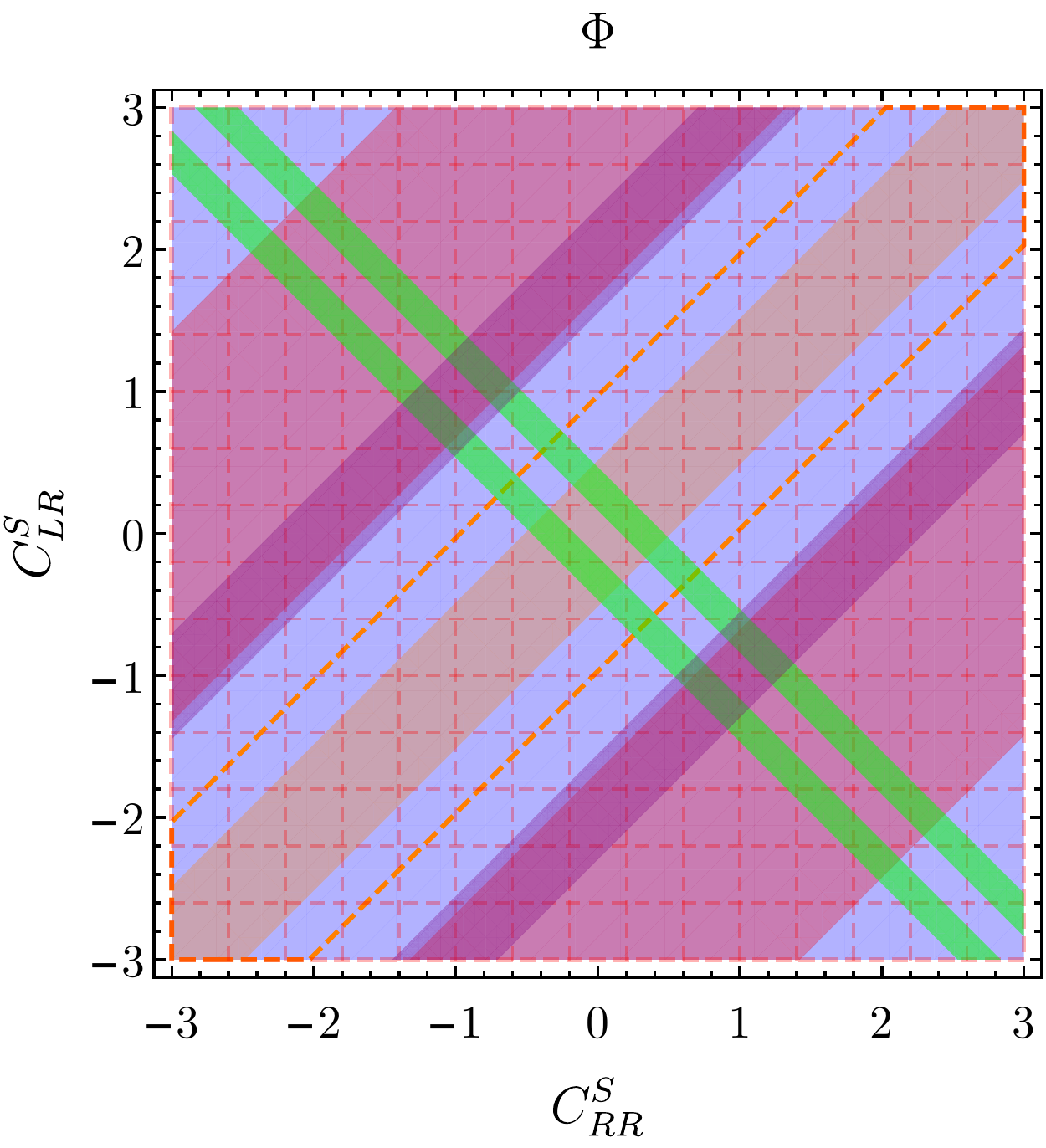} \quad 
        \includegraphics[scale=0.7]{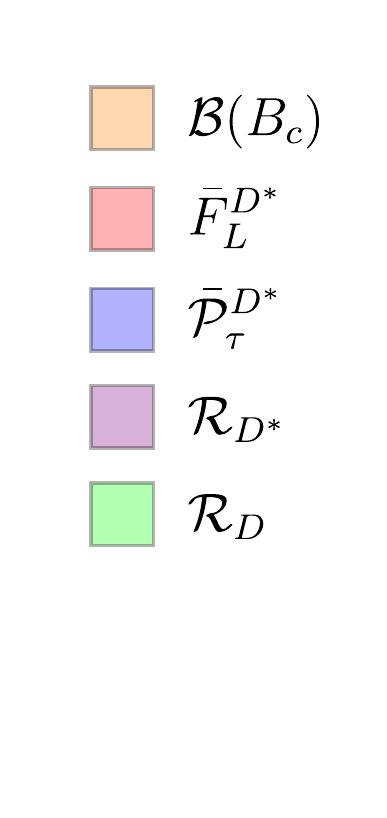}
    \includegraphics[scale=0.45]{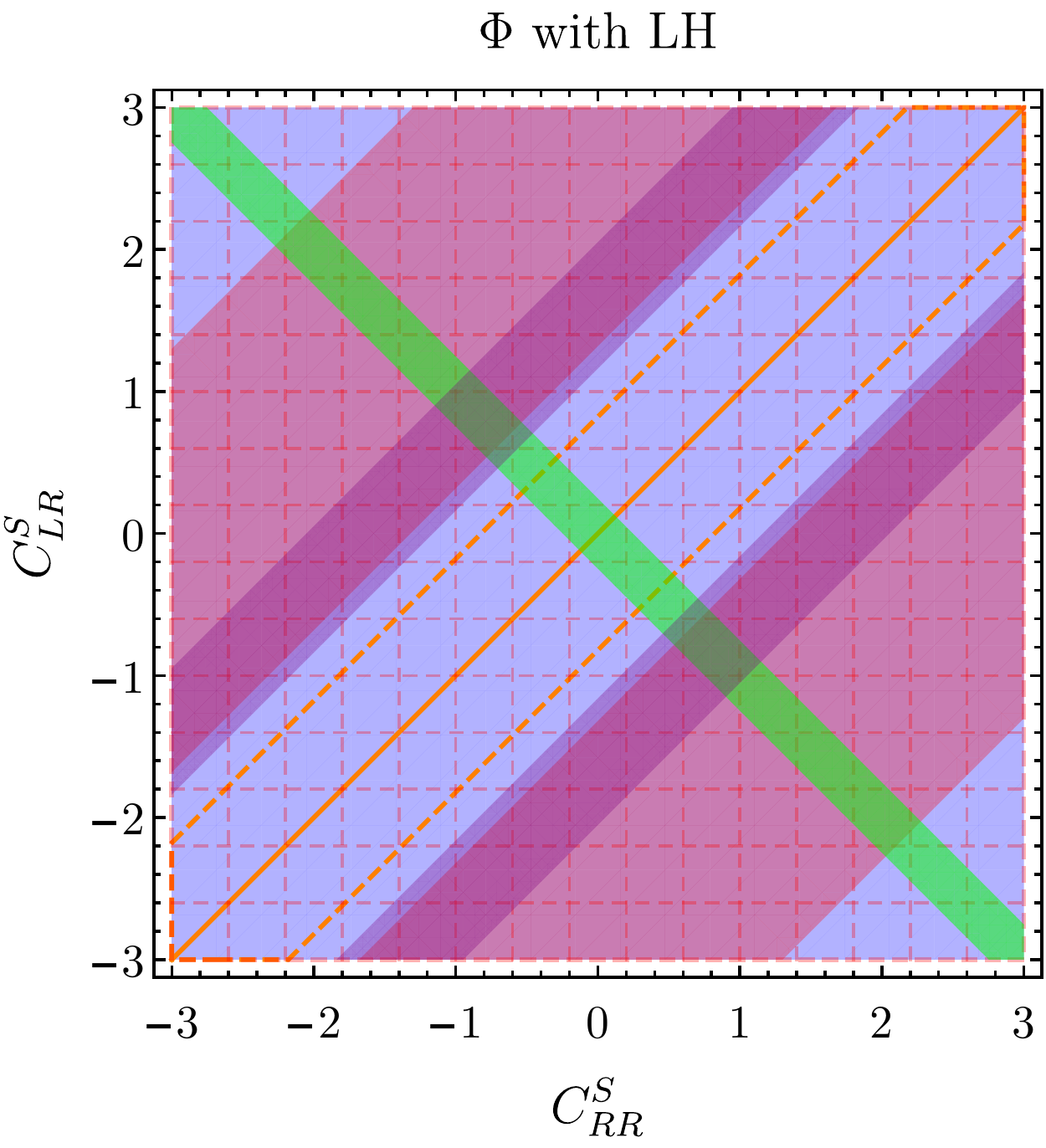}
    \caption{Fit-independent plot of \textit{Scenario 4}, displaying the regions allowed at $1\sigma$.  On the left panel only the RHN Wilson coefficients shown are switched on (\textit{Scenario 4a)}, whereas on the right panel we set the left-handed  neutrino Wilson coefficients entering in \textit{Scenario 4b} to their best-fit values. The dashed orange line shows the more relaxed bound $\mathcal{B}(B_c \to \tau \bar{\nu}) \le 30\%$, and the red grid shows the parameter space consistent with the experimental measurement of $ \FLint{}$ at  $2\sigma$.}
    \label{fig:case4}
\end{figure}

In the left panel of Fig.~\ref{fig:case4} we show the two-dimensional parameter space where the different observables entering in the fit are satisfied at $1\sigma$. As the figure shows, there is no overlap at this given probability. In this case, not even relaxing the leptonic $B_c$ decay upper bound to $30\%$ and the $F_L^{D^*}$ experimental measurement to $2\sigma$, an overlap in the parameter space is achieved.
%

\subsubsection[\texorpdfstring{Scenario 4b: $\Phi$ }{dum2}]{\texorpdfstring{\boldmath Scenario 4b:  $ \Phi$}{dum2}}

The Two Higgs Doublet Models are the simplest examples of UV physics generating this scenario. 
In addition to RHN operators, a second scalar doublet with the same quantum numbers as the SM one generates LHN Wilson coefficients.
The preferred solution of this scenario corresponds to vanishing right-handed 
Wilson coefficients, which eliminates the degeneracy under $C^S_{LR} \leftrightarrow C^S_{RR}$. Owing to the interference with the SM-like contribution, an analogous symmetry does not exist for the left-handed coefficients and, therefore, we find in this case a single solution with $\mathcal{B}(B_c \to \tau \bar{\nu}) \leq 10\%$:
\beqn
\no
&\chi^2/\text{d.o.f.}& =43.56/55\, , \\ \no
&C^S_{RL}& = 0.21^{+0.03}_{-0.11}\, , \qquad C^S_{LL} = -0.11^{+0.07}_{-0.08}\, , \\ 
&C^S_{RR}& = 0.0 \pm 0.3\, , \qquad C^S_{LR} = \phantom{-} 0.0 \pm 0.3\, .
\eeqn
With the relaxed limit $\mathcal{B}(B_c \to \tau \bar{\nu}) \leq 30\%$, the splitting between scalar operators is larger and the $\chi^2$ slightly improves:
\beqn
\no
&\chi^2/\text{d.o.f.}& =40.03/55\, , \\ \no
&C^S_{RL}& = 0.407^{+0.032}_{-0.137}\, , \qquad C^S_{LL} = -0.329^{+0.146}_{-0.080}\, , \\ 
&C^S_{RR}& = 0.00 \pm 0.45\, , \qquad C^S_{LR} = \phantom{-} 0.00 \pm 0.45\, .
\eeqn
The right panel of Fig.~\ref{fig:case4} shows the two dimensional parameter space where the observables entering in the fit are satisfied at $1\sigma$. In this figure, the LHN operators are fixed at their best-fit values. As it can be seen, there is no overlap at this given significance level.
The non-existing overlap is also reflected in Table~\ref{tab:predictions} and 
Fig.~\ref{fig:case4}, where one can see that scalar solutions cannot satisfy ${\cal R}_{D^*}$, nor $\FLint{}$. The later is also shown in a very intuitive way in Fig.~\ref{fig:IndFitPlot}.

\FloatBarrier

\subsubsection[\texorpdfstring{Scenario 5a: $U_{1 \mu}$ }{dum2}]{\texorpdfstring{\boldmath Scenario 5a: $U_{1 \mu}$}{dum2}}
\begin{figure}[th]
    \centering
    \includegraphics[scale=0.45]{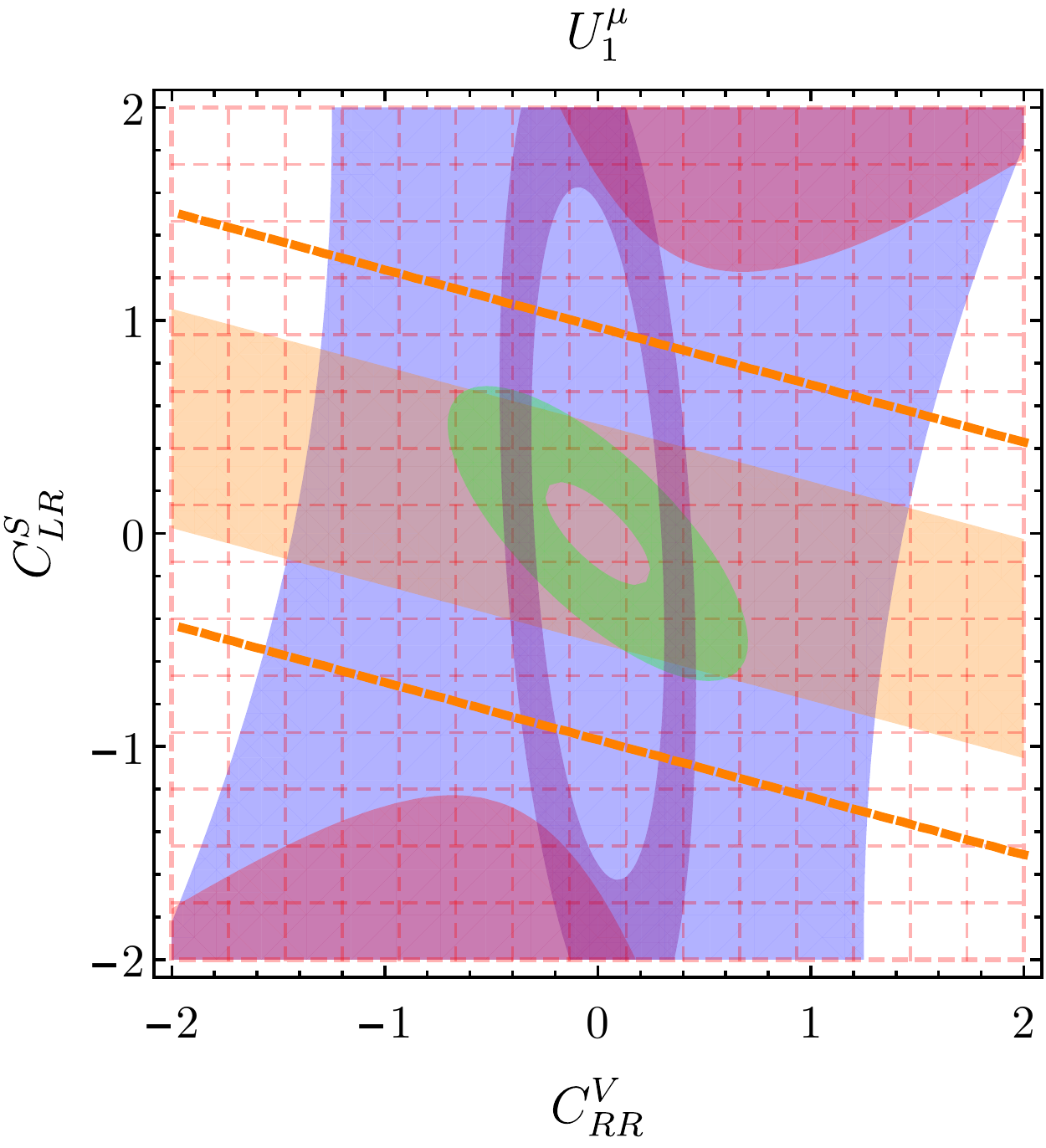} \quad
            \includegraphics[scale=0.7]{img/Leg.pdf}
    \includegraphics[scale=0.45]{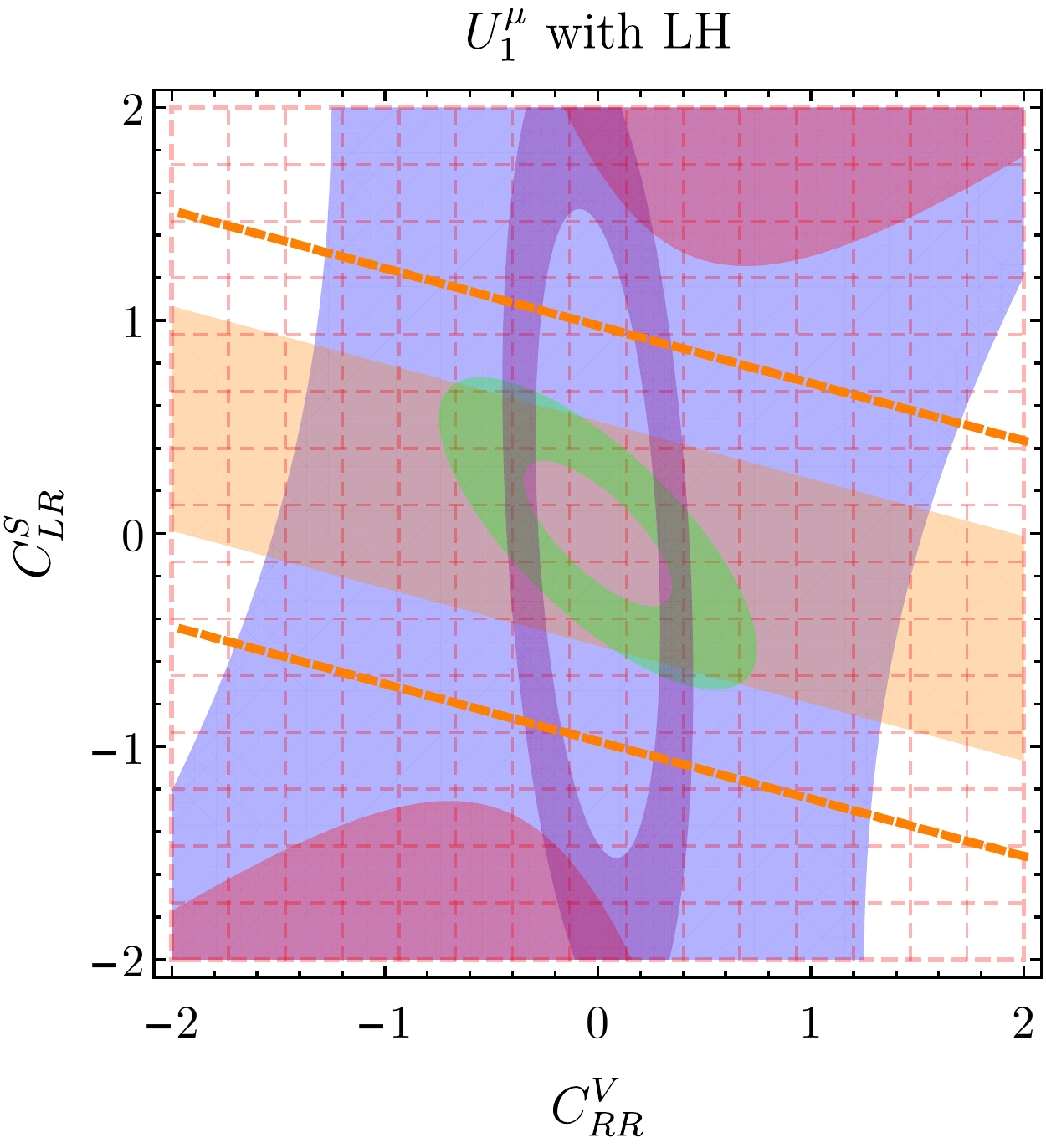}
    \caption{Fit-independent plot of \textit{Scenario 5}, showing the regions allowed at $1\sigma$. On the left panel only the right handed neutrino Wilson coefficients  are switched on (\textit{Scenario 5a}), whereas on the right panel we set the LHN Wilson coefficients entering in \textit{Scenario 5b} to their best-fit values. The dashed orange line shows the more relaxed bound $\mathcal{B}(B_c \to \tau \bar{\nu}) \le 30\%$, and the red grid indicates the parameter space consistent with the experimental measurement of $ \FLint{}$ at 2$\sigma$.}
    \label{fig:case5}
\end{figure}

%
The presence of the vector leptoquark $U_{1 \mu} \sim (3,1,2/3)$ at the high-energy scale will contribute to both left and right-handed operators at the $m_b$ scale. This vector leptoquark can be UV-completed in Pati-Salam based unification theories \cite{Pati:1974yy,Barbieri:2015yvd, DiLuzio:2017vat, Bordone:2017bld, Blanke:2018sro, Calibbi:2017qbu} for instance. Considering only the RHN operators, the preferred solution is compatible with a non-zero value of $C^V_{RR}$ while $C_{LR}^S = 0$ at $0.4\sigma$, i.e.
\beqn
\no
&\chi^2/\text{d.o.f.}&= 39.39/57\, , \\
&C^V_{RR}& = 0.39^{+0.07}_{-0.08}\, ,  \qquad C^S_{LR} = -0.1^{+0.2}_{-0.5}\, .
\eeqn
Since the scalar coefficient is suppressed, the ${\cal B}(B_c \to \tau \bar{\nu})$ limit is not saturated.
Furthermore, all the observables included in the fit agree at $1\sigma$, except $ \FLint{}$ which is compatible with the experimental value at $2\sigma$,   
as illustrated in the left-panel of Fig.~\ref{fig:case5}.
%


\subsubsection[\texorpdfstring{Scenario 5b: $U_{1 \mu}$ }{dum2}]{\texorpdfstring{\boldmath Scenario 5b: $U_{1 \mu}$}{dum2}}
%
Including the contributions to LHN operators, the value of the $\chi^2$ remains almost constant with respect to \textit{Scenario 5a}, $\Delta \chi^2 = -0.02$ for 2 new d.o.f., and the left-handed Wilson coefficients are compatible with zero within $1\sigma$: 
\beqn
\no
&\chi^2/\text{d.o.f.}& = 39.37/55\, , \\ \no
&C^V_{LL}& = 0.01^{+0.10}_{-0.65}\, ,  \qquad C^S_{RL} =  -0.03^{+0.07}_{-0.45}\, , \\
&C^V_{RR}& = 0.38^{+0.60}_{-1.40}\, ,  \; \, \qquad C^S_{LR} = -0.01^{+0.56}_{-0.54}\, . 
\eeqn
This indicates that the best solution for a leptoquark with these quantum numbers involves only RHN operators.

The right panel in Fig.~\ref{fig:case5} shows the small changes on the allowed regions, in comparison with \textit{Scenario 5a} (left panel). Again, all observables are satisfied at $1\sigma$, except for $ \FLint{}$.

\subsubsection[\texorpdfstring{Scenario 6:$\tilde{R}_2$ }{dum2}]{\texorpdfstring{\boldmath Scenario 6: $\tilde{R}_2$}{dum2}}
\begin{figure}[tbh]
    \centering
    \includegraphics[scale=0.55]{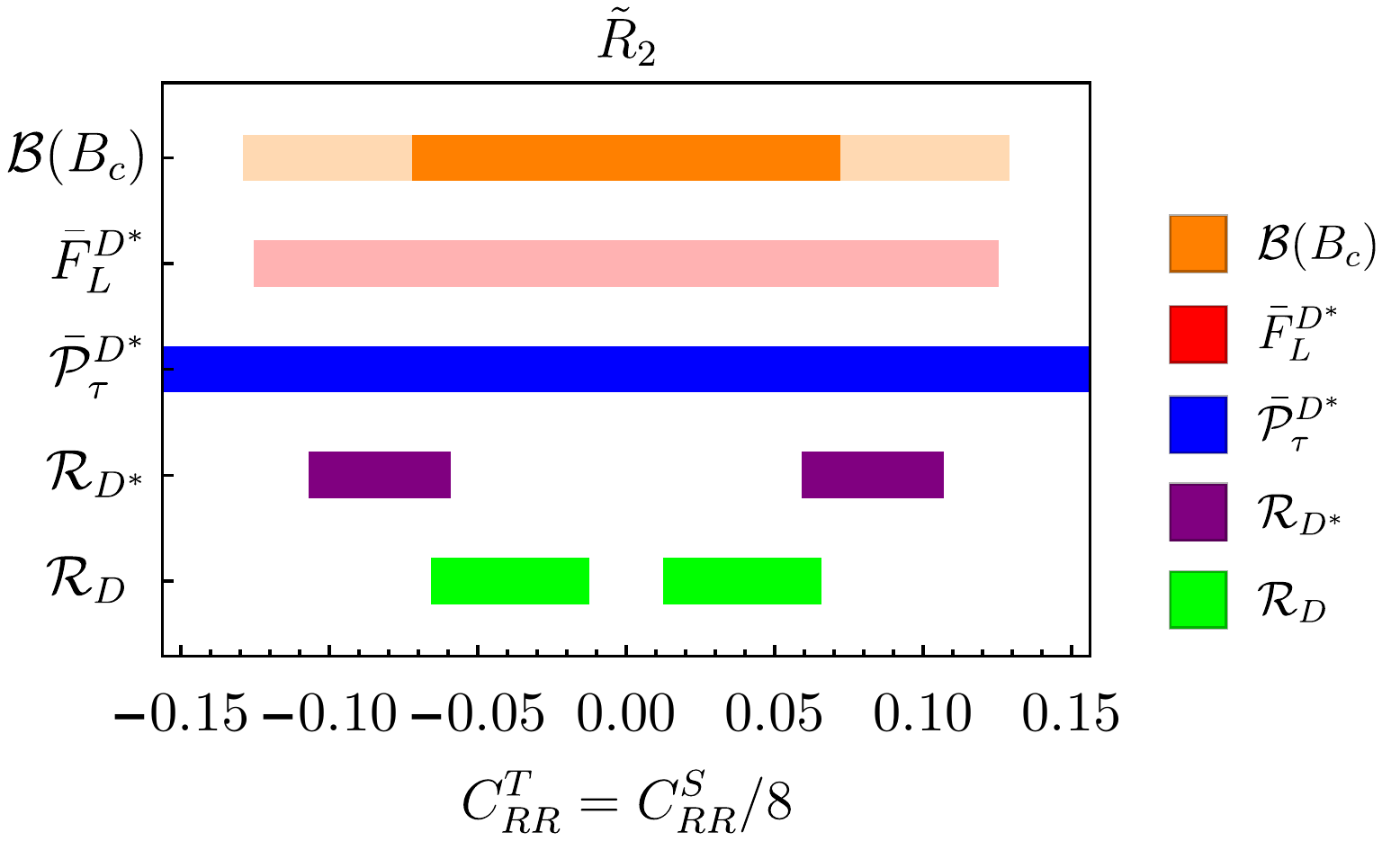}
    \caption{Fit-independent plot of \textit{Scenario 6}, showing the regions allowed at $1\sigma$ (dark colours), for different values of $C_{RR}^T$. 
    There is no allowed region for $ \FLint{}$ at $1\sigma$.
    The light orange and red shaded areas correspond to the more relaxed 30\% bound on the leptonic $B_c$ decay and the $2\sigma$ region for $ \FLint{}$, respectively.
    }
 \label{fig:case6}
\end{figure}

%
This scenario considers the solely presence of the scalar leptoquark $\tilde{R}_2\sim (3,2,1/6)$ \cite{Becirevic:2016yqi}.  It is genuine from the perspective of having RHNs, since it does not mediate any interaction involving left-handed ones. 
The global fit gives:
\beqn
\no
&\chi^2/\text{d.o.f.}& =44.20/58\, , \\
&C^T_{RR}& = 0.054^{+0.009}_{-0.011}\, .
\label{eq:Scen6}
\eeqn
In this case, there is only one free parameter, since the two relevant coefficients, $C^T_{RR}$ and $C^S_{RR}$, are correlated by the Fierz identities. Therefore, one can study the predictions of the fitted observables as a function of only one free parameter in a fit-independent manner, as we show in Fig~\ref{fig:case6}. The region with larger overlap in this figure corresponds to the minimum listed in Eq.~(\ref{eq:Scen6}) and its flipped solution. As in previous scenarios, it is not possible to reproduce the experimental value of $\FLint{}$ at $1\sigma$. However, agreement can be find when ${\cal B}(B_c \to \tau \bar{\nu})\leq 30\%$ and $\FLint{}$ is considered at $2\sigma$.

\FloatBarrier

\subsubsection[\texorpdfstring{Scenario 7a: $S_1$ }{dum2}]{\texorpdfstring{\boldmath Scenario 7a: $S_1$}{dum2}}
\begin{figure}[ht]
    \centering
    \includegraphics[scale=0.45]{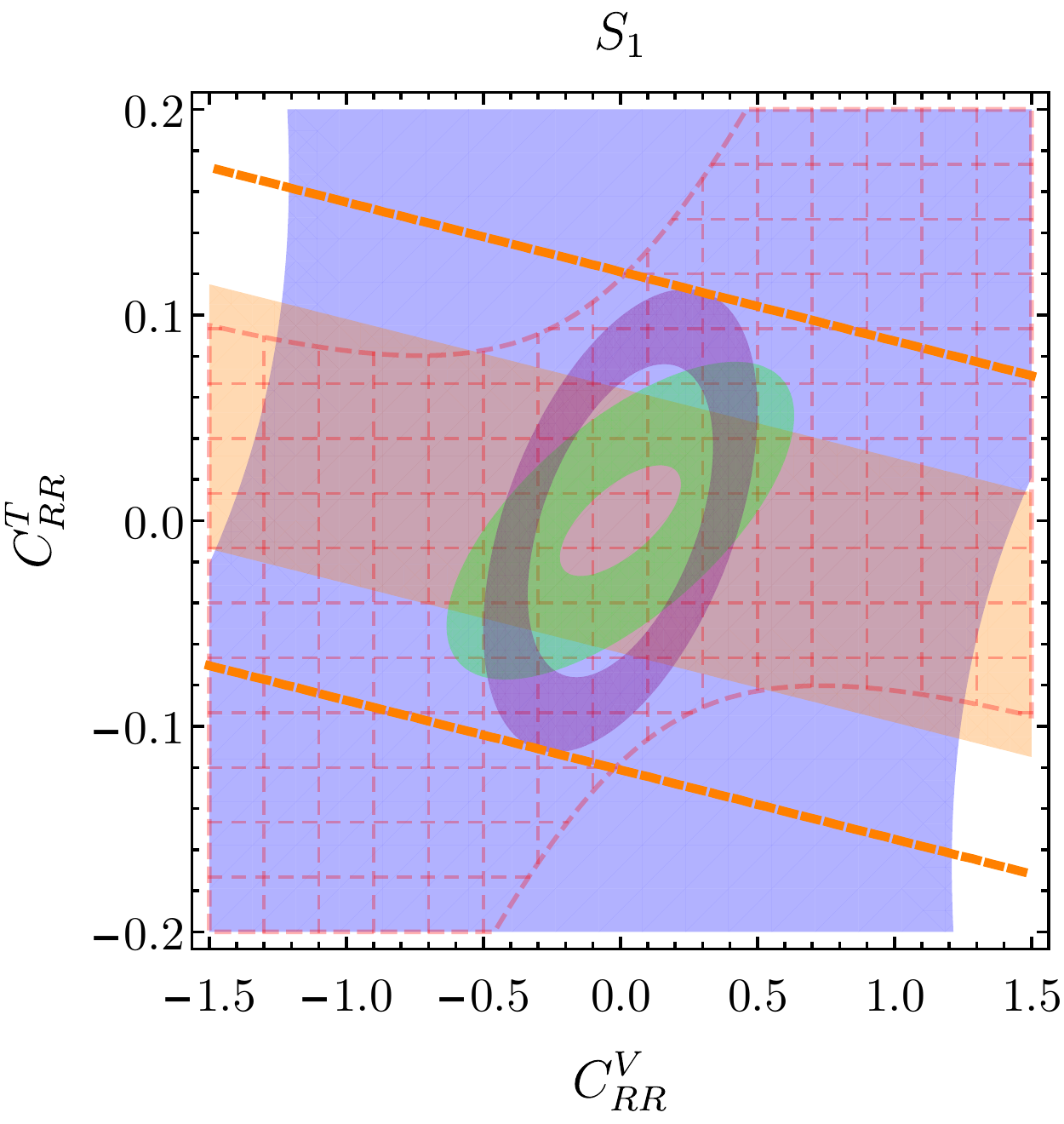} \quad
            \includegraphics[scale=0.7]{img/Leg.pdf}
    \includegraphics[scale=0.45]{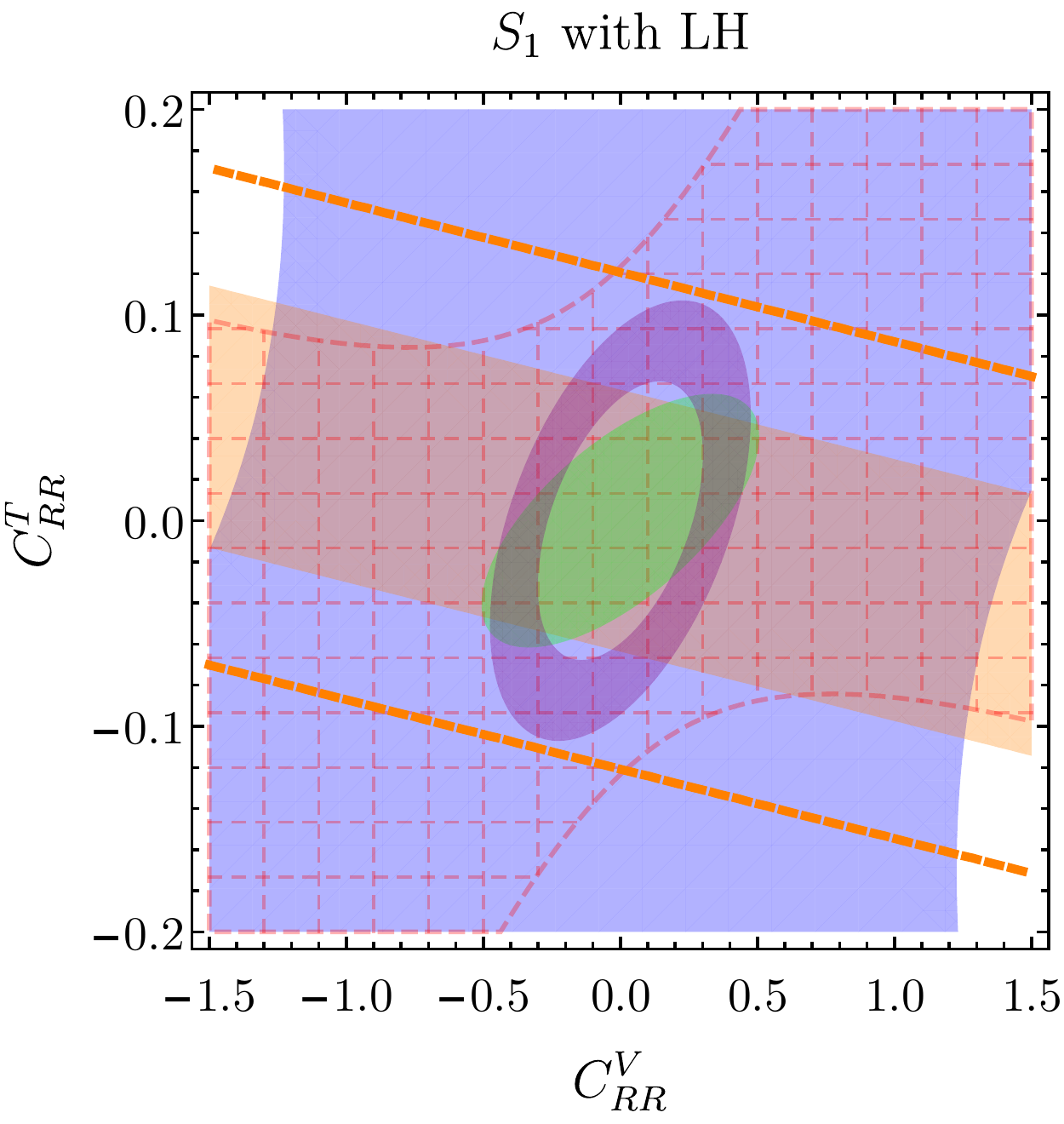}
    \caption{Fit-independent plot of \textit{Scenario 7}, showing the regions allowed at $1\sigma$. On the left panel only the RHN Wilson coefficients are considered (\textit{Scenario 7a}), whereas on the right panel we set the LHN Wilson coefficients entering in \textit{Scenario 7b} to their best fit values. The dashed orange line corresponds to the more relaxed bound $\mathcal{B}(B_c \to \tau \bar{\nu} )\le 30\%$, and the red grid shows the parameter space consistent with the experimental measurement of $ \FLint{}$ at 2$\sigma$.}
    \label{fig:case7}
\end{figure}

%
The scalar leptoquark $S_1 \sim (\bar{3},1,1/3)$ is considered in this scenario.
For \textit{Scenario 7a} we obtain a solution dominated by a single Wilson coefficient, $C^V_{RR}$, being $C^T_{RR}$ compatible with zero within $1\sigma$:
\beqn
\no
&\chi^2/\text{d.o.f.}& = 39.21/57\, , \\
&C^V_{RR}& = 0.422^{+0.071}_{-0.126}\, , \qquad C^T_{RR} = 0.022^{+0.032}_{-0.037}\, .
\eeqn
The left panel of Fig.~\ref{fig:case7} shows the regions in the two-dimensional parameter space where the experimental observables
can be reproduced at $1\sigma$. Again, at this level of precision, the longitudinal $D^*$ polarization cannot be accommodated together with the other measurements,
although it is possible to find overlap between all experimental data when the value of $\FLint{}$ is taken at 2$\sigma$, shown in the figure as a red grid.
\FloatBarrier

\subsubsection[\texorpdfstring{Scenario 7b: $S_1$ }{dum2}]{\texorpdfstring{\boldmath Scenario 7b: $S_1$}{dum2}}
%
Adding the left-handed operators that contribute in the presence of $S_1$, we find a solution
compatible with vanishing left-handed Wilson coefficients ($\Delta \chi^2 = -0.15$ for 2 d.o.f.) and a slightly shifted value of $C^V_{RR}$:
\
\beqn
\no
&\chi^2/\text{d.o.f.}& = 39.06/55\, , \\ \no
&C^V_{LL}& = 0.034^{+0.11}_{-0.70}\, , \qquad  \; C^T_{LL} = 0.010^{+0.037}_{-0.041}\, ,  \\
&C^V_{RR}& = 0.367^{+0.68}_{-1.41^{\dagger}}\, , 
\qquad C^T_{RR} = 0.004^{+0.048}_{-0.055^{\dagger}}\, .
\eeqn

For the RHN coefficients, $C^V_{RR}$ and $C^T_{RR}$, the $\chi^2$ distribution turns out to be very flat between the two flipped minima, which no longer can be separated. This implies a very broad negative $1\sigma$ interval for $C^V_{RR}$, reaching its flipped minimum $C^{V'}_{RR}  = -0.367$.

As in the case of the vector leptoquark $U_1^\mu$ (\textit{Scenarios 5a} and \textit{5b}), the preferred solution for an $S_1$ leptoquark involves only RHN operators.

\FloatBarrier

\subsubsection[\texorpdfstring{Scenario 8: $\tilde{V}_2^\mu$ }{dum2}]{\texorpdfstring{\boldmath Scenario 8:  $\tilde{V}_2^\mu$}{dum2}}

\begin{figure}[ht]
    \centering
    \includegraphics[scale=0.55]{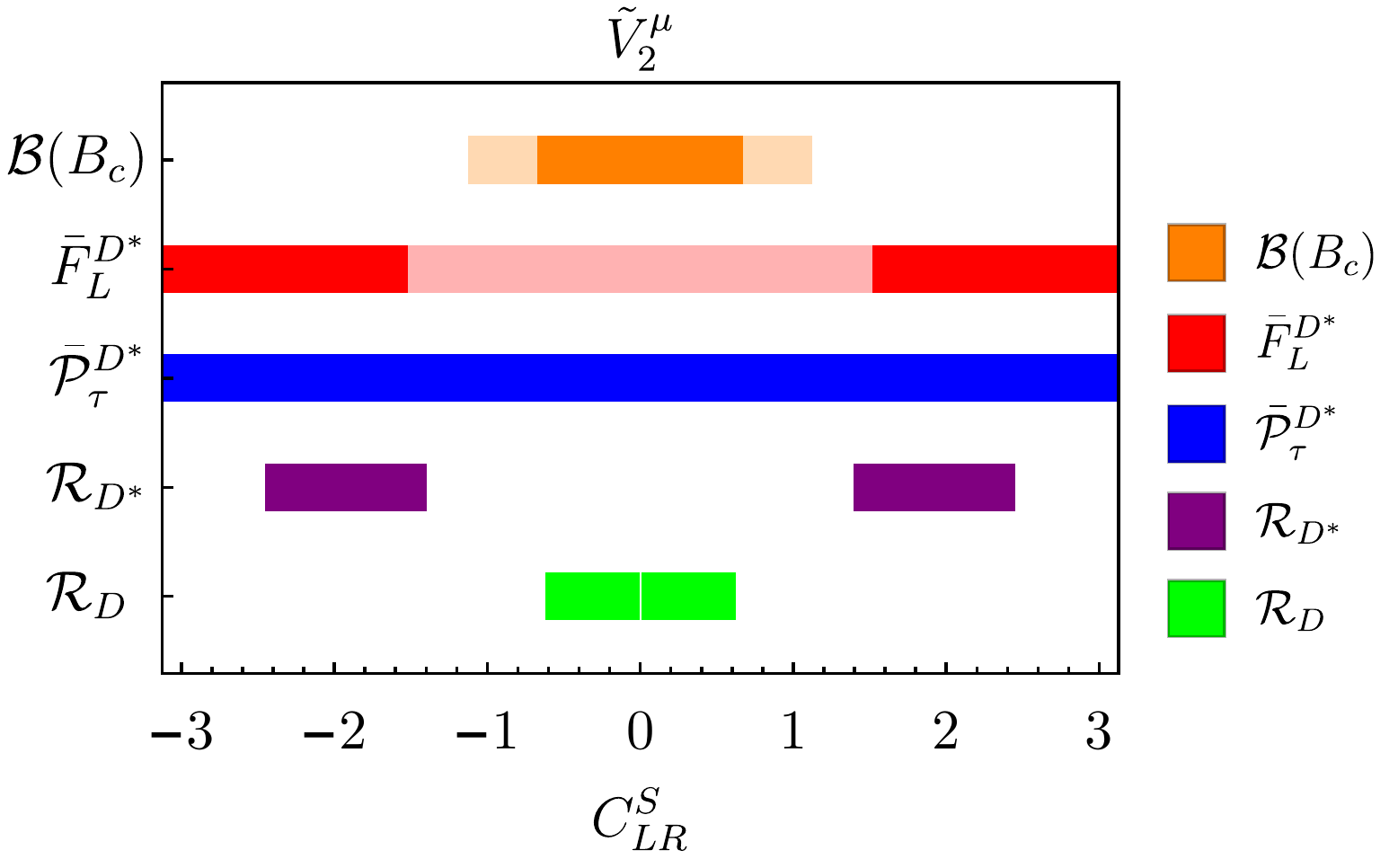}
     \caption{Fit-independent plot of \textit{Scenario 8}.  Dark colours correspond to the regions satisfying the experimental constraints at $1\sigma$ and a 10\% upper limit on $\mathcal{B}(B_c \to \tau \bar{\nu})$, for a given value of $C^S_{LR}$. Lighter orange and red shaded areas correspond to the more relaxed 30\% bound on the leptonic $B_c$ decay
     and the $2\sigma$ region for $ \FLint{}$, respectively.}
    \label{fig:case8}
\end{figure}

This is another genuine scenario
of RHNs, since it does not generate any $b \to c$ transition involving $\nu_L$ operators. The vector leptoquark $\tilde{V}_2^\mu \sim (\bar{3},2,-1/6)$ only contributes to the Wilson coefficient $C^S_{LR}$. This allows us to study the parameter space preferred by the experiment from a fit-independent point of view.
As Fig.~\ref{fig:case8} shows, there is no overlap among the different experimental constraints at the $1\sigma$ level, nor even considering a more relaxed $30\%$ bound for the leptonic decay ${\cal B}(B_c \to \tau \bar{\nu})$ and the experimental value of $\FLint{}$ at $2\sigma$.
Numerically, the fit provides the following minimum:
\beqn
\no
&\chi^2/\text{d.o.f.}& = 47.32/57\, , \\
&C^S_{LR}& = 0.418^{+0.097}_{-0.125}\, .
\eeqn

\FloatBarrier

\subsection{Comments on the fit results}
\begin{figure}[htb]
    \centering
    \includegraphics[scale=0.65]{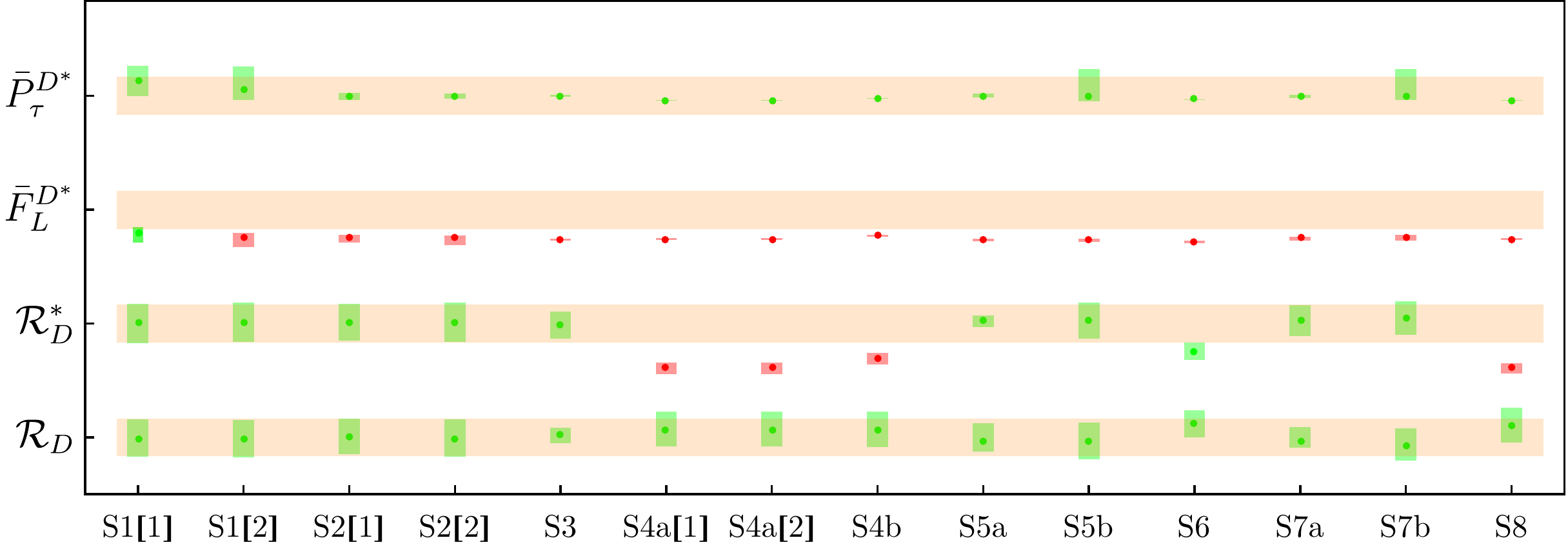}
    \caption{Predictions for the fitted observables, normalized to their measured values, with their $1\sigma$ experimental uncertainties shown as orange bands. For these predictions ${\cal B}(B_c \to \tau \bar{\nu})\leq 10 \%$ is taken. The green and red regions indicate the predictions arising from each NP scenario
    that are in agreement or not with the experimental value, respectively, at the $1\sigma$ level. The labels within brackets specify the minimum within a given scenario. The numerical values of these predictions are listed in Table~\ref{tab:predictions}. }
    \label{fig:preds}
\end{figure}

%
Table~\ref{tab:pulls} summarizes the fit quality of the results obtained in the different scenarios analysed, quantified through the corresponding $\chi^2/\text{d.o.f.}$, the pull with respect to the SM, and the $p$-value. The resulting predictions in each scenario for the observables included in the fit are also given in Table~\ref{tab:predictions}, and compared with their experimental measurements in Fig~\ref{fig:preds}. Several conclusions can be extracted from these results:
\begin{itemize}
    \item In general, it is difficult to reproduce the experimental value of the longitudinal $D^*$ polarization within its $1\sigma$ range. From Fig.~\ref{fig:preds} and  Table~\ref{tab:predictions} we can see that the only solutions reproducing all the experimental values (marked with a \textcolor{Darkgreen}{\cmark}) are 
    \textit{Scenario 1a} with either a 10\% ({\it Min 1}) or 30\% ({\it Min 1} and {\it Min 2}) upper limit on 
    $\mathcal{B}(B_c \to \tau \bar{\nu})$,
    and \textit{Scenario 4b} with a 30\%. 
    
    \item All solutions exhibit pulls between 1.2 and 3.7 with respect to the SM fit, showing a clear preference for NP contributions.
    
    \item The largest pull with respect to the SM fit is obtained in \textit{Scenario 3},
     which only contributes to the $C^V_{RR}$ coefficient. Note that $C^V_{RR}$ plays 
     a similar role than $C^V_{LL}$
     in the observables involving $b \to c$ transitions. Therefore, the preference of the fit for this scenario can be easily understood, since 
     a SM-like modification was the best fit solution in absence of RHN~\cite{Murgui:2019czp}.
    
    \item \textit{Scenarios 4a, 4b, 6} and \textit{8}, involving only scalar (and tensor) operators, have the largest $\chi^2$ value. As Table~\ref{tab:predictions} and Fig.~\ref{fig:preds} show, \textit{Scenarios 4a, 4b} and \textit{8} fail badly reproducing the experimental value of ${\cal R}_{D^*}$.
    
    \item {\it Scenarios 4a, 4b, 6, 8} and {\it Scenario 2 Min2}, are disfavoured by the $q^2$ differential distributions of the $B \to D^{(*)}$ decay with respect to the SM, as the corresponding 
    Pull$_{\text{SM}}$ in Table~\ref{tab:pulls} shows. 
    
    \item Those solutions further away from the SM (larger pulls) present higher $p$-values, as Table~\ref{tab:pulls} shows.
    
    \item In scenarios with several operators, the best fits correspond to solutions where all Wilson coefficients but one are compatible with zero. The non-zero Wilson coefficient is typically $C^V_{RR}$ (\textit{Scenarios 5a, 5b, 7a} and \textit{7b}).

    \item When scenarios with and without LHN operators (``{\it b}'' and ``{\it a}'' variants, respectively) are compared, the fit indicates 
    a preference for solutions with all left-handed Wilson coefficients compatible with zero within $1\sigma$.

\end{itemize}

Comparing our results with similar fits previously done in the literature, we can quantify the impact of adding the differential $q^2$ distributions and considering recently measured observables such as $ \FLint{}$ or $\bar{{\cal P}}_\tau^{D^*}$, together with the update of some experimental measurements. 
Ref.~\cite{Robinson:2018gza} analysed all mediators that can contribute to the $b \to c \tau \bar{\nu}_R$ transition, except the $\tilde{V}_2^\mu$ vector leptoquark, but only included in the fit the values of ${\cal R}_D$ and ${\cal R}_{D^*}$.
The global minimum obtained in Ref.~\cite{Robinson:2018gza} for an extra gauge boson $V'$ (\textit{Scenario~3}) agrees with ours, while the two minima obtained for our \textit{Scenario 4a} deviate more from the SM solution than ours. The latter is due to the fact that the $B_c \to \tau \bar{\nu}$ constraint, which has a strong impact on solutions involving scalar Wilson coefficients, was not taken into account in the fit. Indeed, Fig.~2 from Ref.~\cite{Robinson:2018gza} shows that their minima are excluded by this constraint, and this is the reason why in our analysis, this $\chi^2$ is 
the most unfavorable among all the scenarios considered. For our \textit{Scenario 5a}, mediated by $U_1^\mu$,  
two minima are observed in Ref.~\cite{Robinson:2018gza} where the furthest one from the SM solution is ruled out by the constraint $\mathcal{B}(B_c \to \tau \bar{\nu})\le 10\%$. This situation is repeated in the scenario mediated by $S_1$, \textit{Scenario 7a}. Finally, in the case of the $\tilde R_2$ mediator (our \textit{Scenario 6}), both minima differ slightly from ours since, again, as their Fig.~2 
shows, they are excluded by the $B_c$ leptonic decay limit; however, taking into account the minimum value of the $\chi^2$ satisfying this constraint, our result is compatible with 
Ref.~\cite{Robinson:2018gza}.

\begin{table}[t]
    \centering
\begin{tabular}{c|c|c|c|c|c|c|c}
Scenario & $\mathcal{B}(B_c \to \tau \bar{\nu})$& $\chi^2/{\text{d.o.f}}$ & \multicolumn{3}{c|}{$\text{Pull}_\text{SM}$} & Pull$_{\text{SM}}$  & $p$-value\\
 & & & 
 $\ptauint\!,\,F_L^{D^*}$ & ${\cal R}_{D,D^*}$ & $d\Gamma/dq^2$ & \\
\hline
SM & 2.16\% &  
$52.87/59$ &  &  &  & &  $69.95\% $ \\
\hline
\textit{Scenario 1, Min 1} & $ < 10\% $ & $37.26/53$ & 0.007 & 2.08 & 0.0414  & 2.4 &  $95.02\% $  \\
\textit{Scenario 1, Min 2} &$ < 10\%$ & $38.86/53$ & 0.001& 2.08 & 0.0006 & 2.2&   $92.68\% $ \\
\textit{Scenario 1, Min 1} & $  <30\%$& $36.42/53$ & 0.022 & 2.08 & 0.0866 & 2.5& $96.00\%$ \\
\textit{Scenario 1, Min 2} & $  <30\%$& $38.54/53$ & 0.011 & 2.08 & 0.000 & 2.2 & 
$93.21\%$ \\
\textit{Scenario 2, Min 1} & $< 10\%$ & $38.54/54$  & 0.006  & 2.32  & 0.0113 & 2.5 & $93.20\%$ \\
\textit{Scenario 2, Min 2} &$ < 10\%$ & $39.05/54$ & 0.004 & 2.32  & 0.0003 & 2.4 &  $93.73\% $ \\
\textit{Scenario 2, Min 1} & $ <30\%$ & $38.33/54$ & 0.035 & 2.32 & 0.0023 & 2.5 &  $94.73\% $ \\
\textit{Scenario 2, Min 2}  & $< 30\% $& $38.80/54$ & 0.025 & 2.32 & $0^\ast$ & 2.4 & $94.09\% $ \\
\textit{Scenario 3}  & $ < 10\%$ & $39.50/58$  & 0.150 & 3.65 & 0.0835 & 3.7
& $97.00\% $ \\
\textit{Scenario 4a, Min 1} & $ < 10\%$ & $49.93/57$ & 0.079 & 2.34& $0^\ast$ & 1.2 &  $73.52\% $ \\
\textit{Scenario 4a, Min 2} & $ < 10\%$ & $49.93/57$ & 0.079 & 2.34 & $0^\ast$ &  1.2&  $73.52\% $ \\
\textit{Scenario 4a, Min 1}  & $< 30\%$ &  $44.49/57$ & 0.311 & 2.66& $0^\ast$ & 2.4 &  $88.62\% $ \\
\textit{Scenario 4a, Min 2} & $ <30\%$ &  $44.49/57$ & 0.311 & 2.66 & $0^\ast$ &  2.4&   $88.62\% $ \\
\textit{Scenario 4b} & $ < 10\%$ & $43.56/55$ & 0.054& 2.07& $0^\ast$ & 1.9 &  $86.70\% $ \\
\textit{Scenario 4b} & $ <30\%$ & $40.03/55$ & 0.218 & 2.52 & $0^\ast$ & 2.5& $93.54\% $ \\
\textit{Scenario 5a} & $ < 10\%$ & $39.39/57$ & $0^\ast$ & 3.22 & 0.0981 &  3.2 &   $96.36\% $\\
\textit{Scenario 5b} & $ < 10\%$ & $39.37/55$ & $0^\ast$ & 3.34 & 0.0060 & 2.6 &   $94.47\% $\\
\textit{Scenario 6} & $ < 10\%$ & $44.20/58$ & $0^\ast$ & 3.34 & $0^\ast$ & 2.9 &  $90.93\% $ \\
\textit{Scenario 7a} & $ < 10\%$ & $39.21/57$ & 0.126  & 3.22& 0.0616 & 3.3 &  $96.53\% $ \\
\textit{Scenario 7b} & $ < 10\%$ &  $39.06/55$ & 0.014  & 2.56  & 0.0112 & 2.7 & $94.87\% $ \\
\textit{Scenario 8} & $ < 10\%$ & $47.32/57$  & 0.259 & 2.56 & $0^\ast$ & 1.9 & $81.60\% $ \\
\end{tabular}
    \caption{Fit quality of the different fits: $\chi^2/{\text{d.o.f}}$, pulls with respect to the SM hypothesis and $p$-values. The~${}^\ast$ symbol indicates that the $\chi^2$ of a given scenario is greater than the SM one.}
    \label{tab:pulls}
\end{table}

Fitting the experimental data with generic NP amplitudes has the unavoidable caveat that the NP contributions can modify the decay distributions and acceptances that have been assumed when performing the measurements. This introduces biases in the extraction of NP parameters, which in some cases can be very significant \cite{Bernlochner:2020tfi}. The inclusion of the measured $q^2$ distributions in our fits helps to reduce this unwanted effect, because it disfavours potential solutions with enhanced rates that have differential distributions very different from the SM ones. Nevertheless, some caution has to be taken to interpret the fitted results, specially when comparing scenarios with close pull values.
The quantitative estimate of the induced bias depends strongly on the experimental set-up and is beyond the scope of a global analysis, including data from several flavour experiments.


\section{Predictions }
\label{sec:predctions}
%
In this section we show the predictions of different observables for the fitted scenarios considered in the previous section. As we will discuss in the following, these results can be used to discriminate between the different scenarios and, in some cases, even distinguish the contribution originated by light RHNs from the SM one.

\subsection{Predictions of integrated observables}
%
In Table~\ref{tab:predictions} we list the predictions of the different integrated observables considered in the fit, i.e. ${\cal R}_D$, ${\cal R}_{D^*}$, $\bar{F}_L^{D^*}$, $\bar{P}_\tau^{D^*}$ and the leptonic branching fraction ${\cal B}(B_c \to \tau \bar{\nu})$, for each of the scenarios considered.
\begin{table}[htb]
    \centering
     \resizebox{\textwidth}{!}{  
    \begin{tabular}{c|c|c|c|c|c}
        \textit{Scenario} & $\mathcal{B}(B_c \to \tau \bar{\nu} )$ & $\rd{}$ & $\rdst{}$ & $ \FLint{}$ & $\ptauint{}$   \\
        \hline  \noalign{\vskip2pt}
        Experiment & -  &$ 0.340 \pm 0.027 \pm 0.013$ & $0.295 \pm 0.011 \pm 0.008$ & $0.60 \pm 0.08 \pm 0.04$ & $ -0.38 \pm 0.51 ^{+0.21}_{-0.16}$  \\
        \hline \noalign{\vskip2pt}
         \textit{Scenario 1, Min 1} & 10\% &  $0.339 \pm 0.030$ \textcolor{Darkgreen}{\cmark} &  $0.295 \pm 0.014 $ \textcolor{Darkgreen}{\cmark} &  $0.494^{+0.025}_{-0.045}$ \textcolor{Darkgreen}{\cmark} &  $\phantom{+}0.06^{+0.43}_{-0.45}$ \textcolor{Darkgreen}{\cmark}  \\[0.5ex]
        \textit{Scenario 1, Min 2} & 10\% &  $0.338\pm 0.030$ \textcolor{Darkgreen}{\cmark}  &  $0.296\pm 0.014$  \textcolor{Darkgreen}{\cmark}  & $0.472^{+0.023}_{-0.044}$ \textcolor{red}{\xmark} &  $-0.20^{+0.67}_{-0.30}$ \textcolor{Darkgreen}{\cmark}  \\[0.5ex]
          \textit{Scenario 1, Min 1} & 30\% & $0.338 \pm 0.030$ \textcolor{Darkgreen}{\cmark}  &  $0.295 \pm 0.014 $ \textcolor{Darkgreen}{\cmark}   &  $0.510^{+0.014}_{-0.043}$ \textcolor{Darkgreen}{\cmark}  &  $\phantom{+}0.08^{+0.32}_{-0.46}$ \textcolor{Darkgreen}{\cmark}   \\[0.5ex]
         \textit{Scenario 1, Min 2} & 30\% &  $0.338\pm{0.030}$ \textcolor{Darkgreen}{\cmark}  &  $0.296\pm 0.014$ \textcolor{Darkgreen}{\cmark}  &  $0.488^{+0.032}_{-0.050}$\textcolor{Darkgreen}{\cmark}   &  $-0.24^{+0.64}_{-0.28}$ \textcolor{Darkgreen}{\cmark}  \\[0.75ex]
        \textit{Scenario 2, Min 1} & 10\% &  $0.341^{+0.029}_{-0.028}$ \textcolor{Darkgreen}{\cmark}  &  $0.296  \pm 0.013$ \textcolor{Darkgreen}{\cmark}  & 
        $0.474^{+0.010}_{-0.024}$ \textcolor{red}{\xmark} & 
        $-0.42^{+0.13}_{-0.07}$ \textcolor{Darkgreen}{\cmark}   \\[0.5ex]
         \textit{Scenario 2, Min 2} & 10\% &  $0.339 \pm 0.030$ \textcolor{Darkgreen}{\cmark}   &  $0.296 \pm 0.014$ \textcolor{Darkgreen}{\cmark}  &$0.471^{+0.012}_{-0.033}$  \textcolor{red}{\xmark}  &  $-0.401^{+0.094}_{-0.064}$ \textcolor{Darkgreen}{\cmark}   \\[0.5ex]
          \textit{Scenario 2, Min 1} & 30\% &  $0.341^{+0.029}_{-0.028}$ \textcolor{Darkgreen}{\cmark}  &  $0.296  \pm 0.013$ \textcolor{Darkgreen}{\cmark}  & $0.489^{+0.011}_{-0.048}$ \textcolor{red}{\xmark} &  $-0.47^{+0.15}_{-0.05}$
          \textcolor{Darkgreen}{\cmark}  \\[0.5ex]
          \textit{Scenario 2, Min 2} & 30\% &  $0.340 \pm 0.030$ \textcolor{Darkgreen}{\cmark}  &  $0.295 \pm 0.014$ \textcolor{Darkgreen}{\cmark}   &$0.484^{+0.015}_{-0.045}$ \textcolor{red}{\xmark} &
          $-0.45^{+0.13}_{-0.07}$
          \textcolor{Darkgreen}{\cmark}   \\[0.75ex]
         \textit{Scenario 3} & 2.5\% &  $0.343	\pm 0.012$ \textcolor{Darkgreen}{\cmark}   &  $0.294 \pm	0.010$ \textcolor{Darkgreen}{\cmark}  & $0.462	\pm 0.004$ \textcolor{red}{\xmark} &  $-0.377^{+0.031}_{-0.033}$ \textcolor{Darkgreen}{\cmark}   \\[0.75ex]
         \textit{Scenario 4a, Min 1} & 10\% &  $0.353^{+0.028}_{-0.027}$  \textcolor{Darkgreen}{\cmark}  & $0.2638^{+0.0034}_{-0.0049}$ \textcolor{red}{\xmark} & $0.4662^{+0.0039}_{-0.0057}$ \textcolor{red}{\xmark} &  $-0.5028^{+0.0051}_{-0.0035}$ \textcolor{Darkgreen}{\cmark}   \\[0.5ex]
         \textit{Scenario 4a, Min 2} & 10\% &  $0.353^{+0.028}_{-0.027}$ \textcolor{Darkgreen}{\cmark}  & $0.2638^{+0.0034}_{-0.0049}$ \textcolor{red}{\xmark} & $0.4662^{+0.0039}_{-0.0057}$ \textcolor{red}{\xmark} &  $-0.5028^{+0.0051}_{-0.0034}$ \textcolor{Darkgreen}{\cmark}  \\[0.5ex]
         \textit{Scenario 4a, Min 1} & 30\% &  $0.348^{+0.028}_{-0.027}$ \textcolor{Darkgreen}{\cmark}  & $0.2699^{+0.0032}_{-0.0058}$ \textcolor{red}{\xmark}& $0.4792^{+0.0041}_{-0.0064}$ \textcolor{red}{\xmark} &  $-0.5144^{+0.0056}_{-0.0032}$ \textcolor{Darkgreen}{\cmark}   \\[0.5ex]
        \textit{Scenario 4a, Min 2} & 30\% &  $0.348 ^{+0.028}_{-0.027}$ \textcolor{Darkgreen}{\cmark}  & $0.2699^{+0.0032}_{-0.0058}$ \textcolor{red}{\xmark} & $0.4792^{+0.0041}_{-0.0064}$ \textcolor{red}{\xmark} &  $-0.5144^{+0.0056}_{-0.0032}$ \textcolor{Darkgreen}{\cmark}  \\[0.75ex]
         \textit{Scenario 4b} & 10\% &  $0.353 \pm 0.028$ \textcolor{Darkgreen}{\cmark}  & $0.2708^{+0.0032}_{-0.0052}$ \textcolor{red}{\xmark} & $0.4815^{+0.0041}_{-0.0068}$ \textcolor{red}{\xmark} & 
         $-0.442^{+0.005}_{-0.026}$
         \textcolor{Darkgreen}{\cmark}  \\[0.5ex]
        \textit{Scenario 4b} & 30\% &  $0.340 \pm 0.028$ \textcolor{Darkgreen}{\cmark}  &  $0.2866^{+0.0030}_{-0.0081}$ \textcolor{Darkgreen}{\cmark}  &  $0.5125^{+0.0044}_{-0.0126}$ \textcolor{Darkgreen}{\cmark}  &  
        $-0.356^{+0.006}_{-0.066}$
        \textcolor{Darkgreen}{\cmark}  \\[0.75ex]
         \textit{Scenario 5a} & 2.2\% &  $0.335^{+0.027}_{-0.017}$ \textcolor{Darkgreen}{\cmark}  &   $0.2966^{+0.0043}_{-0.0042} $\textcolor{Darkgreen}{\cmark}   & $0.4611^{+0.0056}_{-0.0070}$ \textcolor{red}{\xmark} &  $-0.364^{+0.048}_{-0.050}$ \textcolor{Darkgreen}{\cmark}  \\[0.5ex]
         \textit{Scenario 5b} & 2.0\% &  $0.334 \pm 0.029$ \textcolor{Darkgreen}{\cmark}  &  $0.297 \pm 0.013$ \textcolor{Darkgreen}{\cmark}  & $0.4609^{+0.0059}_{-0.0083}$ \textcolor{red}{\xmark} &   $-0.38^{+0.77}_{-0.16}$ \textcolor{Darkgreen}{\cmark}  \\[0.75ex]
         \textit{Scenario 6} & 7.6\% &  $0.361^{+0.022}_{-0.021}$ \textcolor{Darkgreen}{\cmark}  &  $0.2748^{+0.0066}_{-0.0059}$ \textcolor{Darkgreen}{\cmark}  & $0.4522 \pm 0.0050$ \textcolor{red}{\xmark}  &  $-0.4800^{+0.0078}_{-0.0076}$ \textcolor{Darkgreen}{\cmark}   \\[0.75ex]
         \textit{Scenario 7a} & 4.6\% &  $0.335^{+0.021}_{-0.011}$ \textcolor{Darkgreen}{\cmark}  &  $ 0.297 \pm 0.011$ \textcolor{Darkgreen}{\cmark}  & 
         $0.468^{+0.007}_{-0.011}$ \textcolor{red}{\xmark} &  $-0.377^{+0.033}_{-0.058}$ \textcolor{Darkgreen}{\cmark}   \\[0.5ex]
         \textit{Scenario 7b} & 4.3\% &  $0.328^{+0.026}_{-0.025}$ \textcolor{Darkgreen}{\cmark}  &  $0.299 \pm 0.012$ \textcolor{Darkgreen}{\cmark}  & $0.471^{+0.014}_{-0.013}$ \textcolor{red}{\xmark} &  $-0.38^{+0.77}_{-0.12}$ \textcolor{Darkgreen}{\cmark}  \\[0.75ex]
         \textit{Scenario 8} & 7.3\% &  $0.359^{+0.028}_{-0.027}$ \textcolor{Darkgreen}{\cmark}  & $0.2629	\pm 0.0036$ \textcolor{red}{\xmark} & $0.4644 \pm 0.0043$ \textcolor{red}{\xmark} &  $-0.5012 \pm	0.0039$ \textcolor{Darkgreen}{\cmark} 
    \end{tabular}
    }
    \caption{Predictions for the fitted observables in the different minima, and their experimental values. }
    \label{tab:predictions}
\end{table}
Those predictions that are in agreement with the measured values at the $1\sigma$ level are marked with a \textcolor{Darkgreen}{\cmark}, while a \textcolor{red}{\xmark} mark indicates disagreement. Only in \textit{Scenarios 1} and \textit{4b} it is possible to simultaneously satisfy all experimental constraints. The second column shows that the upper bound on the $B_c$ leptonic decay is always saturated in \textit{Scenarios 1, 2}, and \textit{4}, which denotes that 
larger pseudoscalar and axial combinations
of the Wilson coefficients would still be preferred.


\subsection{Predictions of angular coefficients}

The three-body differential distribution in $B\to D\tau\bar\nu$ and the
full four-body angular analysis of $B\to D^*\tau\bar\nu\to (D\pi)\tau\bar\nu$
provide a multitude of observables that could be experimentally accessible. The presence of neutrinos in the final state makes the measurement troublesome, compared to the case of well-known neutral-current transitions like $B\to K^* \mu \bar{\mu}$. Nevertheless, measuring the distribution of the secondary $\tau$ decay, some information on the angular coefficients $J_i$ and $I_i$, defined in Eqs.~\eqref{eq:Ddistang} and \eqref{eq:dGamma}, could be obtained in the near future. As it can be seen from their explicit analytic expressions in Eqs.~\eqref{eq:Jfun} and \eqref{eq:Idef}, these $q^2$-dependent functions can be very sensitive to the NP Wilson coefficients present in the theory. In this section, we provide the predictions of such observables in some relevant NP scenarios considered in this work. 

\begin{figure}[htb]
    \centering
\includegraphics[scale=0.43]{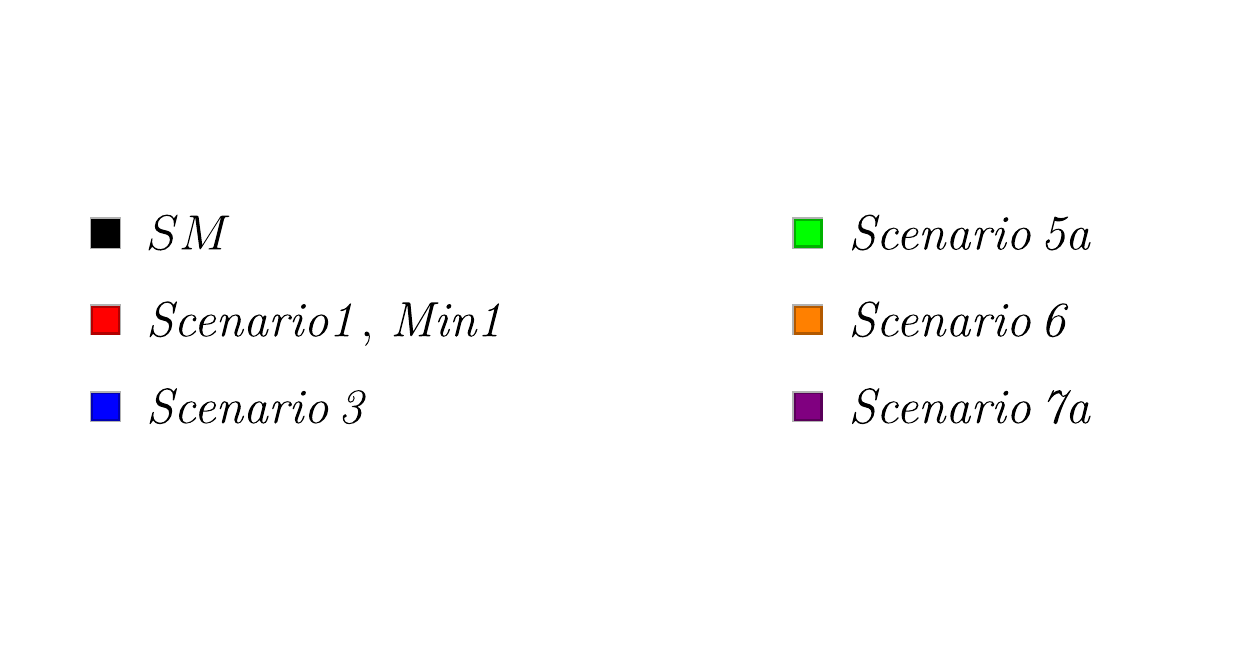}
    \includegraphics[scale=0.42]{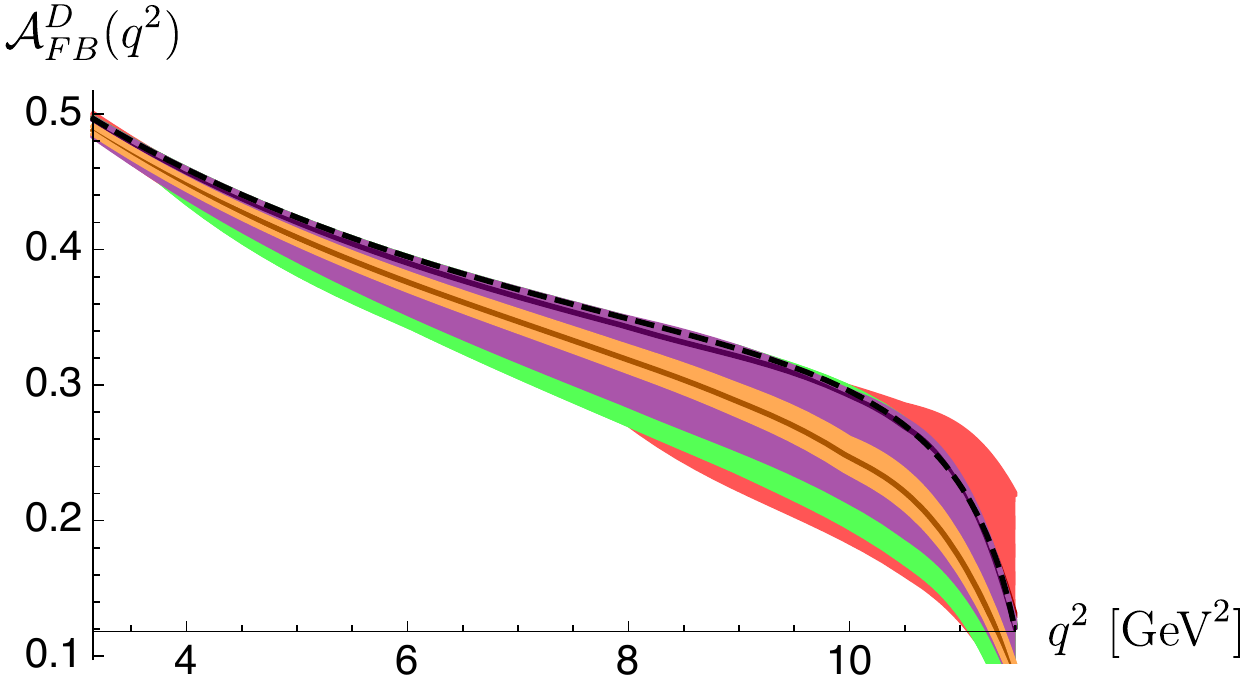}
    \includegraphics[scale=0.42]{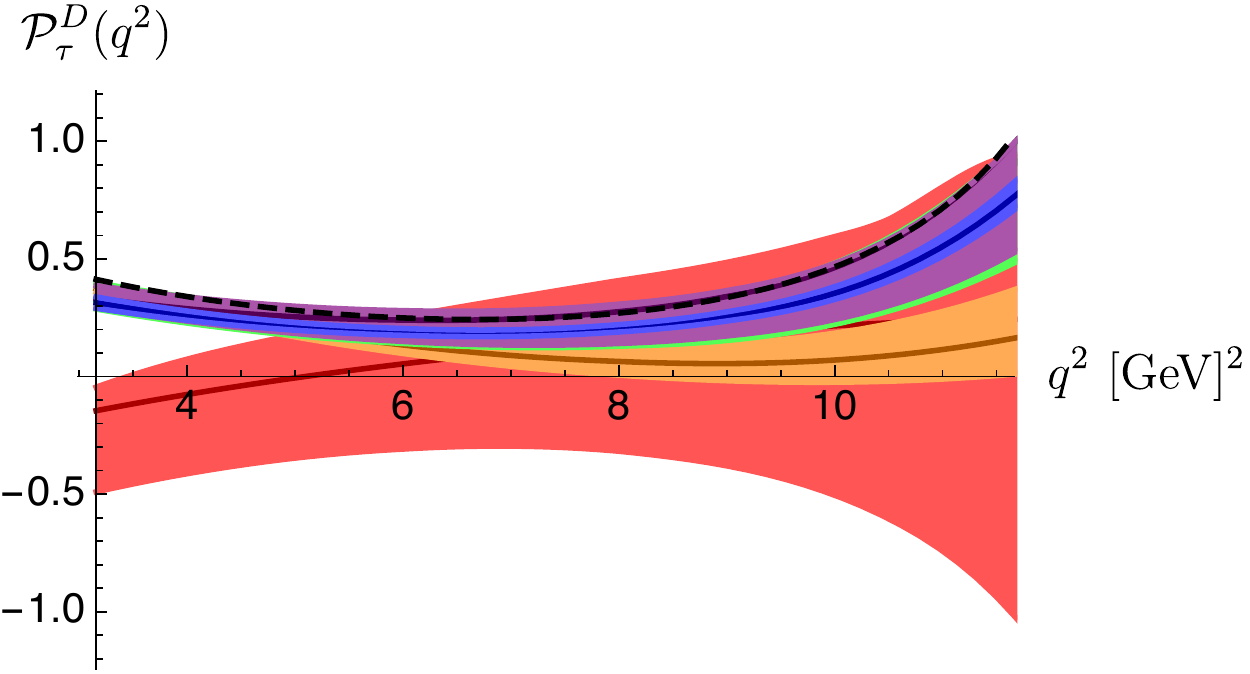}
    \includegraphics[scale=0.42]{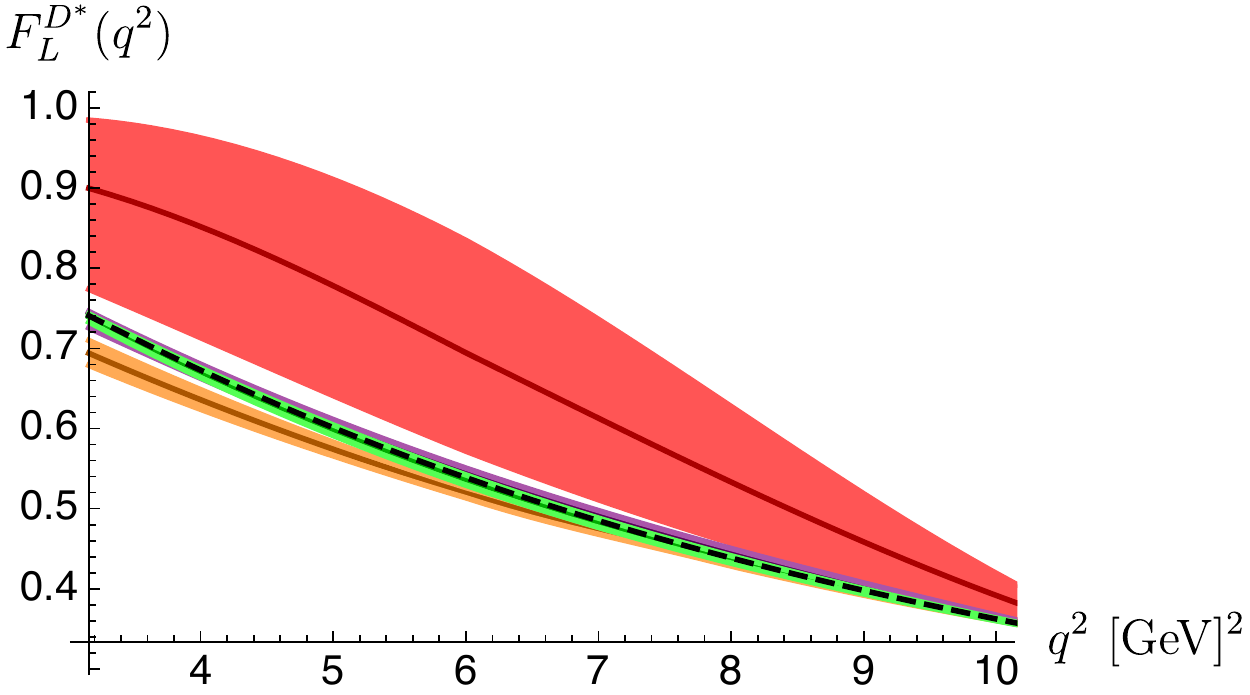}
    \includegraphics[scale=0.42]{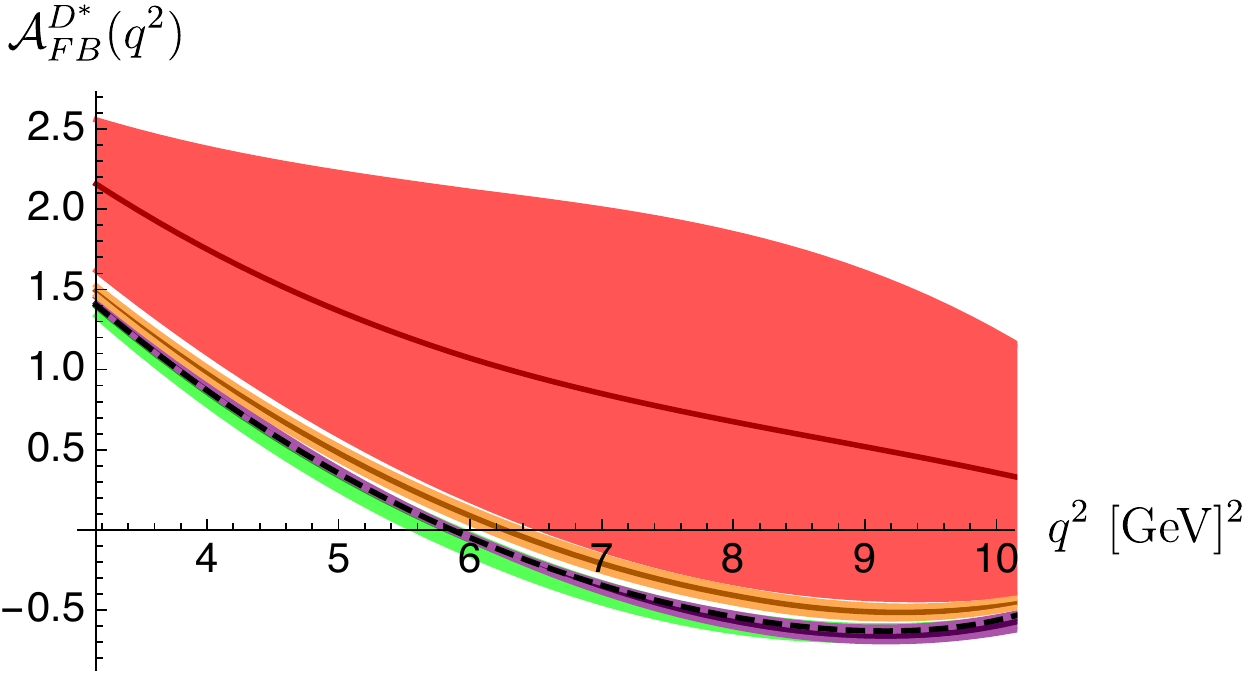}
     \includegraphics[scale=0.42]{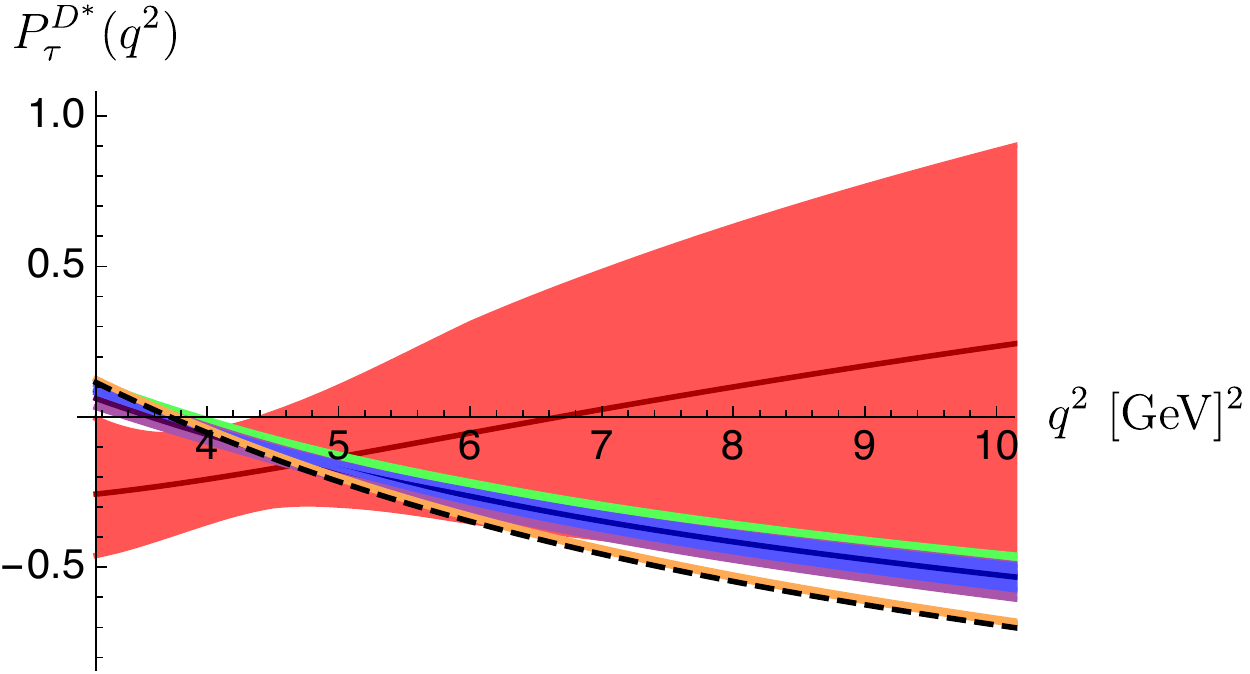}
      \caption{Dependence on $q^2$ of the forward-backward asymmetries ${\cal A}_{FB}^{D}$ and ${\cal A}_{FB}^{D^{*}}$, the longitudinal polarizations ${\cal P}_{\tau}^{D^{(*)}}$  and the longitudinal polarization fraction $F_L^{D^{*}}$, for the best-fit scenarios.}
    \label{fig:AFB_FLD}
\end{figure}

Fig.~\ref{fig:AFB_FLD} shows the predictions for the forward-backward asymmetries ${\cal A}_{FB}^{D^{(*)}}$ defined in Eqs.~\eqref{eq:DAFB} and (\ref{eq:AFB}), the lepton polarization asymmetries of Eqs.~\eqref{eq:PtauD} and \eqref{eq:PtauDstar} and the longitudinal $D^*$ polarization $F_L^{D^*}$ defined in Eq.~\eqref{eq:FLD}, as functions of $q^2$. 
For simplicity we have illustrated the four NP scenarios with largest pulls with respect to the SM.
Note that \textit{Scenario 3}, which contains the single Wilson coefficient $C^V_{RR}$, will always give the same predictions as the SM scenario for the forward-backward asymmetries, $F_L^{D^*}(q^2)$ and the angular coefficients $\bar{I}_i(q^2)$. Therefore, this scenario is only included in the $\tau$ polarization asymmetries. Error bands in these plots correspond only to the uncertainties arising from the fitted Wilson coefficients. These uncertainties have been obtained by minimizing the $\chi^2$, imposing $O_{i} = O_{i,\text{min}} + \Delta O_{i,\text{min}}$, and taking the value of the observable $O_{i}$ for which $\chi^2 = \chi_{\text{min}}^2 +1$. 
Other smaller errors such as FF parameters or additional inputs are not taken into account. Therefore the SM predictions, plotted as dotted black lines, do not present any uncertainties. 

From these plots, we can see that scenarios with a larger number of Wilson coefficients also have larger uncertainties (\textit{Scenario 1, Min 1}), as expected because of the 
wider allowed range of variation of their Wilson coefficients.
The forward-backward asymmetry ${\cal A}_{FB}^{D}$ could be useful to distinguish \textit{Scenario 6} from the SM, but the large uncertainties make difficult to discriminate it from other scenarios or to differentiate the SM from  {\it Scenarios 1, 5a} and {\it 7}. A precise measurement of ${\cal A}_{FB}^{D^*}$ would allow to distinguish {\it Scenarios 1} and 
{\it 6} from the rest of NP scenarios, which partly overlap with the SM prediction. A similar situation occurs for $F_L^{D^*}$, where clear differences manifest at low values of $q^2$ while the different scenarios considered tend to overlap at high $q^2$. The $\tau$ polarizations ${\cal P}_{\tau}^{D^{(*)}}$ are useful to distinguish {\it Scenario~3} from the SM, since these are the only observables that are sensitive to a single shift in $C^V_{RR}$. Moreover, in \textit{Scenario 1} ${\cal P}_\tau^{D}$ and ${\cal P}_\tau^{D^{*}}$ exhibit a quite different dependence on $q^2$ compared to the other scenarios, which could be exploited to distinguish it at low $q^2$ values. In the high $q^2$ region, ${\cal P}_\tau^{D^{*}}$ also allows to discriminate 
\textit{Scenario 1} from the other possibilities.

\begin{figure}[ht]
    \centering
   \hskip .15cm \includegraphics[scale=0.4]{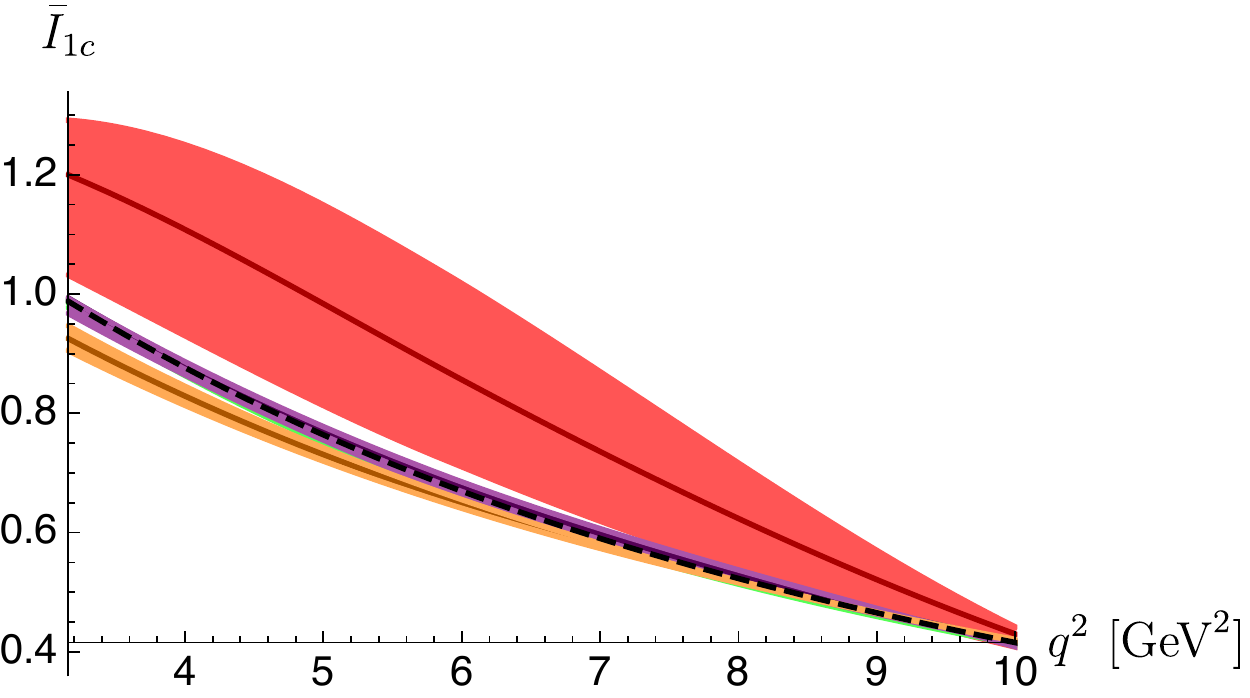}
   \hskip .1cm
    \includegraphics[scale=0.4]{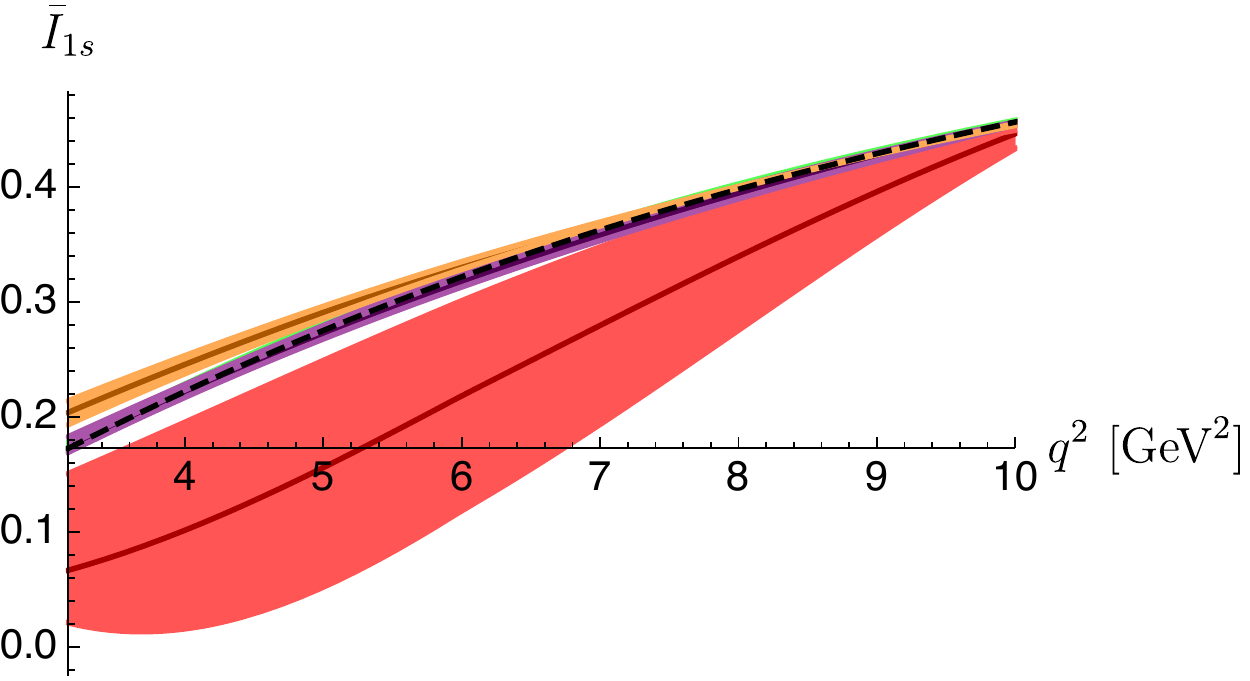}   \hskip .1cm  \includegraphics[scale=0.4]{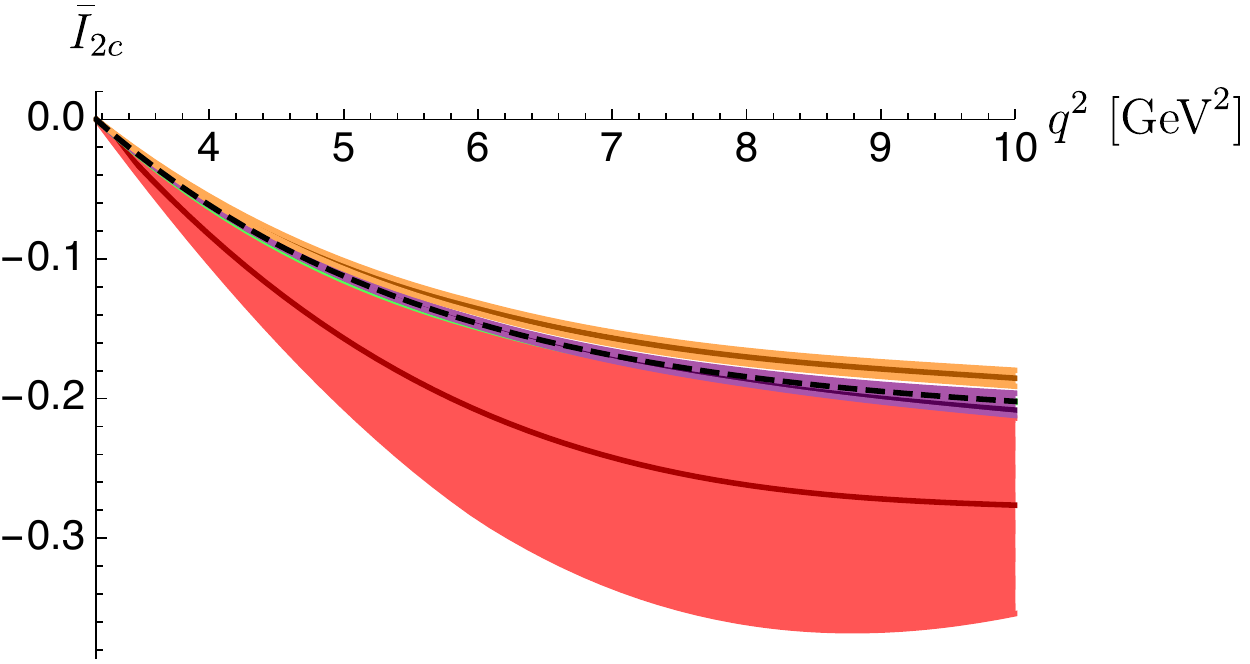}\hskip .5cm
    \includegraphics[scale=0.4]{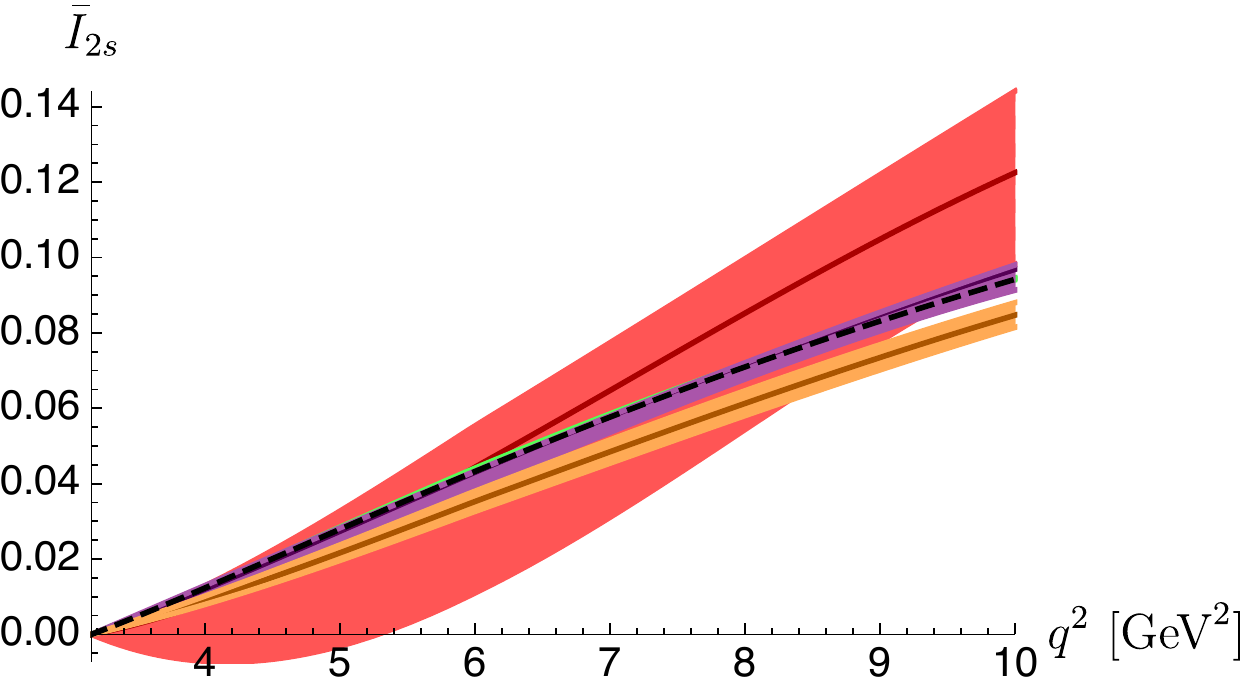}
     \includegraphics[scale=0.4]{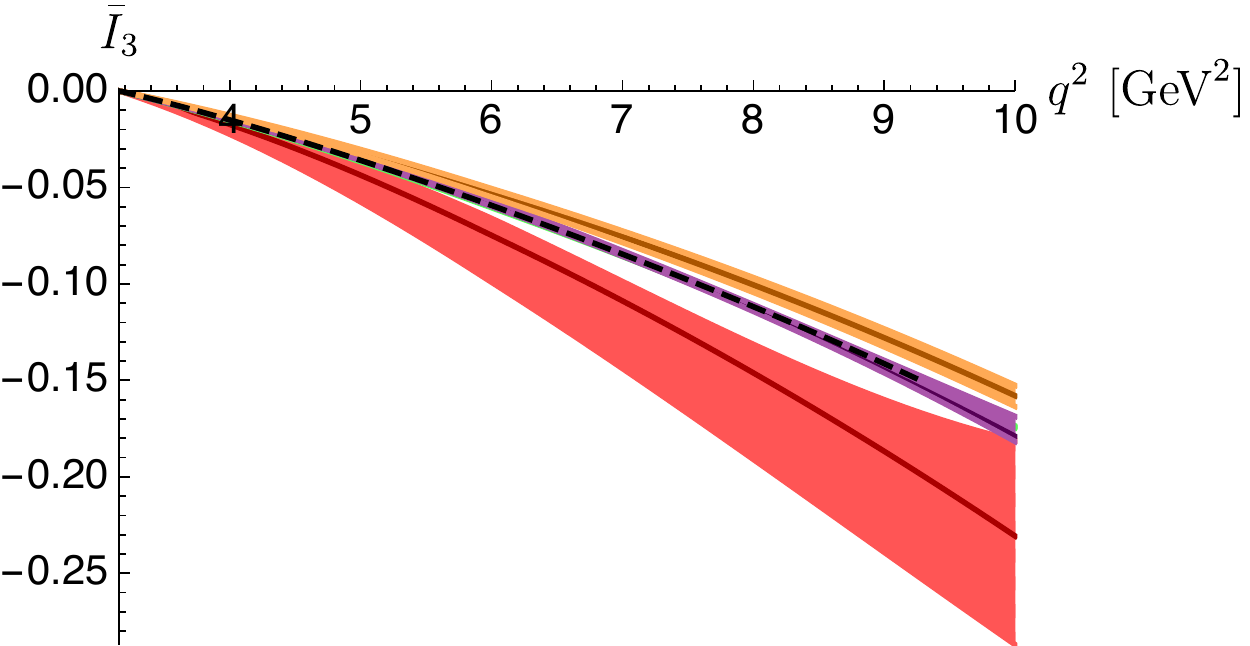}
      \includegraphics[scale=0.4]{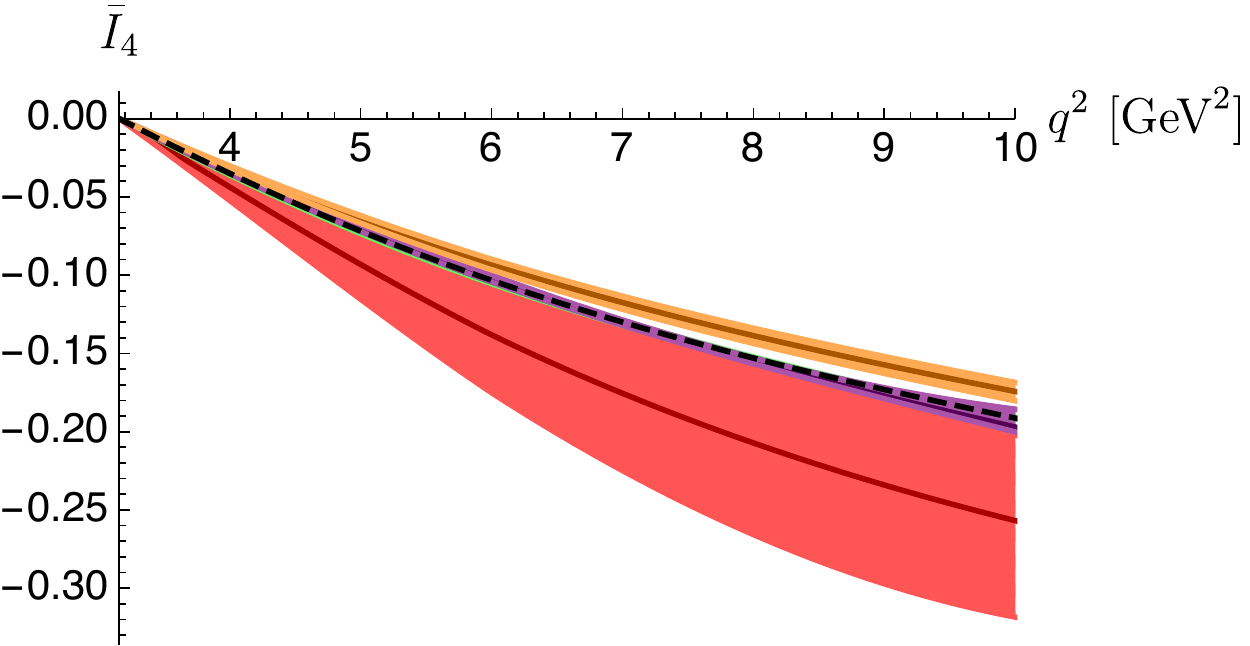}
   \hskip 1.5cm \includegraphics[scale=0.4]{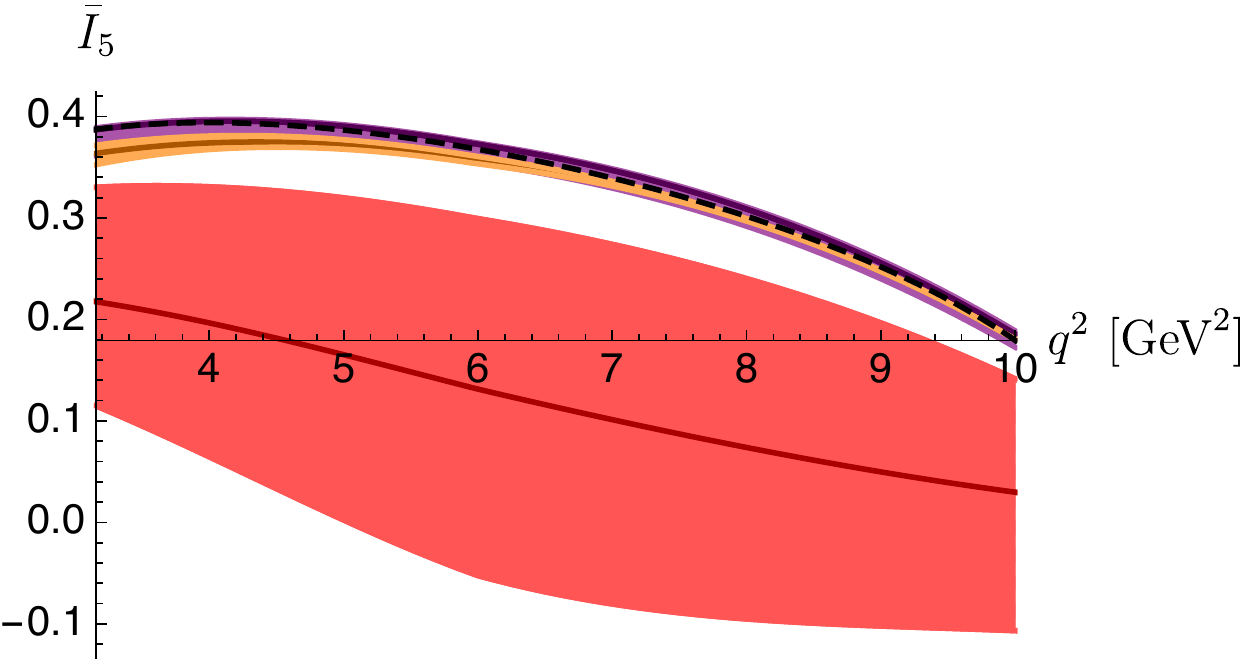}\hskip .35cm
    \includegraphics[scale=0.4]{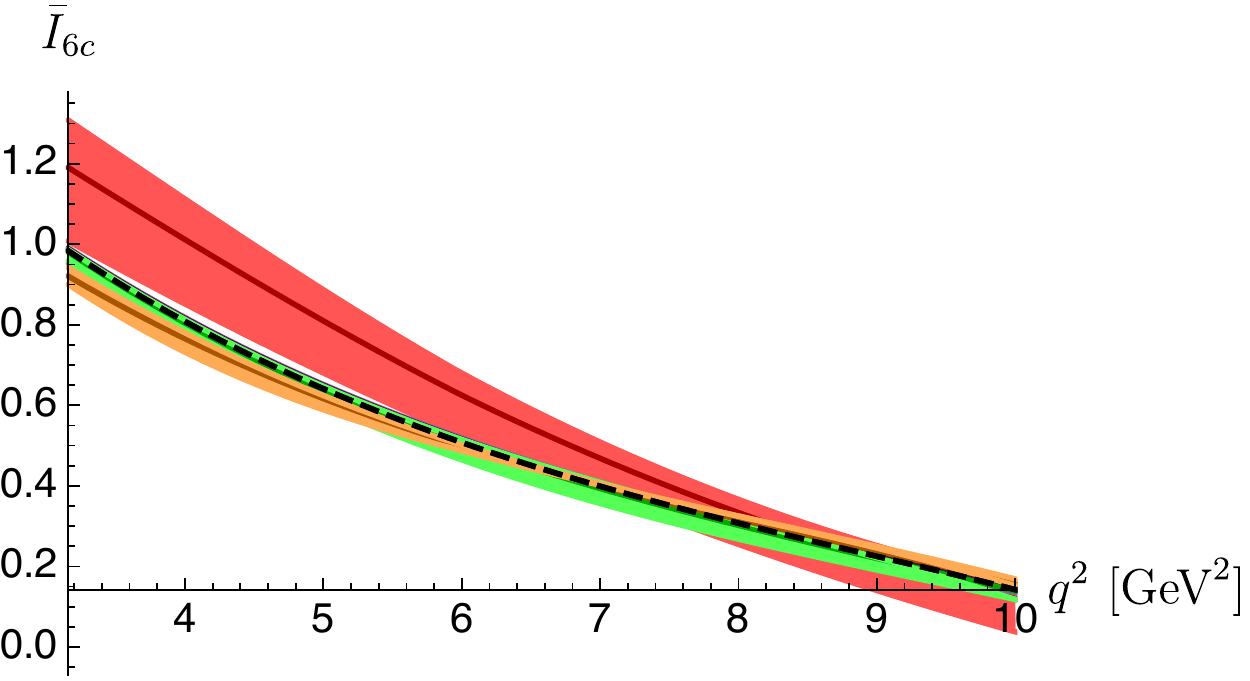}\hskip .2cm
    \includegraphics[scale=0.4]{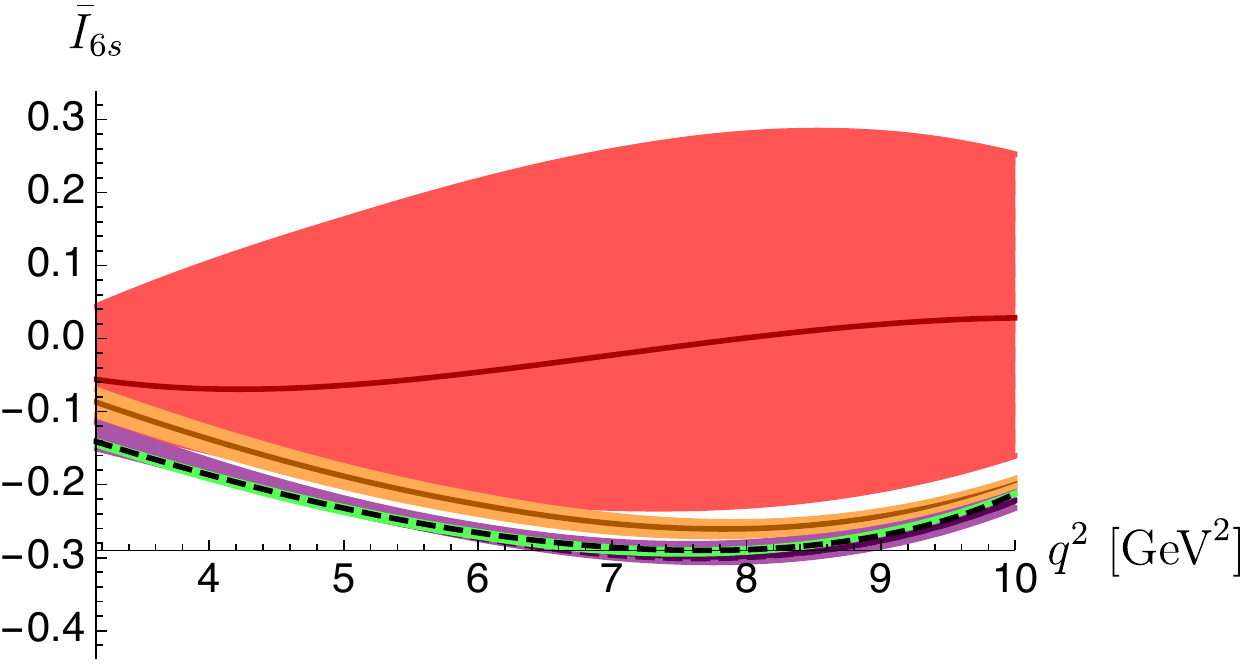}\hskip .2cm
    \caption{$\bar {I}_{i}$, defined in Eq.~\eqref{eq:Inorm}, for different scenarios. 
    Same colour legend as in Fig.~\ref{fig:AFB_FLD}.}
    \label{fig:Is_errorWilson}
\end{figure}

In Fig.~\ref{fig:Is_errorWilson} we plot the $B\to D^* \tau \bar \nu$ angular coefficients, as functions of $q^2$, normalized by the decay width:
\begin{equation}
    \bar{I}_i(q^2)\, \equiv\, \frac{I_i(q^2)}{\Gamma_f(q^2)} \, .
    \label{eq:Inorm}
\end{equation}
The CP-odd quantities $I_7$, $I_8$ and $I_9$ are identically zero in our case, because
we have only considered real Wilson coefficients in our fits. It is interesting to notice that despite the large uncertainties \textit{Scenario 1, Min 1} can be easily  distinguished from the SM predictions and from other minima (for instance looking at $\bar{I}_{1s}$ or $\bar{I}_5$). However, being able to distinguish other scenarios would be more complicated, unless the current errors on the Wilson coefficients are sizable reduced. There is always an overlap between the SM predictions, \textit{Scenario 7a} and \textit{Scenario 5a}.  \textit{Scenario 6} is close to \textit{Scenarios 5a, 7a} and the SM predictions, but it is still possible to distinguish it looking at low  ($\bar{I}_{1s}$, $\bar{I}_{5}$) or high  ($\bar{I}_{2s}$, $\bar{I}_{2c}$, $\bar{I}_3$ and $\bar{I}_4$) $q^2$ values.

Using the symmetries of the angular distribution, Ref.~\cite{Alguero:2020ukk} has proposed an alternative measurement of $F_L^{D^*}(q^2)$, which is only valid in (CP-conserving) scenarios without tensor couplings. In those scenarios, a difference between the two measurements would signal the presence of RHN contributions~\cite{Alguero:2020ukk}.

\section{Conclusions}
\label{sec:conclu}

Using an EFT approach, we have explored the impact of various NP operators on the recently observed anomalies in $b\to c  \tau \bar{\nu}$ transitions. In particular, the focus of this work has been to identify the role of NP operators which can arise due to the presence of RHN in the theory. This has been achieved through a global-fit analysis of all available $b\to c  \tau \bar{\nu}$ data until date: $\rdrdst{}, \ptauint{}, \FLint{}$ and the $q^2$ differential distributions of $B \to D^{(*)}$. Previous analyses only studied the integrated rates and did not include the polarization information ($\ptauint{}, \FLint{}$) and the $q^2$ distributions measured by the BaBar and Belle collaborations, which play an important role in discarding many proposed NP explanations.

We have also studied the differential $B\to D\tau\bar\nu$ decay distribution and have
derived the full four-body angular distribution of the decay $\bar B \to D^* (\to D \, \pi) \tau \bar \nu$, for the most general dimension-six effective Hamiltonian, which includes
(axial)vector, (pseudo)scalar and tensor operators for both the left- and right-handed leptonic currents. The rich dynamical information embodied in the coefficients of these angular distributions could be, in principle, experimentally accessed. From these distributions, we have constructed different observables and have analysed their predicted values within the NP scenarios emerging from our fits. In the next few paragraphs, we briefly summarize the key findings of our analysis.

NP contributions have been assumed to be present only in operators involving charged leptons of the third generation, which is well justified since potential NP effects in $b\to c \, \ell \, \bar{\nu}$ transitions ($\ell = e,\mu$) are known to be negligible~\cite{Jung:2018lfu}. The NP couplings have been also assumed to be real, due to the absence of any evidence of $CP$ violation in these channels. After investigating the separate impact of individual Wilson coefficients, we have performed multi-dimensional fits to the data within eleven different scenarios. The first and the second case include all five RHN operators with and without a SM-like NP contribution, respectively, whereas the 
remaining scenarios correspond to `simplified models' obtained by integrating a single mediator above the EW scale: namely, a scalar boson $\Phi$, a vector boson  $V^\mu$, two scalar leptoquarks $S_1$ and $\tilde{R}_2$, and two vector leptoquarks $U_1^\mu$ and $\tilde{V}_2^\mu$. In those cases where the tree-level exchange of a mediator generates both $\nu_L$ and $\nu_R$ operators, we have further analysed two model variants with and without the 
$\nu_L$ contributions.

Among all scenarios analysed, the vector boson $V^\mu$ (\textit{Scenario 3}) seems to be the preferred option, in terms of the pulls from the SM hypothesis, as shown in Table~\ref{tab:pulls}. The next two possibilities are the scalar leptoquark $S_1$ (\textit{Scenario 7}) and the vector leptoquark $U_1^\mu$ (\textit{Scenario 5}), switching on the RHN couplings only,  which can also provide good agreement to the data. However, it is important to note that none of these three possibilities can generate values of the longitudinal $D^*$ polarization within its current $1\sigma$ experimental range; they can only reach agreement with the $ \FLint{}$ measurement at the $2\sigma$ level. Interestingly, the $ \FLint{}$ data can only be explained at $1\sigma$ in very few cases, namely, with all RHN operators plus the SM-like contribution (\textit{Scenario 1}), or with a scalar boson $\Phi$, switching on both $\nu_L$ and $\nu_R$ operators (\textit{Scenario 4b}) and with a relaxed upper limit of 30\% on $\mathcal{B}(B_c \to \tau \bar{\nu})$. However, these scenarios are not the best choices in explaining the $\rdrdst$ measurements in terms of pull, as reflected in Table~\ref{tab:pulls}. Nevertheless, they do reduce the $\rdrdst$ deviation significantly, and bear very important information about simultaneous agreement of all observables considered in this work. Due to the large uncertainty of the current $\ptauint{}$ measurement, all scenarios are compatible (within $\pm 1 \sigma$) with it. The $\mathcal{R}_D$ measurement is also easily accommodated in all the NP scenarios that we have analysed.

Measurements of additional observables such as polarizations and angular distributions could help to disentangle the dynamical origin of the current anomalies. In particular, we have displayed the information contained in the three-body and four-body angular distributions of $B\to D\tau\bar\nu$ and 
$B\to D^* (\to D\pi) \tau \bar{\nu}$, respectively, 
and their sensitivity to the different NP scenarios analysed. The experimental measurement of these distributions is of course very challenging because of the presence of undetected neutrinos, and one would need to further analyse the decay products of the tau in order to recover the accessible information.

\section*{Acknowledgements}

This work has been supported in part by the Spanish Government and ERDF funds from
the EU Commission [Grant FPA2017-84445-P] and the Generalitat Valenciana [Grant Prometeo/2017/053].
The work of Clara Murgui has been supported by a La Caixa--Severo Ochoa scholarship.
The work of Ana Pe\~nuelas is funded by Ministerio de Ciencia, Innovación y Universidades, Spain [Grant FPU15/05103].
Rusa Mandal also acknowledges the support from the Alexander von Humboldt Foundation through a postdoctoral research fellowship.

\appendix

\section{Form factors }
\label{app:FF}

The hadronic matrix elements can be parametrized by showing explicitly their Lorentz structure as~\cite{Sakaki:2013bfa,Murgui:2019czp}
\begin{eqnarray}
\langle D(p_D) | \bar{c} \gamma_\mu b | \bar{B} (p_B)\rangle &=& \left[ (p_B+p_D)_\mu - \frac{m_B^2-m_D^2}{q^2}q_\mu \right] F_1(q^2) + q_\mu \frac{m_B^2-m_D^2}{q^2} \, F_0(q^2)\, , 
\no \\
\langle D(p_D) |\bar{c}b| \bar{B}(p_B)\rangle &=& \frac{m_B^2-m_D^2}{m_b-m_c} \, F_0(q^2)\, , 
\no\\
\langle D(p_D) | \bar{c} \gamma_\mu \gamma_5 b | \bar{B} (p_B) \rangle &=& \langle D(p_D) | \bar{c} \gamma_5 b | \bar{B} (p_B) \rangle = 0\, , 
\no \\
\langle D (p_D) | \bar{c} \sigma_{\mu \nu} b | \bar{B} (p_B) \rangle &=& -i(p_{B\mu} \, p_{D\nu} - p_{D\mu}\, p_{B\nu})\, \frac{2 F_T(q^2)}{m_B+m_D}\, , 
\no \\
\langle D(p_D) | \bar{c} \sigma_{\mu \nu} \gamma_5 b | \bar{B} (p_B)\rangle &=& -\epsilon_{\mu \nu \alpha \beta}\, p_B^\alpha \, p_D^\beta \, \frac{2F_T(q^2)}{m_B+m_D}\, ,
\end{eqnarray}
for the process involving $B \to D$, and the $B \to D^*$ hadronic matrix elements as
\begin{eqnarray}
\langle D^*(p_{D^*},\lambda_M) | \bar{c} \gamma_\mu  b | \bar{B}(p_B) \rangle &=& -i \, \epsilon_{\mu \nu \rho \sigma} \, \epsilon^{\nu*}(\lambda_M) \, p_B^\rho \, p_{D^*}^\sigma \frac{2 \, V(q^2)}{m_B+m_{D^*}}\, ,
\no \\
\langle D^*(p_{D^*},\lambda_M) | \bar{c} \gamma_\mu \gamma_5 b | \bar{B} (p_B) \rangle &=& (m_B + m_{D^*})A_1(q^2) \left( \epsilon_\mu^* (\lambda_M) - q_\mu \frac{(\epsilon^*(\lambda_M)\cdot q)}{q^2}\right) \no \\ 
&&+ q_\mu (\epsilon^* (\lambda_M) \cdot q) \frac{2m_{D^*}}{q^2}A_0(q^2) \no \\
&& -\frac{\epsilon^* (\lambda_M) \cdot q}{m_B + m_{D^*}}A_2(q^2) \left((p_B+p_{D^*})_\mu -q_\mu \frac{m_B^2-m_{D^*}^2}{q^2} \right) , 
\no \\
\langle D^*(p_{D^*},\lambda_M) | \bar{c} b | \bar{B} (p_B) \rangle &=& 0\, , 
\no \\
\langle D^*(p_{D^*},\lambda_M) | \bar{c} \gamma_5 b | \bar{B} (p_B) \rangle &=& -(\epsilon^* (\lambda_M) \cdot q) \frac{2m_{D^*}}{m_b+m_c}A_0(q^2)\, , 
\no \\
\langle D^* (p_{D^*},\lambda_M) | \bar{c} \sigma_{\mu \nu} b | \bar{B}(p_B) \rangle &=& \epsilon_{\mu \nu \rho \sigma}\left \{ -\epsilon^{\rho*}(\lambda_M)(p_B+p_{D^*})^\sigma T_1(q^2)\right. \no \\
&&+ 2 \frac{(\epsilon^* (\lambda_M)\cdot q)}{q^2}p_B^\rho \, p_{D^*}^\sigma \left( T_1(q^2)-T_2(q^2) - \frac{q^2}{m_B^2-m_{D^*}^2}T_3(q^2)\right) \no \\
&& \left.+ \epsilon^{*\rho}(\lambda_M) q^\sigma \frac{m_B^2-m_{D^*}^2}{q^2}(T_1(q^2)-T_2(q^2))  \right\}.
\end{eqnarray}

The FFs $F_0(q^2)$, $F_1(q^2)$ and $F_T(q^2)$ appearing in the $B \to D$ matrix elements are defined as
\begin{eqnarray}
F_1(q^2) &=& \frac{1}{2\sqrt{m_B m_D}}\,\big[ (m_B + m_D)\, h_+ (q^2) - (m_B - m_D)\, h_- (q^2)\big] \, ,\no \\
F_0(q^2) &=& \frac{1}{2\sqrt{m_B m_D}}\left[\frac{(m_B+m_D)^2-q^2}{m_B + m_D}\, h_+ (q^2)- \frac{(m_B-m_D)^2-q^2}{m_B - m_D}\, h_- (q^2)\right] ,\\
F_T(q^2)&=& \frac{m_B + m_D}{2\sqrt{m_B m_D}}\, h_T(q^2) \, , \no
\end{eqnarray}
while the $B \to D^*$ helicity amplitudes involve the following FFs for vector, axial and pseudoscalar currents, 
\begin{eqnarray}
V(q^2) &=& \frac{m_B+m_{D^*}}{2\sqrt{m_B m_{D^*}}}\, h_V(q^2), 
\no\\
A_1(q^2) &=& \frac{(m_B + m_{D^*})^2-q^2}{2\sqrt{m_B m_{D^*}}(m_B + m_{D^*})}\, h_{A_1}(q^2), 
\no\\
A_2(q^2)&=& \frac{m_B +m_{D^*}}{2\sqrt{m_B m_{D^*}}} \left[ h_{A_3}(q^2) + \frac{m_{D^*}}{m_B}h_{A_2}(q^2)\right],  
\\
A_0(q^2)&=& \frac{1}{2\sqrt{m_B m_{D^*}}} \left[ \frac{(m_B+m_{D^*})^2-q^2}{2m_{D^*}}\, h_{A_1}(q^2)-\frac{m_B^2-m_{D^*}^2+q^2}{2m_B}\, h_{A_2}(q^2)
\right. \no\\ &&\hskip 1.9cm\left. \mbox{}
-\frac{m_B^2-m_{D^*}^2-q^2}{2m_{D^*}}\, h_{A_3}(q^2)\right] ,
\no
\end{eqnarray}
and for the tensor matrix elements,
\begin{eqnarray}
T_1(q^2)&=& \frac{1}{2\sqrt{m_B m_{D^*}}}\left[ (m_B + m_{D^*})\, h_{T_1}(q^2)-(m_B - m_{D^*})\, h_{T_2}(q^2) \right] , 
\no \\
T_2(q^2) &=& \frac{1}{2\sqrt{m_B m_{D^*}}}\left[ \frac{(m_B + m_{D^*})^2-q^2}{m_B + m_{D^*}}\, h_{T_1}(q^2)-\frac{(m_B-m_{D^*})^2-q^2}{m_B - m_{D^*}}\, h_{T_2}(q^2)\right] , 
\\
T_3(q^2)&=&\frac{1}{2\sqrt{m_B m_{D^*}}} \left[ (m_B-m_{D^*})\, h_{T_1}(q^2)-(m_B+m_{D^*})\, h_{T_2}(q^2)-2\,\frac{m_B^2-m_{D^*}^2}{m_B}\, h_{T_3}(q^2)\right] .
\no
\end{eqnarray}
The reduced functions $\hat h_i(q^2) = h_i(q^2)/\xi(q^2)$ take the form \cite{Bernlochner:2017jka}
\beqn
\no
\hat{h}_{+} &=& 1 + \hat{\alpha}_s \left[ C_{V_1} + \frac{\omega + 1}{2} \left( C_{V_2} +  C_{V_3} \right) \right] + \left( \varepsilon_c + \varepsilon_b \right)  \hat{L}_1 \, , \\ \no
\hat{h}_{-} &=&  \hat{\alpha}_s \, \frac{\omega + 1}{2}  \left( C_{V_2} - C_{V_3} \right) + \left( \varepsilon_c - \varepsilon_b \right) \hat{L}_4 \, , \\ 
\hat{h}_{T} &=& 1 +  \hat{\alpha}_s  \left(C_{T_1} - C_{T_2} + C_{T_3} \right) + \left( \varepsilon_c + \varepsilon_b \right) \left( \hat{L}_1 - \hat{L}_4 \right) , 
\eeqn
for $B \to D$, and
\beqn
\no
\hat{h}_{V} &=&  1 + \hat{\alpha}_s C_{V_1} + \varepsilon_c \left( \hat{L}_2 - \hat{L}_5 \right) + \varepsilon_b \left(  \hat{L}_1 - \hat{L}_4 \right)  , \\ \no
\hat{h}_{A_1} &=&  1 + \hat{\alpha}_s  C_{A_1} + \varepsilon_c \left( \hat{L}_2 - \hat{L}_5\, \frac{\omega -1}{\omega + 1} \right) + \varepsilon_b \left(  \hat{L}_1 - \hat{L}_4\, \frac{\omega -1}{\omega + 1} \right) , \\ \no
\hat{h}_{A_2} &=&   \hat{\alpha}_s  C_{A_2} + \varepsilon_c \left( \hat{L}_3 + \hat{L}_6 \right) , \\ \no
\hat{h}_{A_3} &=&  1 + \hat{\alpha}_s \left( C_{A_1} + C_{A_3} \right) + \varepsilon_c \left( \hat{L}_2 - \hat{L}_3 + \hat{L}_6 - \hat{L}_5 \right) + \varepsilon_b \left(  \hat{L}_1 - \hat{L}_4 \right)  , \\
\hat{h}_{T_1} &=&  1 + \hat{\alpha}_s \left[ C_{T_1} + \frac{\omega -1 }{2} \left(C_{T_2} - C_{T_3} \right)  \right] + \varepsilon_c \hat{L}_2 +  \varepsilon_b \hat{L}_1 \, , \\ \no
\hat{h}_{T_2} &=&  \hat{\alpha}_s \, \frac{\omega +1}{2} \left(C_{T_2} + C_{T_3} \right) + \varepsilon_c \hat{L}_5 -  \varepsilon_b \hat{L}_4 \, , \\ \no
\hat{h}_{T_3} &=&   \hat{\alpha}_s C_{T_2} +  \varepsilon_c \left( \hat{L}_6 - \hat{L}_3 \right) ,
\eeqn
for $B \to D^*$. The explicit expressions of the $\omega(q^2)$-dependent factors $\hat L_{1...6}$
and the $\mathcal{O}(\alpha_s)$ corrections $C_i$ can be found in Ref.~\cite{Bernlochner:2017jka}. Note that corrections of order $\Lambda^2_\text{QCD}/m^2_c$ are included via the subleading Isgur-Wise functions $l_{1,2}(\omega)$. The detailed parametrization of the different FFs can be found in Ref.~\cite{Bernlochner:2017jka,Jung:2018lfu}.

\section{Helicity Amplitudes}
\label{app:HelAmp}
We compile here the whole set of relevant helicity amplitudes, following the standard formalism for semileptonic $B$ decays \cite{Hagiwara:1989cu,Hagiwara:1989zt,Korner:1989qb}. 
\subsection{Leptonic amplitudes}
The leptonic helicity amplitudes are defined as,
\beqn
\no
L_{V \mp A, \lambda}^{\lambda_{\tau}, \LR}
(q^2, \theta_{\tau}, \phi ) &=& \epsilon_{\mu} (\lambda)\, \langle \tau  (\lambda_{\tau})   
\bar{\nu}(\lambda_{\nu})
|\bar{\tau} \gamma^{\mu} (1 \mp \gamma_5) \nu|  0 \rangle \, , \\ \no
L_{S\mp P}^{\lambda_{\tau}, \LR} (q^2, \theta_{\tau}, \phi ) &=&  \langle \tau  (\lambda_{\tau}) \bar{\nu}(\lambda_{\nu})|\bar{\tau} (1 \mp \gamma_5) \nu|  0 \rangle \, , \\ 
L_{T \mp T5, \lambda, \lambda'}^{\lambda_{\tau}, \LR} (q^2, \theta_{\tau}, \phi ) &=& - L_{T \mp T5, \lambda', \lambda}^{\lambda_{\tau}, \LR}\, =\,  -i\, \epsilon_{\mu} (\lambda)\, \epsilon_{\nu} (\lambda')\,   \langle \tau  (\lambda_{\tau}) \bar{\nu}(\lambda_{\nu})|\bar{\tau} \sigma^{\mu \nu} (1 \mp \gamma_5) \nu|  0 \rangle \, ,
\label{eq:lep_helicity_def}
\eeqn
where $\lambda_{\tau}$ denotes the sign of the tau lepton helicity and $\epsilon_{\mu}(\lambda)$ are the polarization vectors of the intermediate virtual boson
($\lambda = t,0,\pm$)
in its rest frame, defined in Appendix A of \cite{Hagiwara:1989cu}.
Notice that the helicity of the neutrino 
is explicitly given in the above equation with the symbols $L$ ($\lambda_\nu=-1/2$) and $R$ ($\lambda_\nu=+1/2$) for left-handed and right-handed neutrinos.

The vectorial leptonic amplitudes for LHNs are given by:
\beqn
\no
L_{V-A, +}^{+, L} (q^2, \theta_{\tau}, \phi ) &=& \sqrt{2}\, m_{\tau} \beta_{\tau}\, \sin \theta_{\tau}\, \mathrm{e}^{- 2 i \phi} \, , \\ \no 
L_{V-A, -}^{+, L} (q^2, \theta_{\tau}, \phi ) &=& -\sqrt{2}\, m_{\tau}  \beta_{\tau}\,  \sin \theta_{\tau} \, , \\ \no
L_{V-A, 0}^{+, L} (q^2, \theta_{\tau}, \phi ) &=& 2\, m_{\tau}   \beta_{\tau}\,  \cos \theta_{\tau} \,\mathrm{e}^{-  i \phi} \, , \\ \no
L_{V-A, t}^{+, L} (q^2, \theta_{\tau}, \phi ) &=& -2\, m_{\tau}   \beta_{\tau}\, \mathrm{e}^{-  i \phi} \, , \\ \no
L_{V-A, \pm}^{-, L} (q^2, \theta_{\tau}, \phi ) &=& \sqrt{2 q^2}\,  \beta_{\tau} \left(1 \pm \cos \theta_{\tau}\right) \mathrm{e}^{ \mp i \phi} \, , \\ \no
L_{V-A, 0}^{-, L} (q^2, \theta_{\tau}, \phi ) &=& - 2\, \sqrt{q^2}\,   \beta_{\tau}\,  \sin \theta_{\tau}  \, , \\
L_{V-A, t}^{-, L} (q^2, \theta_{\tau}, \phi ) &=& 0\, , 
\eeqn
where $\beta_{\tau} = \sqrt{1 - m_{\tau}^2/q^2}$. The scalar leptonic amplitudes for LHNs are:
\beqn
\no
L_{S-P}^{+, L} (q^2, \theta_{\tau}, \phi ) &=& - 2 \sqrt{q^2} \beta_{\tau}\, \mathrm{e}^{ - i \phi} \, , \\
L_{S-P}^{-, L} (q^2, \theta_{\tau}, \phi ) &=& 0\, .
\eeqn
The tensor leptonic amplitudes for LHNs take the form:
\beqn
\no
L_{T-T5, +0}^{+, L} (q^2, \theta_{\tau}, \phi ) &=& \sqrt{2 q^2} \beta_{\tau} \,\sin \theta_{\tau}  \,\mathrm{e}^{ - 2 i \phi} \, , \\ \no 
L_{T-T5, -0}^{+, L} (q^2, \theta_{\tau}, \phi ) &=& \sqrt{2 q^2} \beta_{\tau} \,\sin{\theta_{\tau}} \, , \\ \no 
L_{T-T5, +-}^{+, L} (q^2, \theta_{\tau}, \phi ) &=& -L_{T-T5, 0t}^{+, L} \,  =\, 2 \sqrt{q^2}  \beta_{\tau} \,\cos \theta_{\tau} \,\mathrm{e}^{ - i \phi} \, , \\ \no
L_{T-T5, +t}^{+, L} (q^2, \theta_{\tau}, \phi ) &=& - \sqrt{2 q^2} \beta_{\tau} \,\sin \theta_{\tau} \,\mathrm{e}^{ - 2 i \phi} \, , \\ \no 
L_{T-T5, -t}^{+, L} (q^2, \theta_{\tau}, \phi )&=&  \sqrt{2 q^2} \beta_{\tau} \,\sin \theta_{\tau}  \, ,  \\ \no
L_{T-T5, \pm 0}^{-, L} (q^2, \theta_{\tau}, \phi ) &=& \pm \sqrt{2} m_{\tau} \beta_{\tau} \left( 1 \pm \cos \theta_{\tau} \right)  \mathrm{e}^{ \mp i \phi} \, , \\  \no
L_{T-T5, +-}^{-, L} (q^2, \theta_{\tau}, \phi ) &=& -L_{T-T5, 0t}^{-, L} \,  =\, - 2 m_{\tau}  \beta_{\tau} \,\sin \theta_{\tau}  \, , \\
L_{T-T5, \pm t}^{-, L} (q^2, \theta_{\tau}, \phi ) &=& - \sqrt{2} m_{\tau} \beta_{\tau} \left( 1 \pm \cos \theta_{\tau} \right)  \mathrm{e}^{ \mp  i \phi} \, .
\eeqn

The right-handed vectorial leptonic amplitudes are given by: 
\beqn
\no
L_{V+A, \pm}^{+, R} (q^2, \theta_{\tau}, \phi ) &=& \sqrt{2 q^2 } \beta_{\tau} \left(1 \mp \cos \theta_{\tau} \right)  \mathrm{e}^{ \mp  i \phi} \, , \\ \no 
L_{V+A, 0}^{+, R} (q^2, \theta_{\tau}, \phi ) &=& 2 \sqrt{ q^2 }  \beta_{\tau} \,\sin \theta_{\tau} \, , \\ \no 
L_{V+A, t}^{+, R} (q^2, \theta_{\tau}, \phi ) &=& 0 \, , \\ \no 
L_{V+A, +}^{-, R} (q^2, \theta_{\tau}, \phi ) &=&  \sqrt{2} m_{\tau}  \beta_{\tau}  \,\sin \theta_{\tau} \, , \\ \no 
L_{V+A, -}^{-, R} (q^2, \theta_{\tau}, \phi ) &=& - \sqrt{2} m_{\tau}  \beta_{\tau} \, \sin \theta_{\tau}   \,\mathrm{e}^{ 2 i \phi} \, , \\ \no
L_{V+A, 0}^{-, R} (q^2, \theta_{\tau}, \phi ) &=& 2 m_{\tau}  \beta_{\tau}  \,\cos \theta_{\tau} \,\mathrm{e}^{ i \phi}  \, , \\ \no
L_{V+A, t}^{-, R} (q^2, \theta_{\tau}, \phi ) &=& - 2 m_{\tau}  \beta_{\tau} \,\mathrm{e}^{ i \phi} \,  . 
\eeqn
The scalar leptonic amplitudes for RHNs are given by:
\beqn
\no
L_{S+P}^{+, R} (q^2, \theta_{\tau}, \phi ) &=& 0 \, , \\
L_{S+P}^{-, R} (q^2, \theta_{\tau}, \phi ) &=& -2 \sqrt{q^2}   \beta_{\tau}  \,\mathrm{e}^{ i \phi} \, . 
\eeqn
Finally, the tensor leptonic amplitudes for RHNs are:
\beqn
\no
L_{T+T5, \pm 0}^{+, R} (q^2, \theta_{\tau}, \phi ) &=& \mp \sqrt{2} m_{\tau} \beta_{\tau} \left( 1 \mp \cos \theta_{\tau} \right) \mathrm{e}^{ \mp  i \phi}  \, , \\ \no 
L_{T+T5, +-}^{+, R} (q^2, \theta_{\tau}, \phi ) &=& L_{T+T5, 0t}^{+, R}   =  -2 m_{\tau}  \beta_{\tau}  \,\sin \theta_{\tau}  \, , \\ \no
L_{T+T5, \pm t}^{+, R} (q^2, \theta_{\tau}, \phi ) &=&  - \sqrt{2} m_{\tau}  \beta_{\tau} \left( 1 \mp \cos \theta_{\tau} \right) \mathrm{e}^{ \mp  i \phi} \, , \\ \no
L_{T+T5, + 0}^{-, R} (q^2, \theta_{\tau}, \phi ) &=& - \sqrt{2 q^2}  \beta_{\tau}  \,\sin \theta_{\tau}  \, , \\ \no 
L_{T+T5, +-}^{-, R} q^2, \theta_{\tau}, \phi )   &=& L_{T+T5, 0t}^{-, R} \,  =\,  - 2\sqrt{ q^2}   \beta_{\tau}  \,\cos \theta_{\tau} \,\mathrm{e}^{   i \phi} \, , \\ \no
L_{T+T5, + t}^{-, R} (q^2, \theta_{\tau}, \phi ) &=&  - \sqrt{2 q^2}   \beta_{\tau} \,\sin \theta_{\tau}   \, , \\ \no 
L_{T+T5, - 0}^{-, R} (q^2, \theta_{\tau}, \phi ) &=& - \sqrt{2 q^2}  \beta_{\tau} \,\sin \theta_{\tau} \,\mathrm{e}^{ 2  i \phi}  \, , \\ \no
L_{T+T5, - t}^{-, R} (q^2, \theta_{\tau}, \phi ) &=&  \sqrt{2 q^2}   \beta_{\tau} \,\sin \theta_{\tau} \,\mathrm{e}^{ 2  i \phi}   \, . \\ 
\eeqn

\subsection{Hadronic amplitudes}

The hadronic helicity amplitudes of  $B \to M \tau \bar{\nu}_\tau \  (M=D, D^*)$ transitions, $H_{i, \lambda}^{\lambda_M}$, are defined through the matrix elements \cite{Sakaki:2013bfa}
\beqn
\no
H_{V_{L,R}, \lambda}^{\lambda_M} &=& \epsilon_{\mu}^*(\lambda) \bra{ M(\lambda_M)}  \bar{c} \gamma^{\mu} (1 \mp \gamma_5) b \ket{\bar{B}} , \\ 
H_{S_{L,R}, \lambda}^{\lambda_M} &=&  \bra{ M(\lambda_M)}  \bar{c} \gamma^{\mu} (1 \mp \gamma_5) b \ket{\bar{B}}  , \\ \no
 H_{T_{L,R},\lambda \lambda'}^{\lambda_M} &=& -  H_{T_{L,R},\lambda' \lambda}^{\lambda_M}\, =\, i \epsilon_\mu^*(\lambda) \,\epsilon_\nu^*(\lambda')  \bra{ M(\lambda_M)}  \bar{c}\sigma^{\mu \nu} (1\mp \gamma_5) b \ket{\bar{B}} , \\ \no 
\label{eq:hadampl}
\eeqn
where $\lambda_M$ ($=s$ for $D$ and $0,\pm 1$ for $D^*$)
and $\lambda$ ($=0,\pm 1,t$)
are the helicities of the $D^{(*)}$ meson and the intermediate boson, respectively, in the $B$ rest frame. The amplitudes for $B \to D $ transitions are:  
\beqn
\no
H_{V, 0}^s (q^2) &\equiv& H_{V_L, 0}^s (q^2)\, =\,  H_{V_R, 0}^s (q^2)\,  =\,  \sqrt{\frac{\lambda_D(q^2)}{q^2}}\, F_1(q^2) \, , \\ \no
H_{V, t}^s(q^2) &\equiv& H_{V_L, t}^s(q^2)\, =\, H_{V_R, t}^s(q^2) \, =\,   \frac{m_B^2-m_D^2}{\sqrt{q^2}}\, F_0(q^2)\, , \\ 
H_S^s(q^2) &\equiv& H_{S_L}^s (q^2)\, =\,  H_{S_R}^s (q^2) \,\simeq\, \frac{m_B^2 - m_D^2}{m_b - m_c}\, F_0(q^2) \, , \\ \no
H_T^s(q^2) &=& H_{T_L+-}^s=H_{T_L0t}^s \, =\, -H_{T_R+-}^s=H_{T_R0t}^s \, =\,  - \frac{\sqrt{\lambda_D(q^2)}}{m_B+m_D}\, F_T(q^2) \, ,
\label{eq:hadampD}
\eeqn
and for $B \to D^*$:
\beqn
\no
H_{V, \pm} (q^2) &\equiv& H_{V_L, \pm}^{\pm} (q^2) \, =\,  - H_{V_R, \mp}^{\mp} (q^2) \, =\,  (m_B+m_{D^*})\, A_1(q^2) \mp \frac{\sqrt{\lambda_{D^*} (q^2)}}{m_B + m_{D^*}}\, V(q^2) \, , 
\\ \no
H_{V,0} (q^2) &\equiv &  H_{V_L,0}^0 (q^2) \, =\, -H_{V_R,0}^0 (q^2)
\\ \no
&=& \frac{m_B+m_{D^*}}{2 m_{D^*} \sqrt{q^2}} \left[- (m_B^2-m_{D^*}^2 - q^2)\, A_1(q^2) + \frac{\lambda_{D^*}(q^2)}{(m_B+m_{D^*})^2 } \, A_2(q^2)\right] , 
\\ \no
H_{V, t} (q^2) &\equiv& H_{V_L, t}^0 (q^2) \, =\, -H_{V_R, t}^0 (q^2) \, =\, - \sqrt{\frac{\lambda_{D^*}(q^2)}{q^2}}\, A_0(q^2) \, , 
\\ 
H_S(q^2) &\equiv& H_{S_R}^0(q^2) \, =\, -  H_{S_L}^0(q^2) \,\simeq\, - \frac{\sqrt{\lambda_{D^*}(q^2)}}{m_b+m_c}\, A_0(q^2)\, , 
\\ \no
 H_{T,0}(q^2)&\equiv& H_{T_L 0 t}^0(q^2)\, =\, H_{T_L + - }^0(q^2) \, =\,  -H_{T_R 0 t}^0(q^2) \, =\, H_{T_R + - }^0 (q^2) \\ \no 
 &=& \frac{1}{2 m_{D^*}} \left[ - (m_B^2+3 \, m_{D^*}^2- q^2)\, T_2(q^2) + \frac{\lambda_{D^*}(q^2)}{m_B^2 - m_{D^*}^2}\, T_3(q^2) \right] , \\ \no 
H_{T\pm}(q^2)&\equiv& H_{T_L \pm 0}^{\pm} (q^2) \, =\, \pm H_{T_L\pm t}^{\pm}(q^2) \, =\, \mp H_{T_R \mp t}^{\mp} (q^2) \, =\, - H_{T_R \mp 0}^\mp (q^2) \\ \no
&=&  \frac{1}{\sqrt{q^2}}\left[\pm (m_B^2-m_{D^*}^2)\, T_2(q^2) + \sqrt{\lambda_{D^*}}\, T_1(q^2)\right] .
\label{eq:hadampDstar}                          
\eeqn

\subsection{Total amplitude}
\label{app:totamp}
Amplitudes corresponding to  left and right-handed neutrinos do not interfere since they correspond to different final states. Using the completeness relation of the polarization vectors $\epsilon_{\mu}(\lambda)$,
\begin{equation}
    \sum_{\lambda} \delta_{\lambda}\;\epsilon^*_{\mu}(\lambda)\, \epsilon_{\nu} (\lambda) \, =\, g_{\mu \nu} \qquad \qquad \text{with} \qquad\qquad  \delta_0 = \delta_{\pm} = - \delta_t = -1 \, ,
\end{equation}
the decay amplitudes for the transitions $B\to D^{(*)}(\lambda_M)\,\tau(\lambda_\tau)\,\nu_{X}$ with $X=L,R$ can be written as
\beqn
\mathcal{M}[B\to D^{(*)}(\lambda_M)\,\tau (\lambda_\tau) \,\nu_{X}] &\! = &\! 
\frac{G_F}{\sqrt{2}}\, V_{cb}\,\sum_{A=L,R}\Biggl\{ 
\left( \delta_{AL}\delta_{XL} + C^V_{AX} \right) \,\sum_\lambda \delta_{\lambda}\, H^{\lambda_M}_{V_A,\lambda}\, L_{V\mp A,\lambda}^{\lambda_\tau,X}\Biggr.
\no\\ &&\hskip 1.2cm\Biggl.\mbox{} +\,
C^S_{AX}\; H_{S_A}^{\lambda_M}\, L_{S\mp P}^{\lambda_\tau,X}
\, +\, C^T_{AX}\;\sum_{\lambda,\lambda'} \delta_{\lambda}\, \delta_{\lambda'}\, H^{\lambda_M}_{T_A,\lambda\lambda'}\, L_{T\mp T5,\lambda\lambda'}^{\lambda_\tau,X} \Biggr\}
\no\\ &\!\equiv &\!
\frac{G_F}{\sqrt{2}}\, V_{cb}\; \mathcal{M}^{\lambda_M,\lambda_\tau}_{L,R}\, ,
\eeqn
where $\lambda_\tau =\pm \frac{1}{2}$ denotes the $\tau$ helicity in the rest frame of the $\tau\nu$ pair. For the $D^*$ final state, $\lambda_{M}=0,\pm 1$ is the $D^*$ helicity in the $B$ rest frame, while $\lambda_{M}=s$ labels the corresponding $D$ pseudoscalar meson.
Note that the terms proportional to $C^T_{LR}$ or $C^T_{RL}$ vanish identically.

 The helicity amplitudes for the transition $B\to D\tau\nu$ are defined as
\be 
\mathcal{M}[B\to D\,\tau (\lambda_\tau) \,\nu_{L,R}]\, \equiv\, 
- \sqrt{2}\, G_F\, V_{cb}\;
\mathcal{M}^{\lambda_\tau}_{L,R}\, ,
\ee
where $\lambda_\tau =\pm \frac{1}{2}$ denotes the $\tau$ helicity in the rest frame of the $\tau \bar{\nu}$ pair.

The decay $B\to D\tau \bar{\nu}$ is characterized by four reduced amplitudes $\mathcal{M}^{\lambda_\tau}_{L,R}$, which are given by:
\beqn 
\mathcal{M}^{+\frac{1}{2}}_{L} & = & 
-2\sqrt{q^2}\,\beta_{\tau} \; \mathrm{e}^{-i\phi}\,\left\{\frac{m_\tau}{\sqrt{q^2}}\, \Big[\mathcal{\tilde A}_t^L + \cos{\theta_\tau}\,\mathcal{\tilde A}_0^L\Big] + \mathcal{\tilde A}_S^L - 2\,  \cos{\theta_\tau}\,\mathcal{\tilde A}_T^L
\right\} ,
\no\\
\mathcal{M}^{-\frac{1}{2}}_{L}& = & 2\sqrt{q^2}\,\beta_{\tau} \; \sin{\theta_\tau}\,\left\{ \mathcal{\tilde A}_0^L -  \frac{2m_\tau}{\sqrt{q^2}}\,\mathcal{\tilde A}_T^L\right\} ,
\no\\
\mathcal{M}^{+\frac{1}{2}}_{R} & = & -2\sqrt{q^2}\,\beta_{\tau} \; \sin{\theta_\tau}\,\left\{ \mathcal{\tilde A}_0^R -  \frac{2m_\tau}{\sqrt{q^2}}\,\mathcal{\tilde A}_T^R\right\} ,
\no\\
\mathcal{M}^{-\frac{1}{2}}_{R} & = & -2\sqrt{q^2}\,\beta_{\tau} \; \mathrm{e}^{i\phi}\,\left\{\frac{m_\tau}{\sqrt{q^2}}\, \Big[ \mathcal{\tilde A}_t^R + \cos{\theta_\tau}\,\mathcal{\tilde A}_0^R\Big] + \mathcal{\tilde A}_S^R - 2\,  \cos{\theta_\tau}\,\mathcal{\tilde A}_T^R
\right\} .
\eeqn 

The $q^2$-dependent functions $\mathcal{\tilde A}_\lambda^X$ ($\lambda=0,t,S,T$; $X=L,R$) have been defined in Eq.~\eqref{eq:tildeAfun}. 

Analogously, we express the $B\to D^*\tau \bar{\nu}$  transition in terms of twelve reduced helicity amplitudes $\mathcal{M}^{\lambda_{D^*},\lambda_\tau}_{L,R}$ (six for each neutrino chirality).
For LHNs they take the form:
\beqn 
\mathcal{M}^{+1,+\frac{1}{2}}_L & = & -\sqrt{q^2}\,\beta_{\tau} \; \sin{\theta_\tau}\,\mathrm{e}^{-2i\phi}\,\left\{
\frac{m_\tau}{\sqrt{q^2}}\,\left( \apaL + \apeL\right) - 2\, (\apaTL + \apeTL) \right\} ,
\no\\
\mathcal{M}^{+1,-\frac{1}{2}}_L & = & -\sqrt{q^2}\,\beta_{\tau} \; (1 + \cos{\theta_\tau})\,\mathrm{e}^{-i\phi}\,\left\{
\apaL + \apeL  - \frac{2m_\tau}{\sqrt{q^2}}\, (\apaTL + \apeTL )\right\} ,
\no\\
\mathcal{M}^{0,+\frac{1}{2}}_L & = & 
-2\, m_\tau\,\beta_{\tau}
\;\mathrm{e}^{-i\phi}\,\left\{
\mathcal{A}_{tP}^L + \cos{\theta_\tau}\, \left[ \azeL - \frac{2\sqrt{q^2}}{m_\tau}\, \azeTL\right]
\right\} ,
\no\\
\mathcal{M}^{0,-\frac{1}{2}}_L & = & 2\sqrt{q^2}\,\beta_{\tau} \; \sin{\theta_\tau}\,\left[ \azeL - \frac{2m_\tau}{\sqrt{q^2}}\, \azeTL\right] ,
\no\\
\mathcal{M}^{-1,+\frac{1}{2}}_L & = & \sqrt{q^2}\,\beta_{\tau} \; \sin{\theta_\tau}\,\left\{\frac{m_\tau}{\sqrt{q^2}}\, 
(\apaL-\apeL ) - 2\, (\apaTL-\apeTL )\right\} ,
\no\\
\mathcal{M}^{-1,-\frac{1}{2}}_L & = & -\sqrt{q^2}\,\beta_{\tau} \; (1-\cos{\theta_\tau})\,\mathrm{e}^{i\phi}\,\left\{
\apaL-\apeL - 2\,\frac{m_\tau}{\sqrt{q^2}}\, (\apaTL-\apeTL )\right\} ,
\eeqn 
while the corresponding amplitudes for RHNs are given by:
\beqn 
\mathcal{M}^{+1,+\frac{1}{2}}_R & = & -\sqrt{q^2}\,\beta_{\tau} \; (1-\cos{\theta_\tau})\,\mathrm{e}^{-i\phi}\,\left\{
\apaR+\apeR -2\, \frac{m_\tau}{\sqrt{q^2}}\, (\apaTR+\apeTR )\right\} ,
\no \\
\mathcal{M}^{+1,-\frac{1}{2}}_R & = & -\sqrt{q^2}\,\beta_{\tau} \; \sin{\theta_\tau}\,\left\{\frac{m_\tau}{\sqrt{q^2}}\, 
(\apaR+\apeR ) - 2\, (\apaTR+\apeTR )\right\} ,
\no\\
\mathcal{M}^{0,+\frac{1}{2}}_R & = & -2\sqrt{q^2}\,\beta_{\tau} \; \sin{\theta_\tau}\,\left[ \azeR - \frac{2m_\tau}{\sqrt{q^2}}\, \azeTR\right] ,
\no\\
\mathcal{M}^{0,-\frac{1}{2}}_R & = & -2\,m_\tau\,\beta_{\tau} \;\mathrm{e}^{i\phi}\,\left\{
\mathcal{A}_{tP}^R + \cos{\theta_\tau}\, \left[ \azeR - \frac{2\sqrt{q^2}}{m_\tau}\, \azeTR\right]
\right\} ,
\no\\
\mathcal{M}^{-1,+\frac{1}{2}}_R & = &-\sqrt{q^2}\,\beta_{\tau} \;  (1 + \cos{\theta_\tau})\,\mathrm{e}^{i\phi}\,\left\{
\apaR - \apeR - 2\,\frac{m_\tau}{\sqrt{q^2}}\, (\apaTR - \apeTR )\right\} ,
\no\\
\mathcal{M}^{-1,-\frac{1}{2}}_R & = & \sqrt{q^2}\,\beta_{\tau} \;\sin{\theta_\tau}\,\mathrm{e}^{2i\phi}\,\left\{
\frac{m_\tau}{\sqrt{q^2}}\,\left( \apaR - \apeR\right) - 2\, (\apaTR - \apeTR) \right\} .
\eeqn 
The transversity amplitudes $\mathcal{A}_\lambda^{L,R}$ 
are listed in Eq.~\eqref{eq:BtoDsAs}.

\section[\texorpdfstring{Propagation and decay of $D^*$}{dum6}]{\texorpdfstring{\boldmath Propagation and decay of $D^*$}{dum6}}
\label{app:DtoDpi}

To compute the full four-body decay amplitude $B\to D^* \tau \bar{\nu} \to (D \pi)\,\tau \bar{\nu} $, we need to describe the propagation and the decay of the vector boson $D^*$ to the $D\pi$ final state. 
The $D^*\to D \pi$ amplitude can be parametrized in the form
\be\label{eq:DsDpi} 
\mathcal{M}^{\lambda_{D^*}}_{D^*\to D \pi}\, =\, g_{D^*D\pi}^{\phantom{|}} \; \varepsilon_\mu (\lambda_{D^*})\, p_D^\mu \, ,
\ee 
with an effective coupling $g_{D^*D\pi}^{\phantom{|}}$ that can be determined from the total decay width,
\be 
\Gamma(D^*\to D \pi)\, =\, C\;\frac{\lambda^{3/2}(m_{D^*}^2, m_D^2, m_\pi^2)}{192\pi m_{D^*}^5}\; |g_{D^*D\pi}^{\phantom{|}}|^2\, ,
\ee 
where $C=1$, $\frac{1}{2}$ for a final $\pi^\pm$, $\pi^0$, respectively. The dependence of the effective amplitude \eqref{eq:DsDpi} on the momentum and polarization vectors fixes the angular structure of the three possible helicity amplitudes:
\be\label{eq:DstHelicity}
\mathcal{M}^{0}_{D^*\to D \pi}\, =\, - g_{D^*D\pi}^{\phantom{|}} \; |\vec{p}_D|\,\cos{\theta_D}
\qquad \text{and}\qquad
\mathcal{M}^{\pm 1}_{D^*\to D \pi}\, =\, \pm\frac{1}{\sqrt{2}}\; g_{D^*D\pi}^{\phantom{|}} \; |\vec{p}_D|\,\sin{\theta_D}\, ,
\ee
with $|\vec{p}_D| = \lambda^{1/2}(m_{D^*}^2, m_D^2, m_\pi^2)/(2 m_{D^*})$ being the
three-momentum of the $D$ meson in the $D^*$ rest frame.

The propagation of the $D^*$ can be described through a Breit-Wigner function. 
Since the decay width of the $D^*$ is much smaller than its mass, we can use the narrow-width approximation,
\bea
\label{eq:BW}
\frac{1}{(m_{D\pi}^2-m_{D^*}^2)^2+ m_{D^*}^2 \Gamma_{D^*}^2} \quad\xrightarrow[]{\Gamma_{D^*}\,\ll\, m_{D^*}}\quad \frac{\pi}{m_{D^*} \Gamma_{D^*}}\; \delta(m_{D\pi}^2-m_{D^*}^2)\, ,
\eea
and write the decay probability of the process $B\to (D\pi)\,\tau \bar{\nu}$ in the form
\be 
\left|\mathcal{M}[B\to (D\pi)\,\tau(\lambda_\tau)\,\nu_X]\right|^2 \, =\, \frac{1}{2}\, G_F^2\, |V_{cb}|^2\,\frac{\pi}{m_{D^*} \Gamma_{D^*}}\; \delta(m_{D\pi}^2-m_{D^*}^2)\;
\left|\sum_{\lambda_{D^*}} \mathcal{M}_X^{\lambda_{D^*},\lambda_\tau}\, \mathcal{M}^{\lambda_{D^*}}_{D^*\to D \pi}
\right|^2\, .
\ee 
Notice that the dependence on $g_{D^*D\pi}^{\phantom{|}}$ cancels out from this expression.
The interferences among the unobservable helicity amplitudes of the intermediate $D^*$ meson generate the different dependences on $\theta_D$, appearing in the four-body angular distribution listed in Eq.~\eqref{eq:dGamma}.

\bibliographystyle{utphys}
\bibliography{referencies}

\providecommand{\href}[2]{#2}\begingroup\raggedright\begin{thebibliography}{10}

\bibitem{Pich:2019pzg}
A.~Pich, ``{Flavour Anomalies},''
  \href{http://dx.doi.org/10.22323/1.350.0078}{{\em PoS} {\bfseries LHCP2019}
  (2019) 078},
\href{http://arxiv.org/abs/1911.06211}{{\ttfamily arXiv:1911.06211 [hep-ph]}}.

\bibitem{Bifani:2018zmi}
S.~Bifani, S.~Descotes-Genon, A.~Romero~Vidal, and M.-H. Schune, ``{Review of
  Lepton Universality tests in $B$ decays},''
  \href{http://dx.doi.org/10.1088/1361-6471/aaf5de}{{\em J. Phys.} {\bfseries
  G46} no.~2, (2019) 023001},
\href{http://arxiv.org/abs/1809.06229}{{\ttfamily arXiv:1809.06229 [hep-ex]}}.

\bibitem{Lees:2012xj}
{\bfseries BaBar} Collaboration, J.~P. Lees {\em et~al.}, ``{Evidence for an
  excess of $\bar{B} \to D^{(*)} \tau^-\bar{\nu}_\tau$ decays},''
  \href{http://dx.doi.org/10.1103/PhysRevLett.109.101802}{{\em Phys. Rev.
  Lett.} {\bfseries 109} (2012) 101802},
\href{http://arxiv.org/abs/1205.5442}{{\ttfamily arXiv:1205.5442 [hep-ex]}}.

\bibitem{Lees:2013uzd}
{\bfseries BaBar} Collaboration, J.~P. Lees {\em et~al.}, ``{Measurement of an
  Excess of $\bar{B} \to D^{(*)}\tau^- \bar{\nu}_\tau$ Decays and Implications
  for Charged Higgs Bosons},''
  \href{http://dx.doi.org/10.1103/PhysRevD.88.072012}{{\em Phys. Rev.}
  {\bfseries D88} no.~7, (2013) 072012},
\href{http://arxiv.org/abs/1303.0571}{{\ttfamily arXiv:1303.0571 [hep-ex]}}.

\bibitem{Aaij:2015yra}
{\bfseries LHCb} Collaboration, R.~Aaij {\em et~al.}, ``{Measurement of the
  ratio of branching fractions $\mathcal{B}(\bar{B}^0 \to
  D^{*+}\tau^{-}\bar{\nu}_{\tau})/\mathcal{B}(\bar{B}^0 \to
  D^{*+}\mu^{-}\bar{\nu}_{\mu})$},''
  \href{http://dx.doi.org/10.1103/PhysRevLett.115.159901,
  10.1103/PhysRevLett.115.111803}{{\em Phys. Rev. Lett.} {\bfseries 115}
  no.~11, (2015) 111803}, \href{http://arxiv.org/abs/1506.08614}{{\ttfamily
  arXiv:1506.08614 [hep-ex]}}.
[Erratum: Phys. Rev. Lett. 115, no. 15, 159901 (2015)].

\bibitem{Huschle:2015rga}
{\bfseries Belle} Collaboration, M.~Huschle {\em et~al.}, ``{Measurement of the
  branching ratio of $\bar{B} \to D^{(\ast)} \tau^- \bar{\nu}_\tau$ relative to
  $\bar{B} \to D^{(\ast)} \ell^- \bar{\nu}_\ell$ decays with hadronic tagging
  at Belle},'' \href{http://dx.doi.org/10.1103/PhysRevD.92.072014}{{\em Phys.
  Rev.} {\bfseries D92} no.~7, (2015) 072014},
\href{http://arxiv.org/abs/1507.03233}{{\ttfamily arXiv:1507.03233 [hep-ex]}}.

\bibitem{Hirose:2016wfn}
{\bfseries Belle} Collaboration, S.~Hirose {\em et~al.}, ``{Measurement of the
  $\tau$ lepton polarization and $R(D^*)$ in the decay $\bar{B} \to D^* \tau^-
  \bar{\nu}_\tau$},''
  \href{http://dx.doi.org/10.1103/PhysRevLett.118.211801}{{\em Phys. Rev.
  Lett.} {\bfseries 118} no.~21, (2017) 211801},
\href{http://arxiv.org/abs/1612.00529}{{\ttfamily arXiv:1612.00529 [hep-ex]}}.

\bibitem{Hirose:2017dxl}
{\bfseries Belle} Collaboration, S.~Hirose {\em et~al.}, ``{Measurement of the
  $\tau$ lepton polarization and $R(D^*)$ in the decay $\bar{B} \rightarrow D^*
  \tau^- \bar{\nu}_\tau$ with one-prong hadronic $\tau$ decays at Belle},''
  \href{http://dx.doi.org/10.1103/PhysRevD.97.012004}{{\em Phys. Rev.}
  {\bfseries D97} no.~1, (2018) 012004},
\href{http://arxiv.org/abs/1709.00129}{{\ttfamily arXiv:1709.00129 [hep-ex]}}.

\bibitem{Aaij:2017uff}
{\bfseries LHCb} Collaboration, R.~Aaij {\em et~al.}, ``{Measurement of the
  ratio of the $B^0 \to D^{*-} \tau^+ \nu_{\tau}$ and $B^0 \to D^{*-} \mu^+
  \nu_{\mu}$ branching fractions using three-prong $\tau$-lepton decays},''
  \href{http://dx.doi.org/10.1103/PhysRevLett.120.171802}{{\em Phys. Rev.
  Lett.} {\bfseries 120} no.~17, (2018) 171802},
\href{http://arxiv.org/abs/1708.08856}{{\ttfamily arXiv:1708.08856 [hep-ex]}}.

\bibitem{Aaij:2017deq}
{\bfseries LHCb} Collaboration, R.~Aaij {\em et~al.}, ``{Test of Lepton Flavor
  Universality by the measurement of the $B^0 \to D^{*-} \tau^+ \nu_{\tau}$
  branching fraction using three-prong $\tau$ decays},''
  \href{http://dx.doi.org/10.1103/PhysRevD.97.072013}{{\em Phys. Rev.}
  {\bfseries D97} no.~7, (2018) 072013},
\href{http://arxiv.org/abs/1711.02505}{{\ttfamily arXiv:1711.02505 [hep-ex]}}.

\bibitem{Abdesselam:2019dgh}
{\bfseries Belle} Collaboration, A.~Abdesselam {\em et~al.}, ``{Measurement of
  $\mathcal{R}(D)$ and $\mathcal{R}(D^{\ast})$ with a semileptonic tagging
  method},'' \href{http://arxiv.org/abs/1904.08794}{{\ttfamily arXiv:1904.08794
  [hep-ex]}}.

\bibitem{Aaij:2017tyk}
{\bfseries LHCb} Collaboration, R.~Aaij {\em et~al.}, ``{Measurement of the
  ratio of branching fractions
  $\mathcal{B}(B_c^+\,\to\,J/\psi\tau^+\nu_\tau)$/$\mathcal{B}(B_c^+\,\to\,J/\psi\mu^+\nu_\mu)$},''
  \href{http://dx.doi.org/10.1103/PhysRevLett.120.121801}{{\em Phys. Rev.
  Lett.} {\bfseries 120} no.~12, (2018) 121801},
\href{http://arxiv.org/abs/1711.05623}{{\ttfamily arXiv:1711.05623 [hep-ex]}}.

\bibitem{Amhis:2019ckw}
{\bfseries HFLAV} Collaboration, Y.~S. Amhis {\em et~al.}, ``{Averages of
  $b$-hadron, $c$-hadron, and $\tau$-lepton properties as of 2018},''
  \href{http://arxiv.org/abs/1909.12524}{{\ttfamily arXiv:1909.12524
  [hep-ex]}}.

\bibitem{Bigi:2016mdz}
D.~Bigi and P.~Gambino, ``{Revisiting $B\to D \ell \nu$},''
  \href{http://dx.doi.org/10.1103/PhysRevD.94.094008}{{\em Phys. Rev.}
  {\bfseries D94} no.~9, (2016) 094008},
\href{http://arxiv.org/abs/1606.08030}{{\ttfamily arXiv:1606.08030 [hep-ph]}}.

\bibitem{Bernlochner:2017jka}
F.~U. Bernlochner, Z.~Ligeti, M.~Papucci, and D.~J. Robinson, ``{Combined
  analysis of semileptonic $B$ decays to $D$ and $D^*$: $R(D^{(*)})$,
  $|V_{cb}|$, and new physics},''
  \href{http://dx.doi.org/10.1103/PhysRevD.95.115008,
  10.1103/PhysRevD.97.059902}{{\em Phys. Rev.} {\bfseries D95} no.~11, (2017)
  115008}, \href{http://arxiv.org/abs/1703.05330}{{\ttfamily arXiv:1703.05330
  [hep-ph]}}.
[erratum: Phys. Rev.D97,no.5,059902(2018)].

\bibitem{Bigi:2017jbd}
D.~Bigi, P.~Gambino, and S.~Schacht, ``{$R(D^*)$, $|V_{cb}|$, and the Heavy
  Quark Symmetry relations between form factors},''
  \href{http://dx.doi.org/10.1007/JHEP11(2017)061}{{\em JHEP} {\bfseries 11}
  (2017) 061},
\href{http://arxiv.org/abs/1707.09509}{{\ttfamily arXiv:1707.09509 [hep-ph]}}.

\bibitem{Jaiswal:2017rve}
S.~Jaiswal, S.~Nandi, and S.~K. Patra, ``{Extraction of $|V_{cb}|$ from $B\to
  D^{(*)}\ell\nu_\ell$ and the Standard Model predictions of $R(D^{(*)})$},''
  \href{http://dx.doi.org/10.1007/JHEP12(2017)060}{{\em JHEP} {\bfseries 12}
  (2017) 060},
\href{http://arxiv.org/abs/1707.09977}{{\ttfamily arXiv:1707.09977 [hep-ph]}}.

\bibitem{Jung:2018lfu}
M.~Jung and D.~M. Straub, ``{Constraining new physics in $b\to c\ell\nu$
  transitions},'' \href{http://dx.doi.org/10.1007/JHEP01(2019)009}{{\em JHEP}
  {\bfseries 01} (2019) 009},
\href{http://arxiv.org/abs/1801.01112}{{\ttfamily arXiv:1801.01112 [hep-ph]}}.

\bibitem{Anisimov:1998uk}
A.~{\relax Yu}. Anisimov, I.~M. Narodetsky, C.~Semay, and B.~Silvestre-Brac,
  ``{The $B_c$ meson lifetime in the light front constituent quark model},''
  \href{http://dx.doi.org/10.1016/S0370-2693(99)00273-7}{{\em Phys. Lett.}
  {\bfseries B452} (1999) 129--136},
\href{http://arxiv.org/abs/hep-ph/9812514}{{\ttfamily arXiv:hep-ph/9812514
  [hep-ph]}}.

\bibitem{Kiselev:2002vz}
V.~Kiselev, ``{Exclusive decays and lifetime of $B_c$ meson in QCD sum
  rules},'' \href{http://arxiv.org/abs/hep-ph/0211021}{{\ttfamily
  arXiv:hep-ph/0211021}}.

\bibitem{Ivanov:2006ni}
M.~A. Ivanov, J.~G. Korner, and P.~Santorelli, ``{Exclusive semileptonic and
  nonleptonic decays of the $B_c$ meson},''
  \href{http://dx.doi.org/10.1103/PhysRevD.73.054024}{{\em Phys. Rev.}
  {\bfseries D73} (2006) 054024},
\href{http://arxiv.org/abs/hep-ph/0602050}{{\ttfamily arXiv:hep-ph/0602050
  [hep-ph]}}.

\bibitem{Hernandez:2006gt}
E.~Hernandez, J.~Nieves, and J.~M. Verde-Velasco, ``{Study of exclusive
  semileptonic and non-leptonic decays of $B_c$ - in a nonrelativistic quark
  model},'' \href{http://dx.doi.org/10.1103/PhysRevD.74.074008}{{\em Phys.
  Rev.} {\bfseries D74} (2006) 074008},
\href{http://arxiv.org/abs/hep-ph/0607150}{{\ttfamily arXiv:hep-ph/0607150
  [hep-ph]}}.

\bibitem{Huang:2007kb}
T.~Huang and F.~Zuo, ``{Semileptonic $B_c$ decays and charmonium distribution
  amplitude},'' \href{http://dx.doi.org/10.1140/epjc/s10052-007-0333-4}{{\em
  Eur. Phys. J.} {\bfseries C51} (2007) 833--839},
\href{http://arxiv.org/abs/hep-ph/0702147}{{\ttfamily arXiv:hep-ph/0702147
  [HEP-PH]}}.

\bibitem{Issadykov:2018myx}
A.~Issadykov and M.~A. Ivanov, ``{The decays $B_{c}\to J/\psi+\bar\ell\nu_\ell$
  and $B_{c}\to J/\psi + \pi(K)$ in covariant confined quark model},''
  \href{http://dx.doi.org/10.1016/j.physletb.2018.06.056}{{\em Phys. Lett.}
  {\bfseries B783} (2018) 178--182},
\href{http://arxiv.org/abs/1804.00472}{{\ttfamily arXiv:1804.00472 [hep-ph]}}.

\bibitem{Wang:2008xt}
W.~Wang, Y.-L. Shen, and C.-D. Lu, ``{Covariant Light-Front Approach for $B_c$
  transition form factors},''
  \href{http://dx.doi.org/10.1103/PhysRevD.79.054012}{{\em Phys. Rev.}
  {\bfseries D79} (2009) 054012},
\href{http://arxiv.org/abs/0811.3748}{{\ttfamily arXiv:0811.3748 [hep-ph]}}.

\bibitem{Hu:2019qcn}
X.-Q. Hu, S.-P. Jin, and Z.-J. Xiao, ``{Semileptonic decays $B_c \to
  (\eta_c,J/\psi) l \bar{\nu}_l $ in the "PQCD + Lattice" approach},''
  \href{http://dx.doi.org/10.1088/1674-1137/44/2/023104}{{\em Chin.\ Phys.\ C}
  {\bfseries 44} no.~2, (2020) 023104},
  \href{http://arxiv.org/abs/1904.07530}{{\ttfamily arXiv:1904.07530
  [hep-ph]}}.

\bibitem{Leljak:2019eyw}
D.~Leljak, B.~Melic, and M.~Patra, ``{On lepton flavour universality in
  semileptonic B$_{c}$ $\rightarrow$ $\eta$$_{c}$, J/$\psi$ decays},''
  \href{http://dx.doi.org/10.1007/JHEP05(2019)094}{{\em JHEP} {\bfseries 05}
  (2019) 094}, \href{http://arxiv.org/abs/1901.08368}{{\ttfamily
  arXiv:1901.08368 [hep-ph]}}.

\bibitem{Azizi:2019aaf}
K.~Azizi, Y.~Sarac, and H.~Sundu, ``{Lepton flavor universality violation in
  semileptonic tree level weak transitions},''
  \href{http://dx.doi.org/10.1103/PhysRevD.99.113004}{{\em Phys.\ Rev.\ D}
  {\bfseries 99} no.~11, (2019) 113004},
  \href{http://arxiv.org/abs/1904.08267}{{\ttfamily arXiv:1904.08267
  [hep-ph]}}.

\bibitem{Tran:2018kuv}
C.-T. Tran, M.~A. Ivanov, J.~G. Körner, and P.~Santorelli, ``{Implications of
  new physics in the decays $B_c \to (J/\psi,\eta_c)\tau\nu$},''
  \href{http://dx.doi.org/10.1103/PhysRevD.97.054014}{{\em Phys. Rev.}
  {\bfseries D97} no.~5, (2018) 054014},
\href{http://arxiv.org/abs/1801.06927}{{\ttfamily arXiv:1801.06927 [hep-ph]}}.

\bibitem{Abdesselam:2019wbt}
{\bfseries Belle} Collaboration, A.~Abdesselam {\em et~al.}, ``{Measurement of
  the $D^{\ast-}$ polarization in the decay $B^0 \to D^{\ast
  -}\tau^+\nu_{\tau}$},'' in {\em {10th International Workshop on the CKM
  Unitarity Triangle}}.
\newblock 3, 2019.
\newblock \href{http://arxiv.org/abs/1903.03102}{{\ttfamily arXiv:1903.03102
  [hep-ex]}}.

\bibitem{Murgui:2019czp}
C.~Murgui, A.~Peñuelas, M.~Jung, and A.~Pich, ``{Global fit to $b \to c \tau
  \nu$ transitions},'' \href{http://dx.doi.org/10.1007/JHEP09(2019)103}{{\em
  JHEP} {\bfseries 09} (2019) 103},
\href{http://arxiv.org/abs/1904.09311}{{\ttfamily arXiv:1904.09311 [hep-ph]}}.

\bibitem{Ligeti:2016npd}
Z.~Ligeti, M.~Papucci, and D.~J. Robinson, ``{New Physics in the Visible Final
  States of $B\to D^{(*)}\tau\nu$},''
  \href{http://dx.doi.org/10.1007/JHEP01(2017)083}{{\em JHEP} {\bfseries 01}
  (2017) 083},
\href{http://arxiv.org/abs/1610.02045}{{\ttfamily arXiv:1610.02045 [hep-ph]}}.

\bibitem{Asadi:2018wea}
P.~Asadi, M.~R. Buckley, and D.~Shih, ``{It’s all right(-handed neutrinos): a
  new $W'$ model for the $ {R}_{D^{{\left(\ast \right)}}} $ anomaly},''
  \href{http://dx.doi.org/10.1007/JHEP09(2018)010}{{\em JHEP} {\bfseries 09}
  (2018) 010},
\href{http://arxiv.org/abs/1804.04135}{{\ttfamily arXiv:1804.04135 [hep-ph]}}.

\bibitem{Greljo:2018ogz}
A.~Greljo, D.~J. Robinson, B.~Shakya, and J.~Zupan, ``{$R(D^{(*)})$ from
  $W^{'}$ and right-handed neutrinos},''
  \href{http://dx.doi.org/10.1007/JHEP09(2018)169}{{\em JHEP} {\bfseries 09}
  (2018) 169},
\href{http://arxiv.org/abs/1804.04642}{{\ttfamily arXiv:1804.04642 [hep-ph]}}.

\bibitem{Robinson:2018gza}
D.~J. Robinson, B.~Shakya, and J.~Zupan, ``{Right-handed Neutrinos and
  $R(D^{(*)})$},'' \href{http://dx.doi.org/10.1007/JHEP02(2019)119}{{\em JHEP}
  {\bfseries 02} (2019) 119},
\href{http://arxiv.org/abs/1807.04753}{{\ttfamily arXiv:1807.04753 [hep-ph]}}.

\bibitem{Azatov:2018kzb}
A.~Azatov, D.~Barducci, D.~Ghosh, D.~Marzocca, and L.~Ubaldi, ``{Combined
  explanations of B-physics anomalies: the sterile neutrino solution},''
  \href{http://dx.doi.org/10.1007/JHEP10(2018)092}{{\em JHEP} {\bfseries 10}
  (2018) 092},
\href{http://arxiv.org/abs/1807.10745}{{\ttfamily arXiv:1807.10745 [hep-ph]}}.

\bibitem{Heeck:2018ntp}
J.~Heeck and D.~Teresi, ``{Pati-Salam explanations of the B-meson anomalies},''
  \href{http://dx.doi.org/10.1007/JHEP12(2018)103}{{\em JHEP} {\bfseries 12}
  (2018) 103},
\href{http://arxiv.org/abs/1808.07492}{{\ttfamily arXiv:1808.07492 [hep-ph]}}.

\bibitem{Asadi:2018sym}
P.~Asadi, M.~R. Buckley, and D.~Shih, ``{Asymmetry Observables and the Origin
  of $R_{D^{(*)}}$ Anomalies},''
  \href{http://dx.doi.org/10.1103/PhysRevD.99.035015}{{\em Phys. Rev.}
  {\bfseries D99} no.~3, (2019) 035015},
\href{http://arxiv.org/abs/1810.06597}{{\ttfamily arXiv:1810.06597 [hep-ph]}}.

\bibitem{Babu:2018vrl}
K.~S. Babu, B.~Dutta, and R.~N. Mohapatra, ``{A theory of $R(D^{*}$, D) anomaly
  with right-handed currents},''
  \href{http://dx.doi.org/10.1007/JHEP01(2019)168}{{\em JHEP} {\bfseries 01}
  (2019) 168},
\href{http://arxiv.org/abs/1811.04496}{{\ttfamily arXiv:1811.04496 [hep-ph]}}.

\bibitem{Bardhan:2019ljo}
D.~Bardhan and D.~Ghosh, ``{$B$ -meson charged current anomalies: The
  post-Moriond 2019 status},''
  \href{http://dx.doi.org/10.1103/PhysRevD.100.011701}{{\em Phys. Rev.}
  {\bfseries D100} no.~1, (2019) 011701},
\href{http://arxiv.org/abs/1904.10432}{{\ttfamily arXiv:1904.10432 [hep-ph]}}.

\bibitem{Shi:2019gxi}
R.-X. Shi, L.-S. Geng, B.~Grinstein, S.~Jäger, and J.~Martin~Camalich,
  ``{Revisiting the new-physics interpretation of the $b\to c\tau\nu$ data},''
  \href{http://dx.doi.org/10.1007/JHEP12(2019)065}{{\em JHEP} {\bfseries 12}
  (2019) 065},
\href{http://arxiv.org/abs/1905.08498}{{\ttfamily arXiv:1905.08498 [hep-ph]}}.

\bibitem{Gomez:2019xfw}
J.~D. Gómez, N.~Quintero, and E.~Rojas, ``{Charged current $b \to c \tau
  \bar{\nu}_\tau$ anomalies in a general $W^\prime$ boson scenario},''
  \href{http://dx.doi.org/10.1103/PhysRevD.100.093003}{{\em Phys. Rev.}
  {\bfseries D100} no.~9, (2019) 093003},
\href{http://arxiv.org/abs/1907.08357}{{\ttfamily arXiv:1907.08357 [hep-ph]}}.

\bibitem{Dutta:2017xmj}
R.~Dutta and A.~Bhol, ``{$B_c \to (J/\psi,\,\eta_c)\tau\nu$ semileptonic decays
  within the standard model and beyond},''
  \href{http://dx.doi.org/10.1103/PhysRevD.96.076001}{{\em Phys. Rev. D}
  {\bfseries 96} no.~7, (2017) 076001},
  \href{http://arxiv.org/abs/1701.08598}{{\ttfamily arXiv:1701.08598
  [hep-ph]}}.

\bibitem{Dutta:2017wpq}
R.~Dutta, ``{Exploring $R_D$, $R_{D^{\ast}}$ and $R_{J/\Psi}$ anomalies},''
  \href{http://arxiv.org/abs/1710.00351}{{\ttfamily arXiv:1710.00351
  [hep-ph]}}.

\bibitem{Dutta:2013qaa}
R.~Dutta, A.~Bhol, and A.~K. Giri, ``{Effective theory approach to new physics
  in $b\to u$ and $b \to c$ leptonic and semileptonic decays},''
  \href{http://dx.doi.org/10.1103/PhysRevD.88.114023}{{\em Phys. Rev.}
  {\bfseries D88} no.~11, (2013) 114023},
\href{http://arxiv.org/abs/1307.6653}{{\ttfamily arXiv:1307.6653 [hep-ph]}}.

\bibitem{Dutta:2016eml}
R.~Dutta and A.~Bhol, ``{$b \to (c,u),\tau\nu$ leptonic and semileptonic decays
  within an effective field theory approach},''
  \href{http://dx.doi.org/10.1103/PhysRevD.96.036012}{{\em Phys. Rev. D}
  {\bfseries 96} no.~3, (2017) 036012},
  \href{http://arxiv.org/abs/1611.00231}{{\ttfamily arXiv:1611.00231
  [hep-ph]}}.

\bibitem{Becirevic:2016yqi}
D.~Bečirević, S.~Fajfer, N.~Košnik, and O.~Sumensari, ``{Leptoquark model to
  explain the $B$-physics anomalies, $R_K$ and $R_D$},''
  \href{http://dx.doi.org/10.1103/PhysRevD.94.115021}{{\em Phys. Rev.}
  {\bfseries D94} no.~11, (2016) 115021},
\href{http://arxiv.org/abs/1608.08501}{{\ttfamily arXiv:1608.08501 [hep-ph]}}.

\bibitem{Cline:2015lqp}
J.~M. Cline, ``{Scalar doublet models confront $\tau$ and b anomalies},''
  \href{http://dx.doi.org/10.1103/PhysRevD.93.075017}{{\em Phys. Rev. D}
  {\bfseries 93} no.~7, (2016) 075017},
  \href{http://arxiv.org/abs/1512.02210}{{\ttfamily arXiv:1512.02210
  [hep-ph]}}.

\bibitem{Sakaki:2014sea}
Y.~Sakaki, M.~Tanaka, A.~Tayduganov, and R.~Watanabe, ``{Probing New Physics
  with $q^2$ distributions in $\bar{B} \to D^{(*)} \tau \bar\nu$},''
  \href{http://dx.doi.org/10.1103/PhysRevD.91.114028}{{\em Phys. Rev.}
  {\bfseries D91} no.~11, (2015) 114028},
\href{http://arxiv.org/abs/1412.3761}{{\ttfamily arXiv:1412.3761 [hep-ph]}}.

\bibitem{Freytsis:2015qca}
M.~Freytsis, Z.~Ligeti, and J.~T. Ruderman, ``{Flavor models for $\bar{B} \to
  D^{(*)} \tau \bar{\nu}$},''
  \href{http://dx.doi.org/10.1103/PhysRevD.92.054018}{{\em Phys. Rev.}
  {\bfseries D92} no.~5, (2015) 054018},
\href{http://arxiv.org/abs/1506.08896}{{\ttfamily arXiv:1506.08896 [hep-ph]}}.

\bibitem{Bhattacharya:2016zcw}
S.~Bhattacharya, S.~Nandi, and S.~K. Patra, ``{Looking for possible new physics
  in $B\to D^{(\ast)}\tau\nu_{\tau}$ in light of recent data},''
  \href{http://dx.doi.org/10.1103/PhysRevD.95.075012}{{\em Phys. Rev.}
  {\bfseries D95} no.~7, (2017) 075012},
\href{http://arxiv.org/abs/1611.04605}{{\ttfamily arXiv:1611.04605 [hep-ph]}}.

\bibitem{Celis:2016azn}
A.~Celis, M.~Jung, X.-Q. Li, and A.~Pich, ``{Scalar contributions to $b\to c
  (u) \tau \nu$ transitions},''
  \href{http://dx.doi.org/10.1016/j.physletb.2017.05.037}{{\em Phys. Lett.}
  {\bfseries B771} (2017) 168--179},
\href{http://arxiv.org/abs/1612.07757}{{\ttfamily arXiv:1612.07757 [hep-ph]}}.

\bibitem{Hagiwara:1989cu}
K.~Hagiwara, A.~D. Martin, and M.~Wade, ``{EXCLUSIVE SEMILEPTONIC B MESON
  DECAYS},'' \href{http://dx.doi.org/10.1016/0550-3213(89)90306-4}{{\em Nucl.\
  Phys.\ B} {\bfseries 327} (1989) 569--594}.

\bibitem{Hagiwara:1989zt}
K.~Hagiwara, A.~D. Martin, and M.~F. Wade, ``{The Semileptonic Decays $B \to M
  \tau\nu$ as a Probe of Hadron Dynamics},''
\href{http://dx.doi.org/10.1007/BF01556008}{{\em Z. Phys.} {\bfseries C46}
  (1990) 299}.

\bibitem{Korner:1989qb}
J.~G. Korner and G.~A. Schuler, ``{Exclusive Semileptonic Heavy Meson Decays
  Including Lepton Mass Effects},''
\href{http://dx.doi.org/10.1007/BF02440838}{{\em Z. Phys.} {\bfseries C46}
  (1990) 93}.

\bibitem{Becirevic:2019tpx}
{Be\v{c}irevi\'c, Damir and Fedele, Marco and Ni\v{s}and\v{z}i\'c, Ivan and
  Tayduganov, Andrey}, ``{Lepton Flavor Universality tests through angular
  observables of $\overline{B}\to D^{(\ast)}\ell\overline{\nu}$ decay modes},''
  \href{http://arxiv.org/abs/1907.02257}{{\ttfamily arXiv:1907.02257
  [hep-ph]}}.

\bibitem{Duraisamy:2013pia}
M.~Duraisamy and A.~Datta, ``{The Full $B \to D^{*} \tau^{-} \bar{\nu_\tau}$
  Angular Distribution and CP violating Triple Products},''
  \href{http://dx.doi.org/10.1007/JHEP09(2013)059}{{\em JHEP} {\bfseries 09}
  (2013) 059}, \href{http://arxiv.org/abs/1302.7031}{{\ttfamily arXiv:1302.7031
  [hep-ph]}}.

\bibitem{Duraisamy:2014sna}
M.~Duraisamy, P.~Sharma, and A.~Datta, ``{Azimuthal $B \to D^{*} \tau^{-}
  \bar{\nu_\tau}$ angular distribution with tensor operators},''
  \href{http://dx.doi.org/10.1103/PhysRevD.90.074013}{{\em Phys. Rev.}
  {\bfseries D90} no.~7, (2014) 074013},
\href{http://arxiv.org/abs/1405.3719}{{\ttfamily arXiv:1405.3719 [hep-ph]}}.

\bibitem{Becirevic:2016hea}
D.~Becirevic, S.~Fajfer, I.~Nisandzic, and A.~Tayduganov, ``{Angular
  distributions of $\bar B \to D^{(\ast)}\ell\bar \nu_\ell$ decays and search
  of New Physics},''
  \href{http://dx.doi.org/10.1016/j.nuclphysb.2019.114707}{{\em Nucl. Phys. B}
  {\bfseries 946} (2019) 114707},
  \href{http://arxiv.org/abs/1602.03030}{{\ttfamily arXiv:1602.03030
  [hep-ph]}}.

\bibitem{Alonso:2016gym}
R.~Alonso, A.~Kobach, and J.~Martin~Camalich, ``{New physics in the kinematic
  distributions of $\bar B\to
  D^{(*)}\tau^-(\to\ell^-\bar\nu_\ell\nu_\tau)\bar\nu_\tau$},''
  \href{http://dx.doi.org/10.1103/PhysRevD.94.094021}{{\em Phys. Rev. D}
  {\bfseries 94} no.~9, (2016) 094021},
  \href{http://arxiv.org/abs/1602.07671}{{\ttfamily arXiv:1602.07671
  [hep-ph]}}.

\bibitem{Hill:2019zja}
D.~Hill, M.~John, W.~Ke, and A.~Poluektov, ``{Model-independent method for
  measuring the angular coefficients of $B^0 \to D^{*-} \tau^+ \nu_{\tau}$
  decays},'' \href{http://dx.doi.org/10.1007/JHEP11(2019)133}{{\em JHEP}
  {\bfseries 11} (2019) 133}, \href{http://arxiv.org/abs/1908.04643}{{\ttfamily
  arXiv:1908.04643 [hep-ph]}}.

\bibitem{Aebischer:2019zoe}
J.~Aebischer, T.~Kuhr, and K.~Lieret, ``{Clustering of $\bar B\to
  D^{(*)}\tau^-\bar\nu_\tau$ kinematic distributions with ClusterKinG},''
  \href{http://dx.doi.org/10.1007/JHEP04(2020)007}{{\em JHEP} {\bfseries 04}
  (2020) 007}, \href{http://arxiv.org/abs/1909.11088}{{\ttfamily
  arXiv:1909.11088 [hep-ph]}}.

\bibitem{Blanke:2018yud}
M.~Blanke, A.~Crivellin, S.~de~Boer, T.~Kitahara, M.~Moscati, U.~Nierste, and
  I.~Nišandžić, ``{Impact of polarization observables and $ B_c\to \tau \nu$
  on new physics explanations of the $b\to c \tau \nu$ anomaly},''
  \href{http://dx.doi.org/10.1103/PhysRevD.99.075006}{{\em Phys. Rev.}
  {\bfseries D99} no.~7, (2019) 075006},
\href{http://arxiv.org/abs/1811.09603}{{\ttfamily arXiv:1811.09603 [hep-ph]}}.

\bibitem{Asadi:2019xrc}
P.~Asadi and D.~Shih, ``{Maximizing the Impact of New Physics in $b\rightarrow
  c \tau \nu$ Anomalies},''
  \href{http://dx.doi.org/10.1103/PhysRevD.100.115013}{{\em Phys. Rev. D}
  {\bfseries 100} no.~11, (2019) 115013},
  \href{http://arxiv.org/abs/1905.03311}{{\ttfamily arXiv:1905.03311
  [hep-ph]}}.

\bibitem{Alonso:2016oyd}
R.~Alonso, B.~Grinstein, and J.~Martin~Camalich, ``{Lifetime of $B_c^-$
  Constrains Explanations for Anomalies in $B\to D^{(*)}\tau\nu$},''
  \href{http://dx.doi.org/10.1103/PhysRevLett.118.081802}{{\em Phys. Rev.
  Lett.} {\bfseries 118} no.~8, (2017) 081802},
\href{http://arxiv.org/abs/1611.06676}{{\ttfamily arXiv:1611.06676 [hep-ph]}}.

\bibitem{Beneke:1996xe}
M.~Beneke and G.~Buchalla, ``{The $B_c$ Meson Lifetime},''
  \href{http://dx.doi.org/10.1103/PhysRevD.53.4991}{{\em Phys. Rev.} {\bfseries
  D53} (1996) 4991--5000},
\href{http://arxiv.org/abs/hep-ph/9601249}{{\ttfamily arXiv:hep-ph/9601249
  [hep-ph]}}.

\bibitem{Akeroyd:2017mhr}
A.~G. Akeroyd and C.-H. Chen, ``{Constraint on the branching ratio of $B_c \to
  \tau \bar{\nu}$ from LEP1 and consequences for $R(D^{(*)})$ anomaly},''
  \href{http://dx.doi.org/10.1103/PhysRevD.96.075011}{{\em Phys. Rev.}
  {\bfseries D96} no.~7, (2017) 075011},
\href{http://arxiv.org/abs/1708.04072}{{\ttfamily arXiv:1708.04072 [hep-ph]}}.

\bibitem{Neubert:1993mb}
M.~Neubert, ``{Heavy quark symmetry},''
  \href{http://dx.doi.org/10.1016/0370-1573(94)90091-4}{{\em Phys. Rept.}
  {\bfseries 245} (1994) 259--396},
\href{http://arxiv.org/abs/hep-ph/9306320}{{\ttfamily arXiv:hep-ph/9306320
  [hep-ph]}}.

\bibitem{Manohar:2000dt}
A.~V. Manohar and M.~B. Wise, ``{Heavy quark physics},''
{\em Camb. Monogr. Part. Phys. Nucl. Phys. Cosmol.} {\bfseries 10} (2000)
  1--191.

\bibitem{Boyd:1994tt}
C.~G. Boyd, B.~Grinstein, and R.~F. Lebed, ``{Constraints on form-factors for
  exclusive semileptonic heavy to light meson decays},''
  \href{http://dx.doi.org/10.1103/PhysRevLett.74.4603}{{\em Phys. Rev. Lett.}
  {\bfseries 74} (1995) 4603--4606},
\href{http://arxiv.org/abs/hep-ph/9412324}{{\ttfamily arXiv:hep-ph/9412324
  [hep-ph]}}.

\bibitem{Boyd:1995sq}
C.~G. Boyd, B.~Grinstein, and R.~F. Lebed, ``{Model independent determinations
  of $\bar B \to D l\bar\nu, D^* l\bar\nu$ form-factors},''
  \href{http://dx.doi.org/10.1016/0550-3213(95)00653-2}{{\em Nucl. Phys.}
  {\bfseries B461} (1996) 493--511},
\href{http://arxiv.org/abs/hep-ph/9508211}{{\ttfamily arXiv:hep-ph/9508211
  [hep-ph]}}.

\bibitem{Boyd:1997kz}
C.~G. Boyd, B.~Grinstein, and R.~F. Lebed, ``{Precision corrections to
  dispersive bounds on form-factors},''
  \href{http://dx.doi.org/10.1103/PhysRevD.56.6895}{{\em Phys. Rev.} {\bfseries
  D56} (1997) 6895--6911},
\href{http://arxiv.org/abs/hep-ph/9705252}{{\ttfamily arXiv:hep-ph/9705252
  [hep-ph]}}.

\bibitem{Isgur:1989ed}
N.~Isgur and M.~B. Wise, ``{WEAK TRANSITION FORM-FACTORS BETWEEN HEAVY
  MESONS},''
\href{http://dx.doi.org/10.1016/0370-2693(90)91219-2}{{\em Phys. Lett.}
  {\bfseries B237} (1990) 527--530}.

\bibitem{Bordone:2019vic}
M.~Bordone, M.~Jung, and D.~van Dyk, ``{Theory determination of $\bar{B}\to
  D^{(*)}\ell^-\bar\nu$ form factors at $\mathcal{O}(1/m_c^2)$},''
  \href{http://dx.doi.org/10.1140/epjc/s10052-020-7616-4}{{\em Eur. Phys. J.}
  {\bfseries C80} no.~2, (2020) 74},
\href{http://arxiv.org/abs/1908.09398}{{\ttfamily arXiv:1908.09398 [hep-ph]}}.

\bibitem{Na:2015kha}
{\bfseries HPQCD} Collaboration, H.~Na, C.~M. Bouchard, G.~P. Lepage,
  C.~Monahan, and J.~Shigemitsu, ``{$B \rightarrow D l \nu$ form factors at
  nonzero recoil and extraction of $|V_{cb}|$},''
  \href{http://dx.doi.org/10.1103/PhysRevD.93.119906,
  10.1103/PhysRevD.92.054510}{{\em Phys. Rev.} {\bfseries D92} no.~5, (2015)
  054510}, \href{http://arxiv.org/abs/1505.03925}{{\ttfamily arXiv:1505.03925
  [hep-lat]}}.
[Erratum: Phys. Rev.D93,no.11,119906(2016)].

\bibitem{Lattice:2015rga}
{\bfseries MILC} Collaboration, J.~A. Bailey {\em et~al.}, ``{$B \to D \tau
  \nu$ form factors at nonzero recoil and $|V_{cb}|$ from 2+1-flavor lattice
  QCD},'' \href{http://dx.doi.org/10.1103/PhysRevD.92.034506}{{\em Phys. Rev.}
  {\bfseries D92} no.~3, (2015) 034506},
\href{http://arxiv.org/abs/1503.07237}{{\ttfamily arXiv:1503.07237 [hep-lat]}}.

\bibitem{Bailey:2014tva}
{\bfseries Fermilab Lattice, MILC} Collaboration, J.~A. Bailey {\em et~al.},
  ``{Update of $|V_{cb}|$ from the $\bar{B}\to D^*\ell\bar{\nu}$ form factor at
  zero recoil with three-flavor lattice QCD},''
  \href{http://dx.doi.org/10.1103/PhysRevD.89.114504}{{\em Phys. Rev.}
  {\bfseries D89} no.~11, (2014) 114504},
\href{http://arxiv.org/abs/1403.0635}{{\ttfamily arXiv:1403.0635 [hep-lat]}}.

\bibitem{Harrison:2017fmw}
{\bfseries HPQCD} Collaboration, J.~Harrison, C.~Davies, and M.~Wingate,
  ``{Lattice QCD calculation of the ${{B}_{(s)}\to D_{(s)}^{*}\ell{\nu}}$ form
  factors at zero recoil and implications for ${|V_{cb}|}$},''
  \href{http://dx.doi.org/10.1103/PhysRevD.97.054502}{{\em Phys. Rev.}
  {\bfseries D97} no.~5, (2018) 054502},
\href{http://arxiv.org/abs/1711.11013}{{\ttfamily arXiv:1711.11013 [hep-lat]}}.

\bibitem{Faller:2008tr}
S.~Faller, A.~Khodjamirian, C.~Klein, and T.~Mannel, ``{$B\to D^{(*)}$ Form
  Factors from QCD Light-Cone Sum Rules},''
  \href{http://dx.doi.org/10.1140/epjc/s10052-009-0968-4}{{\em Eur. Phys. J.}
  {\bfseries C60} (2009) 603--615},
\href{http://arxiv.org/abs/0809.0222}{{\ttfamily arXiv:0809.0222 [hep-ph]}}.

\bibitem{Neubert:1992wq}
M.~Neubert, Z.~Ligeti, and Y.~Nir, ``{QCD sum rule analysis of the subleading
  Isgur-Wise form-factor $\chi_2 (v\cdot v')$},''
  \href{http://dx.doi.org/10.1016/0370-2693(93)90728-Z}{{\em Phys. Lett.}
  {\bfseries B301} (1993) 101--107},
\href{http://arxiv.org/abs/hep-ph/9209271}{{\ttfamily arXiv:hep-ph/9209271
  [hep-ph]}}.

\bibitem{Neubert:1992pn}
M.~Neubert, Z.~Ligeti, and Y.~Nir, ``{The Subleading Isgur-Wise form-factor
  $\chi_3 (v\cdot v')$ to order $\alpha_s$ in QCD sum rules},''
  \href{http://dx.doi.org/10.1103/PhysRevD.47.5060}{{\em Phys. Rev.} {\bfseries
  D47} (1993) 5060--5066},
\href{http://arxiv.org/abs/hep-ph/9212266}{{\ttfamily arXiv:hep-ph/9212266
  [hep-ph]}}.

\bibitem{Ligeti:1993hw}
Z.~Ligeti, Y.~Nir, and M.~Neubert, ``{The Subleading Isgur-Wise form-factor
  $\xi_3 (v\cdot v')$ and its implications for the decays $\bar B\to
  D^*\ell\bar\nu$},'' \href{http://dx.doi.org/10.1103/PhysRevD.49.1302}{{\em
  Phys. Rev.} {\bfseries D49} (1994) 1302--1309},
\href{http://arxiv.org/abs/hep-ph/9305304}{{\ttfamily arXiv:hep-ph/9305304
  [hep-ph]}}.

\bibitem{Capdevila:2017iqn}
B.~Capdevila, A.~Crivellin, S.~Descotes-Genon, L.~Hofer, and J.~Matias,
  ``{Searching for New Physics with $b\to s\tau^+\tau^-$ processes},''
  \href{http://dx.doi.org/10.1103/PhysRevLett.120.181802}{{\em Phys. Rev.
  Lett.} {\bfseries 120} no.~18, (2018) 181802},
\href{http://arxiv.org/abs/1712.01919}{{\ttfamily arXiv:1712.01919 [hep-ph]}}.

\bibitem{Capdevila:2018jhy}
B.~Capdevila, U.~Laa, and G.~Valencia, ``{Anatomy of a six-parameter fit to the
  $b\to s \ell^+\ell^-$ anomalies},''
  \href{http://dx.doi.org/10.1140/epjc/s10052-019-6944-8}{{\em Eur.\ Phys.\ J.\
  C} {\bfseries 79} no.~6, (2019) 462},
  \href{http://arxiv.org/abs/1811.10793}{{\ttfamily arXiv:1811.10793
  [hep-ph]}}.

\bibitem{Pati:1974yy}
J.~C. Pati and A.~Salam, ``{Lepton Number as the Fourth Color},''
  \href{http://dx.doi.org/10.1103/PhysRevD.10.275}{{\em Phys.\ Rev.\ D}
  {\bfseries 10} (1974) 275--289}. [Erratum: Phys.Rev.D 11, 703--703 (1975)].

\bibitem{Barbieri:2015yvd}
R.~Barbieri, G.~Isidori, A.~Pattori, and F.~Senia, ``{Anomalies in $B$-decays
  and $U(2)$ flavour symmetry},''
  \href{http://dx.doi.org/10.1140/epjc/s10052-016-3905-3}{{\em Eur. Phys. J.}
  {\bfseries C76} no.~2, (2016) 67},
\href{http://arxiv.org/abs/1512.01560}{{\ttfamily arXiv:1512.01560 [hep-ph]}}.

\bibitem{DiLuzio:2017vat}
L.~Di~Luzio, A.~Greljo, and M.~Nardecchia, ``{Gauge leptoquark as the origin of
  B-physics anomalies},''
  \href{http://dx.doi.org/10.1103/PhysRevD.96.115011}{{\em Phys. Rev.}
  {\bfseries D96} no.~11, (2017) 115011},
\href{http://arxiv.org/abs/1708.08450}{{\ttfamily arXiv:1708.08450 [hep-ph]}}.

\bibitem{Bordone:2017bld}
M.~Bordone, C.~Cornella, J.~Fuentes-Martin, and G.~Isidori, ``{A three-site
  gauge model for flavor hierarchies and flavor anomalies},''
  \href{http://dx.doi.org/10.1016/j.physletb.2018.02.011}{{\em Phys. Lett.}
  {\bfseries B779} (2018) 317--323},
\href{http://arxiv.org/abs/1712.01368}{{\ttfamily arXiv:1712.01368 [hep-ph]}}.

\bibitem{Blanke:2018sro}
M.~Blanke and A.~Crivellin, ``{$B$ Meson Anomalies in a Pati-Salam Model within
  the Randall-Sundrum Background},''
  \href{http://dx.doi.org/10.1103/PhysRevLett.121.011801}{{\em Phys. Rev.
  Lett.} {\bfseries 121} no.~1, (2018) 011801},
  \href{http://arxiv.org/abs/1801.07256}{{\ttfamily arXiv:1801.07256
  [hep-ph]}}.

\bibitem{Calibbi:2017qbu}
L.~Calibbi, A.~Crivellin, and T.~Li, ``{Model of vector leptoquarks in view of
  the $B$-physics anomalies},''
  \href{http://dx.doi.org/10.1103/PhysRevD.98.115002}{{\em Phys. Rev. D}
  {\bfseries 98} no.~11, (2018) 115002},
  \href{http://arxiv.org/abs/1709.00692}{{\ttfamily arXiv:1709.00692
  [hep-ph]}}.

\bibitem{Bernlochner:2020tfi}
F.~U. Bernlochner, S.~Duell, Z.~Ligeti, M.~Papucci, and D.~J. Robinson, ``{Das
  ist der HAMMER: Consistent new physics interpretations of semileptonic
  decays},'' \href{http://arxiv.org/abs/2002.00020}{{\ttfamily arXiv:2002.00020
  [hep-ph]}}.

\bibitem{Alguero:2020ukk}
M.~Algueró, S.~Descotes-Genon, J.~Matias, and M.~Novoa~Brunet, ``{Symmetries
  in $B \to D^* \ell \nu$ angular observables},''
  \href{http://arxiv.org/abs/2003.02533}{{\ttfamily arXiv:2003.02533
  [hep-ph]}}.

\bibitem{Sakaki:2013bfa}
Y.~Sakaki, M.~Tanaka, A.~Tayduganov, and R.~Watanabe, ``{Testing leptoquark
  models in $\bar B \to D^{(*)} \tau \bar\nu$},''
  \href{http://dx.doi.org/10.1103/PhysRevD.88.094012}{{\em Phys. Rev.}
  {\bfseries D88} no.~9, (2013) 094012},
\href{http://arxiv.org/abs/1309.0301}{{\ttfamily arXiv:1309.0301 [hep-ph]}}.

\end{thebibliography}\endgroup

\end{document}